\documentclass{aa}

\usepackage{graphicx}	
\usepackage{bm}	
\usepackage{wasysym}    
\usepackage{natbib}
\usepackage[version=4]{mhchem}
\bibpunct{(}{)}{;}{a}{}{,} 
\usepackage{placeins}

\usepackage{txfonts,textcomp}
\usepackage[colorlinks=true,allcolors=blue]{hyperref}

\newcommand\Msun{\text{M}_{\astrosun}} 
\newcommand\Zsun{\text{Z}_{\astrosun}} 

\newcommand\HI{\ion{H}{I}\xspace} 
\newcommand\HII{\ion{H}{II}\xspace} 
\newcommand\HeI{\ion{He}{I}\xspace} 
\newcommand\HeII{\ion{He}{II}\xspace} 
\newcommand\HeIII{\ion{He}{III}\xspace} 
\newcommand\CI{\ion{C}{I}\xspace} 
\newcommand\CII{\ion{C}{II}\xspace} 

\newcommand\OI{\ion{O}{I}\xspace} 
\newcommand\OII{\ion{O}{II}\xspace} 
\newcommand\arepo{\textsc{arepo}\xspace}
\newcommand\areport{\mbox{\textsc{arepo-rt}}\xspace}
\newcommand\orcid[1]{\protect\href{http://orcid.org/#1}{\protect\includegraphics[height=12pt]{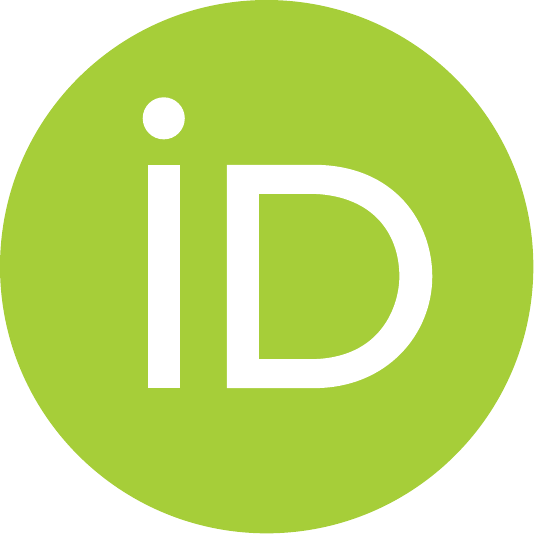}}}
  
\defcitealias{2020MNRAS.495.4170T}{T20}
\defcitealias{2003ApJS..146..417L}{LH03}
\defcitealias{Schaerer02}{S02}
\defcitealias{Gessey-Jones2022}{GJ22}
\defcitealias{2019MNRAS.482.1304E}{E19}
\defcitealias{2017MNRAS.471.2151H}{H17}
\defcitealias{2023MNRAS.522.3092L}{L23}
\defcitealias{2018ApJS..237...13L}{LC18}

\begin{document}

   \title{RIGEL: Simulating dwarf galaxies at solar mass resolution with radiative transfer and feedback from individual massive stars 
   }
   \authorrunning{Y. Deng et al.}
   \titlerunning{RIGEL}

   \subtitle{}

  \author{Yunwei~Deng\inst{1}\orcid{0000-0002-7478-6427}
          \and Hui~Li\inst{1}\orcid{0000-0002-1253-2763}
          \and Boyuan~Liu\inst{2}\orcid{0000-0002-4966-7450}
          \and Rahul~Kannan\inst{3}\orcid{0000-0001-6092-2187}
          \and Aaron~Smith\inst{4}\orcid{0000-0002-2838-9033}
          \and Greg~L.~Bryan\inst{5}\orcid{0000-0003-2630-9228}
          }

   \institute{
             1. Department of Astronomy, Tsinghua University, Haidian DS 100084, Beijing, China\\
             \email{\href{mailto:hliastro@tsinghua.edu.cn}{hliastro@tsinghua.edu.cn}}\\
             2. Institute of Astronomy, University of Cambridge, Madingley Road, Cambridge, CB3 0HA, UK\\
             3. Department of Physics and Astronomy, York University, 4700 Keele Street, Toronto, ON M3J 1P3, Canada\\
             4. Department of Physics, The University of Texas at Dallas, Richardson, Texas 75080, USA\\
             5. Department of Astronomy, Columbia University, New York, NY 10027, USA
             }

   \date{Received XXX XX, XXXX; accepted YYY YY, YYYY}

  \abstract
   {Feedback from stars in the form of radiation, stellar winds, and supernovae is crucial to regulating the star formation activity of galaxies. Dwarf galaxies are especially susceptible to these processes, making them an ideal test bed for studying the effects of stellar feedback in detail. Recent numerical models have aimed to resolve the interstellar medium (ISM) in dwarf galaxies with a very high resolution of several solar masses. However, when it comes to modeling the radiative feedback from stars, many models opt for simplified approaches instead of explicitly solving radiative transfer (RT) because of the computational complexity involved.}
   { We introduce the Realistic ISM modeling in Galaxy Evolution and Lifecycles (RIGEL) model, a novel framework to self-consistently model the effects of stellar feedback in the multiphase ISM of dwarf galaxies with explicit RT on a star-by-star basis.}
   {The RIGEL model integrates detailed implementations of feedback from individual massive stars into the state-of-the-art radiation-hydrodynamics code, \areport.  It forms individual massive stars from the resolved multiphase ISM by sampling the initial mass function and tracks their evolution individually. The lifetimes, photon production rates, mass-loss rates, and wind velocities of these stars are determined by their initial masses and metallicities based on a library that incorporates a variety of stellar models. The RT equations are solved explicitly in seven spectral bins accounting for the infrared to \HeII ionizing bands, using a moment-base scheme with the M1 closure relation. 
   The thermochemistry model tracks the nonequilibrium H, He chemistry as well as the equilibrium abundance of \CI, \CII, \OI, \OII, and CO in the irradiated ISM to capture the thermodynamics of all ISM phases, from cold molecular gas to hot ionized gas.}
   {We evaluated the performance of the RIGEL model using $1\,\Msun$ resolution simulations of isolated dwarf galaxies. We found that the star formation rate (SFR) and interstellar radiation field (ISRF) show strong positive correlations with the metallicity of the galaxy. Photoionization and photoheating can reduce the SFR by an order of magnitude by removing the available cold, dense gas fuel for star formation. The presence of ISRF also significantly changes the thermal structure of the ISM. Radiative feedback occurs immediately after the birth of massive stars and rapidly disperses the molecular clouds within 1\,Myr. As a consequence, radiative feedback reduces the age spread of star clusters to less than 2\,Myr, prohibits the formation of massive star clusters, and shapes the cluster initial mass function to a steep power-law form with a slope of $\sim-2$. The mass-loading factor (measured at $z=1$\,kpc) of the fiducial galaxy has a median of $\eta_M\sim50$, while turning off radiative feedback reduces this factor by an order of magnitude.
   }
   {We demonstrate that RIGEL effectively captures the nonlinear coupling of early radiative feedback and supernova feedback in the multiphase ISM of dwarf galaxies. This novel framework enables the utilization of a comprehensive stellar feedback and ISM model in cosmological simulations of dwarf galaxies and various galactic environments spanning a wide dynamic range in both space and time.}


   \maketitle
%

\section{Introduction}
Star formation on galactic scales is remarkably inefficient \citep{1998ApJ...498..541K,2008AJ....136.2846B,2023ApJ...945L..19S}, with only a small fraction of the total baryonic mass converted into stars in the Universe \citep{2003MNRAS.339.1057Y,2009ApJ...696..620C,2013MNRAS.428.3121M,2018ARA&A..56..435W}. This inefficiency is particularly pronounced in dwarf galaxies ($M_\star<10^9\,\Msun$), which exhibit the highest dark-to-stellar mass ratios \citep[e.g.][]{1997MNRAS.290..533D,1998ARA&A..36..435M,2020ApJ...904...45H:Hayashi,2022MNRAS.517.4714E:Eftekhari}
The low star formation efficiency (SFE) in these low-mass systems is commonly attributed to various physical processes, including radiation, stellar winds, and supernovae (SNe) from massive stars, collectively referred to as ``stellar feedback.'' They deposit mass, momentum, and energy into the cold interstellar medium (ISM), transforming it into a multiphase structure before potentially expelling it into the circumgalactic medium (CGM) or even the intergalactic medium (IGM) \citep[e.g.][]{1977ApJ...218..148M,1986ApJ...303...39D,2011MNRAS.417..950H,2012MNRAS.421.3522H}. 

Stellar feedback reduces galaxy-wide SFE by strongly shaping the evolution of small-scale star-forming regions. Feedback from young massive OB stars rapidly disrupts their natal giant molecular clouds (GMCs) and prevents further gravitational collapse and local star formation \citep[][]{2018MNRAS.475.3511G,2019MNRAS.487..364L,2021ApJ...911..128K}. In particular, pre-SN (early) feedback mechanisms, especially radiation, play a crucial role as they contribute to the feedback budget that is comparable to or even larger than that of SN explosions (\citealt{2013ApJ...770...25A,2014MNRAS.437.3529K,2015MNRAS.448.3248G,2020MNRAS.491.3702H,2021MNRAS.505.3470J}, see \citealt{2024arXiv240319843S:Schinnerer} for a review from the observational perspective). Moreover, recent observations indicate that radiative and stellar wind feedback disperses GMCs on very short timescales, well before SN events occur \citep{2019Natur.569..519K,2022MNRAS.509..272C,2023ASPC..534....1C}, making GMCs short-lived objects where stars and clusters formed within a very short time span of a few million years \citep{2012MNRAS.420.1457H,2015MNRAS.449.1106H,2021MNRAS.504..487K}. 
On a larger scale, stellar feedback drives turbulence, stabilizing the galactic disk against runaway vertical collapse by balancing momentum injection with turbulent dissipation \citep[][]{2013MNRAS.433.1970F,2022ApJ...936..137O}. 
Consequently, the interplay between these local and global effects regulates the galaxy-wide star formation \citep{2011MNRAS.417.1318D,2011MNRAS.417..950H,2017ApJ...845..133S,2017MNRAS.465.1682H} and establishes an extended gas depletion timescale of 1--3\,Gyr in star-forming galaxies \citep[][]{2008AJ....136.2846B,2008AJ....136.2782L,2023ApJ...945L..19S}. 

Over the past few decades, cosmological simulations have been able to reproduce the inefficiency of cosmic star formation histories and various observed galaxy statistics, including the galaxy luminosity function, clustering, color bimodality, and scaling relations between galaxy mass, size, or metallicity hyperplanes (e.g., \citealt{2014MNRAS.444.1518V,2015MNRAS.446..521S,2015MNRAS.454...83W,2018MNRAS.473.4077P,2023MNRAS.524.2594K}, see \citealt{2020NatRP...2...42V} for a review). A crucial factor in achieving these successes is the incorporation of sophisticated 
sub-grid models that approximately capture the complex baryonic physics involved, including stellar feedback originating within the multiphase ISM. However, these models face limitations in resolving the small-scale structure ($\lesssim1$\,kpc) of star-forming regions, as they do not explicitly model the physical processes within the ISM.
While higher-resolution simulations enhance numerical accuracy, they struggle to naturally reproduce the observed complex multiphase ISM structure without incorporating more realistic feedback and ISM models \citep[e.g.][]{2013ApJ...770...25A,2019MNRAS.489.4233M}.  

Recent advances in computational capabilities have enabled simulations of the cold ($\lesssim10^4$\,K) dense ($\gtrsim1$\,cm$^{-3}$) star-forming ISM in galaxies, attempting to resolve the physics including the cooling layers, \HII regions, Sedov-Taylor expansion, dust, and chemical processes, etc., with significantly enhanced fidelity to the underlying stellar feedback mechanisms \citep[e.g.][]{2014MNRAS.445..581H,2018MNRAS.480..800H,2023MNRAS.519.3154H,2015MNRAS.451...34R,2015ApJ...804...18A,2019MNRAS.489.4233M,2020MNRAS.499.5732K}. In these simulations, stellar feedback is typically modeled by averaging stellar properties over a stellar initial mass function (IMF), assuming that the IMF is fully sampled by stellar particles with masses of $\gtrsim10^3\,\Msun$. The feedback contribution of each stellar particle is then determined by the IMF-averaged properties rescaled by its mass. This IMF-averaged method is probably appropriate for massive galaxies, such as the Milky Way (MW). However, in dwarf galaxies, this approach tends to unrealistically ``smear'' the stochastic mass, metal, and energy injections from the rare massive stars in space and time. This leads to an exacerbated ``over-cooling'' problem, overly effective photoionization feedback, and an incorrect baryonic cycle within these smaller galaxies \citep{2018MNRAS.480.1666S,2020MNRAS.492....8A,2021MNRAS.502.5417S}. 

For the above reasons, some state-of-the-art simulations follow the so-called ``star-by-star'' approach that explicitly samples individual massive stars from the IMF and tracks their feedback in simulations of both ISM boxes and entire dwarf galaxies (e.g., \citealt{2016A&A...588A..21R:Revaz}; \citealt{2017MNRAS.471.2151H}, H17; \citealt{2019MNRAS.482.1304E}, E19; \citealt{2020MNRAS.492....8A,2020ApJ...891....2L,2021MNRAS.506.3882S,2021ApJ...920...44H,2021MNRAS.501.5597G,2021MNRAS.504.1039R,2022MNRAS.516.5914C:Calura,2023MNRAS.521.2196A,2023MNRAS.526.1408S}; \citealt{2023MNRAS.522.3092L}, L23). 
Notably, some of these dwarf galaxy simulations have reached unprecedented resolutions, albeit at the cost of modeling radiative feedback through approximate prescriptions. For example, \cite{2016MNRAS.458.3528H} simulated dwarf galaxies with a gas mass of $4\times10^7\,\Msun$ and a resolution of $4\,\Msun$ by assuming a constant far-ultraviolet (FUV) radiation field. \citetalias{2017MNRAS.471.2151H} improved upon their model by adopting optically thin prescriptions for a variable interstellar radiation field (ISRF) and Str\"omgren type approximation for photoionization feedback \citep{2012MNRAS.421.3488H,2018MNRAS.480..800H} to study the nonlinear combination of feedback from radiation and SN explosions. Similar approximate treatments were then further developed and implemented by \cite{2019MNRAS.483.3363H}, \cite{2021MNRAS.506.3882S}, and \cite{2023MNRAS.526.1408S}. Recently, based on the \citetalias{2017MNRAS.471.2151H} feedback model, the GRIFFIN project simulated dwarf galaxies in isolation (\citealt{2022MNRAS.509.5938H}; \citetalias{2023MNRAS.522.3092L}) and mergers \citep{2019ApJ...879L..18L,2020ApJ...891....2L}, providing systematic studies of various numerical and physical effects on the formation and population of star clusters. 

Although these models have demonstrated some successes in reproducing self-regulated star formation, multiphase ISM structures, galactic outflows, and cluster mass functions, their treatments of stellar feedback and ISM thermochemistry remain relatively simplified. Firstly, these models have calibrated effective photoionization feedback with idealized tests of \HII region expansion \citep[e.g.][]{2015MNRAS.453.1324B}. However, it is unclear whether these approximations for radiative feedback are sufficiently accurate in the wide variety of realistic environments encountered in simulations \citep{2015MNRAS.451...34R,2019MNRAS.487.1717F,Kannan2020a}. The lack of long-range radiative effects in these effective models hampers their ability to accurately predict the thermal structure of CGM and IGM, leading to systematic errors in outflow properties and loading factors \citep[e.g.][]{2018ApJ...865L..22E}. To more accurately capture stellar feedback and ISM thermochemistry, it is crucial to solve radiative transfer (RT) equations on the fly. Sophisticated numerical algorithms have been developed to make fully coupled radiation-hydrodynamics computationally feasible \citep[e.g.][]{2008MNRAS.387..295A,2009MNRAS.396.1383P,2011MNRAS.414.3458W,2013MNRAS.436.2188R,2013ApJS..206...21S,2018MNRAS.475.2822J,2019MNRAS.485..117K,2020ApJ...905...27S,Chan2021,Peter2023}. A few models have already incorporated explicit RT solvers that, so far, are only affordably realized for simulations of small dwarf galaxies with a modest resolution ($\gtrsim100\,\Msun$) or limited time coverage ($\lesssim500\,$Myr, e.g., \citetalias{2019MNRAS.482.1304E}; \citealt{2020MNRAS.491.1656A,2022arXiv221104626K,2023MNRAS.525.3806M,2024A&A...681A..28A}). Additionally, these works may be limited by inadequate spatial and temporal resolution, which can result in inaccurate ionization feedback \citep{2024MNRAS.527..478D}. Due to the improved efficiency and accuracy of \areport \citep{2019MNRAS.485..117K,2024MNRAS.527..478D}, now on-the-fly RT simulations are already obtainable for intermediate-sized dwarf galaxies (e.g., $M_\text{gas}=4\times10^7\,\Msun$) with a high resolution of $\sim1\,\Msun$.

In addition, only a few models (e.g., \citetalias{2019MNRAS.482.1304E}) have taken into account the critical aspect of metallicity dependence on stellar properties such as lifetimes, photon production rates, mass-loss rates, and wind velocities. Notably, the ionizing photon production rate of OB stars tends to decrease with increasing metallicity (e.g., \citealt{Schaerer02}, S02; \citealt{2003ApJS..146..417L}, LH03), as metal-poor stars are generally more compact and hotter. Conversely, wind mass-loss rates are known to increase with metallicity \citep[e.g.][]{2001A&A...369..574V,2021MNRAS.504.2051V}, a consequence of boosted radiation pressure due to increased opacity in their atmospheres. These metallicity-dependent factors can introduce orders-of-magnitude differences in the feedback budget of stars formed in either enriched or pristine environments, which must be reflected in cosmological simulations.

Finally, increases in simulation resolution must also be accompanied by more detailed cooling models capable of accurately capturing the multiphase structure of the ISM at smaller scales. The coupling between galactic star formation and outflow properties is highly sensitive to the ISM phase structure and the associated cooling rates, which have a complex dependence on metallicity, ultraviolet (UV) radiation fields, and nonequilibrium chemistry \citep[e.g.][]{2016MNRAS.458..270R}. However, \cite{2023ApJS..264...10K} emphasized that the tabulated cooling models or fitting functions widely used for low-temperature ($<10^4$\,K) cooling in many low-resolution galaxy formation simulations \citep[e.g.][]{1995ApJ...440..634R,2017MNRAS.466.2217S,2018MNRAS.480..800H,2020MNRAS.497.4857P} have significant discrepancies with those developed with detailed atomic or molecular physics by the ISM community \citep[e.g.][]{2003ApJ...587..278W,2017ApJ...843...38G,2019ApJ...881..160B,2023ApJS..264...10K}. Although the former models are computationally efficient and capable of reproducing the cosmic star formation history and galaxy populations, they are inadequate at capturing the thermodynamics of the cold neutral ISM, especially when also modeling strong local radiation fields. 
 
In this paper, we introduce the Realistic ISM modeling in Galaxy Evolution and Lifecycles (RIGEL) model, a self-consistent model for simulating stellar feedback and the multiphase ISM in dwarf galaxies. Our model brings together on-the-fly RT, detailed photochemistry and cooling mechanisms, and metallicity-dependent feedback from individual massive stars. In this model, gravity, hydrodynamics, and the radiation field are explicitly modeled using the radiation hydrodynamics (RHD) code \areport. The photochemistry model integrates nonequilibrium thermochemistry for H and He, along with the equilibrium abundances of key species like \CI, \CII, \OI, and CO, within irradiated ISM contexts. The physically motivated cooling model includes accurate prescriptions for all ISM phases, from hot ionized gas to cold molecular gas. Properties of
massive stars, such as their lifetimes, photon production rates, mass-loss rates, and wind velocities, are determined by their initial masses and metallicities based on a library that incorporates a variety of stellar models. The remainder of the paper is organized as follows. In Section~\ref{sec:RIGEL}, we give an overview of the RIGEL model. In Section~\ref{sec:isolateddwarf}, we present the results of a suite of simulations of an isolated dwarf galaxy, exploring variations in numerical resolution ($1\,\Msun$ and $10\,\Msun$), metallicity ($0.02\,\Zsun$ and $0.2\,\Zsun$), and the effects of turning RT on or off. We discuss the implications and caveats of our model in Section~\ref{sec:discussion}. Lastly, we give a summary in Section~\ref{sec:summary}.
 
\section{The RIGEL model}
\label{sec:RIGEL}
\subsection{Gravity, hydrodynamics, and radiative transfer}
\label{sec:solver}
The simulations presented in this work were performed with the moving-mesh hydrodynamic code \arepo (\citealt{2010MNRAS.401..791S,2016MNRAS.455.1134P}, see \citealt{2020ApJS..248...32W:Weinberger} for the public release \footnote{\url{https://arepo-code.org}}). Gravity was solved with the tree particle-mesh method \citep{1995ApJS...98..355X,2000ApJS..128..561B,2002JApA...23..185B}, whereby the gravitational potential is split into short- and long-range contributions. The short-range gravity was calculated with a hierarchical octree method \citep{1986Natur.324..446B}, and the long-range gravity with the fast-Fourier-transformation method on a Cartesian mesh. 
Hydrodynamics was handled using a quasi-Lagrangian finite-volume method, which employs unstructured meshes generated via a Voronoi tessellation from discrete mesh-generating points. These volume-filling cells drift as they follow the local gas velocity, ensuring a pseudo-Lagrangian solution to the fluid dynamics. We employed a mass refinement scheme that maintains the cell masses around a target gas mass ($m_{\rm gas}$) throughout the simulation domain by refining a cell into two if its mass exceeds twice the target mass or merging a cell with its neighbors if its mass falls below half of the target mass.

The radiation field was explicitly modeled and coupled to gas hydrodynamics via the moment-based RHD solver \areport \citep{2019MNRAS.485..117K}. The RT equations were solved by combining its zeroth and first moments with the M1 closure relation \citep{1984JQSRT..31..149L}. The radiation spectra were discretized into seven broad bands, including
infrared (IR, $0.1-1$\,eV), optical (Opt., $1-5.8$\,eV), far-ultraviolet (FUV, $5.8-11.2$\,eV), Lyman–Warner (LW, $11.2-13.6$\,eV), hydrogen ionizing (EUV1, $13.6-24.6$\,eV), \HeI ionizing (EUV2, $24.6-54.4$\,eV), and \HeII ionizing (EUV3, $54.4-\infty$\,eV) bands, to provide a comprehensive account for the photoionization or dissociation of various species and the dust-radiation coupling  (see Table~\ref{tab:RadiationBins} for a summary and Section~\ref{sec:rad} for details). To avoid extremely small time steps, we use the reduced speed of light approximation \citep{2001NewA....6..437G}, where the selection of a reduced speed of light, $\tilde{c}$, is problem-dependent \citep[e.g.][]{2013MNRAS.436.2188R}. We also employed the spatial resolution correction methods outlined by \cite{2024MNRAS.527..478D} to obtain convergent ionization feedback from the unresolved \HII regions. The coupled system of gravity, hydrodynamics, and RT was integrated using a two-stage second-order Runge-Kutta scheme \citep[i.e., Heun's method,][]{2016MNRAS.455.1134P}. Although omitted in this paper, magnetohydrodynamics implemented by \arepo \citep{2011MNRAS.418.1392P:Pakmor} can also be included in the simulations with RIGEL.

\subsection{Chemistry and cooling model}
\label{sec:chemistry}
Our model for chemistry and cooling closely follows \citet{2020MNRAS.499.5732K}, with an update to the low-temperature ($\lesssim10^4$\,K) cooling by explicitly modeling the abundance of C and O species to better capture the thermodynamics in the warm and cold ISM. We tracked the time-dependent evolution of five primordial species — H, \ce{H+}, \ce{H2}, He, \ce{He+}, and He$^{2+}$ — through \areport. We refer the readers to Section~2.1 in \cite{2020MNRAS.499.5732K} for details of the primordial thermochemical network.
When solving the nonequilibrium equations, we used a semi-implicit time integration approach based on the method outlined in \cite{2009MNRAS.396.1383P} and further improved by applying a photon absorption limiter to correct the temporal resolution issues due to large time steps \citep{2024MNRAS.527..478D}. If the internal energy or one of the abundances changes by more than 10\% during a time step, the equations will be solved implicitly by calling the functions from the SUNDIALS CVODE package \citep{hindmarsh2005sundials}.

We also followed the enrichment and evolution of seven individual metal species including C, N, O, Ne, Mg, Si, and Fe.\footnote{We adopted the (proto-)solar abundances of these elements from \cite{2009ARA&A..47..481A}  \citep[see also][]{2019arXiv191200844L:Lodders,2023MNRAS.519.3154H}, where the mass fractions of H, C, N, O, Ne, Mg, Si, and Fe are $\{0.7154,0.0025,0.0007,0.0061,0.0013,0.0008,0.0007,0.0014\}$. Therefore, the solar metallicity is $\Zsun=0.0136$ by summing up the solar abundances of the last seven elements, and we adopt the solar helium mass fraction $Y =0.2710$.} The metals released by the stars diffused and mixed numerically within the ISM, following the gas motion and mass exchange among gas cells as passive tracers. We defined the (absolute) metallicity, $Z$, of the ISM as the total mass fraction of these seven elements. Unlike \cite{2020MNRAS.499.5732K}, which treats metal cooling as tabulated cooling rates from {\sc cloudy}, we modeled high-temperature ($\gtrsim10^5$\,K) and low-temperature ($\lesssim10^4$\,K) metal cooling separately. High-temperature metal cooling continued to be modeled using a pre-calculated {\sc cloudy} table (see Appendix~\ref{sec:CIE}). For low-temperature metal cooling, we tracked the equilibrium abundances of C, \ce{C+}, CO, O, and \ce{O+} to model the important metal cooling processes including fine structure lines of \CII~$158$\,$\mu$m, \CI$^*$ 610\,$\mu$m, \CI$^*$ 370\,$\mu$m, \OI$^*$ 63\,$\mu$m, and \OI$^*$ 146\,$\mu$m, as well as the CO rotational lines.
The total abundances of these C and O species were directly obtained from the C and O abundances tracked by the gas particles. For dust involved in extinction and chemical reactions, we adopted a constant dust-to-metal ratio of 0.5, and thus $Z_\text{g} = Z_\text{d} = 0.5Z$, where $Z_\text{g}$ and $Z_\text{d}$ are the gas and dust phase metallicity, respectively. Cosmic ray (CR) chemistry was also included by assuming constant ionization and heating rates (equation~\ref{equ:CR}) with 0.01 times the MW value. Specifically, the CR ionization rate for the neutral hydrogen atom is $\zeta_\text{cr,\HI}=1.78\times10^{-18}\,\text{s}^{-1}$, and $\zeta_\text{cr}$ is used for $\zeta_\text{cr, \HI}$ if not specified.
 
To model the cooling by the atomic and molecular lines of C and O species in the warm and cold ISM, we derived the equilibrium abundances of C, \ce{C+}, CO, O, and \ce{O+}, assuming that they are in a formation–destruction balance under the given abundance of primordial species and radiation field tracked by the RT module following the methods outlined by \cite{2023ApJS..264...10K}. For C, we calculated the equilibrium \ce{C+} fraction by
\begin{equation}
    \frac{x_\text{\ce{C+}}}{x_\text{C,tot}}=\frac{\zeta_\text{pi,C}+\zeta_\text{cr,C}}{\zeta_\text{pi,C}+\zeta_\text{cr,C}+\alpha_\text{gr,\ce{C+}}n+\alpha_\text{rr+dr,\ce{C+}}n_e+k_{\ce{C+}-\ce{H2}}n_\text{H2}}\,,
\end{equation}
where we have assumed that $x_\text{C} = x_\text{C,tot} - x_\text{\ce{C+}}$, since $x_\text{CO}$ is negligible in C/\ce{C+} transition regions.

In our model, C is ionized by photoionization and CRs. The photoionization rate, $\zeta_\text{pi,C}$ = $\zeta^\text{Draine}_\text{pi,C}\chi_\text{LW}f_\text{s,C}+520\times2x_\text{\ce{H2}}\zeta_\text{cr}$, where $\chi_\text{LW}$ is the LW band radiation intensity (tracked by \areport) normalized by the \cite{1978ApJS...36..595D} value of $1.3\times10^{-14}$erg\,cm$^{-3}$, $\zeta^\text{Draine}_\text{pi,C}=3.43\times10^{-10}$\,s$^{-1}$ from \cite{1978ApJS...36..595D}, and $f_\text{s,C}$ is the self- and cross-shielding factor of C obtained following \cite{1985ApJ...291..747T} and \cite{2017ApJ...843...38G}. The second term accounts for the ionization by CR-generated LW-band photons in molecular gas, assuming that these photons are absorbed locally by gas \citep{2017A&A...602A.105H}. The CR ionization rate of C is proportional to that of H by $\zeta_\text{cr,C}=3.85\zeta_\text{cr}$. The rate coefficients for the formation channels of C, including gas phase and grain surface recombination ($\alpha_\text{rr+dr,\ce{C+}}$ and $\alpha_\text{gr,\ce{C+}}$) and the reaction of $\ce{C+}+\ce{H2}\rightarrow \text{CH}_2^+$ ($k_{\ce{C+}-\ce{H2}}$), were taken from \cite{2017ApJ...843...38G}.

The \ce{O+} abundance was determined by assuming $x_\text{\ce{O
+}}/x_\text{O,tot}=x_\text{\HII}$ because of the almost identical ionization potentials of O (13.62 eV) and H (13.60 eV) and the  charge exchange between oxygen and hydrogen (see Section~14.7.1 in \citealt{2011piim.book.....D}). Once the $x_\text{\ce{C+}}$ and $x_\text{\ce{O+}}$ were determined, we determined the CO fraction $x_\text{\ce{CO}}$ by
\begin{equation}
\frac{x_\text{\ce{CO}}}{x_\text{C,tot}}=\frac{2x_\text{\ce{H2}}\left[1-\max{\left(x_\text{\ce{C+}}/x_\text{C,tot},x_\text{\ce{O+}}/x_\text{O,tot}\right)}\right]}{1+(n_\text{\ce{CO},crit}/n)^2}\,,
\end{equation}
where $n_\text{\ce{CO},crit}$ is the ``critical density'' above which  ${x_\text{\ce{CO}}}/{x_\text{C,tot}}>0.5$. We adopted the equilibrium  fit in \cite{2017ApJ...843...38G},
\begin{equation}
n_\text{\ce{CO},crit} =(4\times10^3Z'_\text{d}\zeta^{-2}_{\text{cr},16})^{\chi_\text{LW}^{1/3}}\left(\frac{50\zeta_{\text{cr},16}}{{Z'}^{1.4}_\text{d}}\right)\,\text{cm}^{-3}\,,
\end{equation}
where $\zeta_{\text{cr},16}=\zeta_{\text{cr}}/(10^{-16}\text{ s}^{-1})$ and $Z'_\text{d}$ is the dust metallicity normalized by the solar value.

\label{sec:cooling}
Based on the chemical abundance of the main ISM coolants, we can calculate the cooling rate of different cooling channels. In our model, the cooling function is split into eight separate terms: primordial cooling from hydrogen and helium species ($\Lambda_\text{pri}$) tracked by the nonequilibrium chemical network and RT (local and ultraviolet background (UVB) photoionization heating is also incorporated into this term), high-temperature ($\gtrsim10^5$\,K) metal collisional excitation lines cooling tracked by the {\sc cloudy} look-up table ($\Lambda_\text{Z,CIE}$), nebular lines cooling in photoionized gas ($\Lambda_\text{Z,neb}$), equilibrium metal cooling in warm and cold gas ($\Lambda_\text{Z,C/O}$, including the cooling by \CI, \CII, \OI, and \ce{CO} lines), cooling due to dust–gas–radiation interacting ($\Lambda_\text{dust}$), and heating due to photoelectric heating ($\Gamma_\text{PE}$) and CRs ($\Gamma_\text{cr}$). The net cooling rate ($\Lambda_\text{net}$) is then given by
\begin{align}
\Lambda_\text{net}&=\Lambda_\text{pri}(n_j, N_\gamma^i,T)+\frac{Z}{\Zsun}\Lambda_\text{Z,CIE}(T,\rho)\,\notag\\
&+\Lambda_\text{Z,C/O}(n_j,n_\text{C/O},N_\gamma^{\text{LW}},T)\,\notag\\
&+\Lambda_\text{Z,neb}(n_\text{\HII},n_e,T, Z_\text{g})+\Lambda_\text{dust}(\rho, T, D,N_\gamma^{\text{IR}},Z_\text{d}) \notag
\\&-\Gamma_\text{PE}(D,T,N_\gamma^{\text{FUV}},Z_\text{g})  -\Gamma_\text{cr}(n_j)
,\end{align}
where $n_j$ is the number density of all the
primordial species ($j\in\{\text{\ce{H2}, \HI, \HII, \HeI, \HeII, \HeIII}\}$), $N_\gamma^i$ is the photon number density of all the photon bins from IR to \HeII ionizing bands, $T$ is the gas temperature, $\rho$ is the density of the gas cell, and $Z$, $Z_\text{g}$, and $Z_\text{d}$ are the total, gas-phase, and dust-phase metallicity, respectively. In Appendix~\ref{sec:cooling_app}, we provide a detailed description of these cooling channels.

\subsection{Star formation and initial mass function sampling}
\label{sec:SFmodel}
In dwarf galaxies where the majority of stellar feedback comes from rare, massive stars, the stochasticity of when and where these massive stars form significantly influences the evolution of the galaxy. To capture this stochasticity, we developed a method to sample individual massive stars from a given IMF explicitly.

Closely following the method in \cite{2019MNRAS.487..364L}, we identified gas particles as star-forming cells if they were cold ($T<T_\text{th}$), dense ($n_\text{H}>n_\text{th}$), contracting ($\nabla\cdot {\bm v}<0$), self-gravitating \citep[$|\nabla\cdot {\bm v}|^2+|\nabla\times {\bm v}|^2<2G\rho$,][]{2013MNRAS.432.2647H}, and marginally Jeans-resolved ( $m_\text{gas}<f^{3}_\text{J,n}M_\text{J}$, where $f_\text{J,n}$ is a free parameter, $M_\text{J}$ is the local Jeans mass of the gas
particle, see below). The choice of $f_\text{J,n}$ has large freedom, and typically this criterion is overwhelmed by the density and temperature threshold in our simulations. A star-forming gas cell can be converted to a star particle with a probability of ${\cal P}_\text{SF}=\Delta t/\tau_\text{ff}$, where $\tau_\text{ff}=(3\pi/32G\rho)^{1/2}$ is the free-fall time of each cell and $\Delta t$ is the length of the time step. We note that we assume the SFE per free-fall time, $\epsilon_\text{ff}$, equals unity as the resolution of our simulation is high enough to resolve the fragmentation of dense star-forming cores. For gas cells with $m_\text{gas}>f_\text{J,s}^3M_\text{J}$, we enforced instantaneous star
formation to avoid the artificial fragmentation \citep{1997ApJ...489L.179T}, where $f_\text{J,s}$ is a free parameter similar to $f_\text{J,n}$. We noticed there are different definitions of $M_\text{J}$ and selections of 
$f_\text{J,s}$ in the literature. We followed the numerical experiments by \citet{2021MNRAS.506.2199G}, which suggest that $f_\text{J,s}=0.5$ is acceptable with the Jeans mass defined as $M_\text{J}=\pi^{1.5}c_\text{s}^{3}G^{-1.5}\rho^{-0.5}$. The parameters used for the simulations in this work are presented in Section~\ref{sec:IC}. 

Newly created star particles inherit the mass and velocity of their progenitor gas cells ($m_\star=m_\text{gas,prog.}$, ${\bm v}_\star ={\bm v}_\text{gas,prog.}$) and are distinguished into two classes by a variable $M_\text{fb}$. This variable determines whether the star particle actually hosts a $M>M_\text{min,SN}$ massive star and, if so, what its mass is. This feedback mass, $M_\text{fb}$, typically differs from the dynamical mass of the star particles, $m_\star$, and we determined the value of $M_\text{fb}$ using the method outlined below.

Given the IMF, $\Phi(M,Z)\equiv {\rm d}N/{\rm d}M$, and the minimum mass of massive stars (which can lead to core-collapse SN), $M_\text{min,SN}$, we first drew a random number to determine whether the star particle hosts a massive star; each star particle with mass $m_\star$ has a possibility, ${\cal P}_\text{host}$, of hosting a massive star,
\begin{equation}
   {\cal P}_\text{host} = m_\star\int_{M_\text{min,SN}}^{M_\text{max,IMF}}\Phi(M,Z){\rm d}M/\int_{M_\text{min,IMF}}^{M_\text{max,IMF}}M\Phi(M,Z){\rm d}M\,,
\end{equation}
where $M_\text{min,IMF}$ and $M_\text{max,IMF}$ are the lower and upper mass limits of the IMF, respectively. If the star particle does not host an $M>M_\text{min, SN}$ star, we set $M_\text{fb}=-1$. We note that in our high-resolution simulations with $m_\star\lesssim 10\ \rm M_{\odot}$, ${\cal P}_\text{host}<1$ always holds for the considered range of IMF variations.

If a star particle does host a massive star, we drew another random number to determine its zero-age main-sequence (ZAMS) mass recorded in $M_\text{fb}$ from the inverse function of the cumulative function of IMF by $M_\text{fb}=F^{-1}(y)$, where $y$ is a random variable with the value of $[0,1]$ and $F^{-1}$ is the inverse function of
\begin{align}
    F(x)\equiv \int_{M_\text{min, SN}}^{x}\Phi(M,Z)dM/\left[\int^{M_\text{max, IMF}}_{M_\text{min, SN}}\Phi(M,Z)dM\right]\ .
\end{align}
If the IMF has a constant slope $\alpha$ (e.g., $\alpha=-2.3$) at the $M>M_\text{min,SN}$ range, $F^{-1}(y)$ has a simple form of 
\begin{equation}
    M_\text{fb} = \left[(M_\text{max}^{\alpha+1}-M_\text{min}^{\alpha+1})y+M_\text{min}^{\alpha+1}\right]^{1/(\alpha+1)}\,.
\end{equation}
For the simulations presented in this work, we adopted the IMF given by \cite{2003PASP..115..763C} and $M_\text{min,IMF}=0.1\,\Msun$, $M_\text{max,IMF}=100\,\Msun$, and $M_\text{min,SN}=8\,\Msun$, respectively. We highlight here that implementing a variable IMF model (e.g., metallicity-dependent IMF, see \citealt{2013pss5.book..115K} for a review) is straightforward under this framework and will be explored in future works. 

We emphasize that, in our high-resolution simulations, $M_\text{fb}$ usually exceeds $m_\star$ when the star particle hosts a massive star. For example, a star particle with $m_\star=1\,\Msun$ can host a $30\,\Msun$ massive star that releases $23\,\Msun$ ejecta when it explodes as an SN. To ensure mass conservation and avoid negative mass values of star particles, the star particles lose their own masses and release masses to the ambient medium in two distinct ways. Regarding the mass loss, we extracted the lost mass from all star particles continuously according to the IMF-averaged mass-loss rate through both the AGB winds and the SN explosion following the treatment outlined by \cite{2013MNRAS.436.3031V}, regardless of whether it hosts a massive star. For mass release, star particles that do not host massive stars ($M_\text{fb}=-1$) can only return mass and metals to ISM mass through the AGB winds by $0.1-8\,\Msun$ stars. Only star particles with $M_\text{fb}\geq8$ have the capability to release mass, metals, and energy through the SN explosion of the massive star they host and provide pre-SN feedback through ionizing radiation and stellar winds. As the actual mass of star particles can be much smaller than the mass of ejecta, these particles do not consistently lose mass when they return mass by the stellar winds and SN ejecta. This inconsistency is counterbalanced by the mass loss resulting from the evolution of IMF-averaged stellar populations in all other star particles, ensuring mass conservation at the scale of star clusters.

\subsection{Zero-age main-sequence mass- and metallicity-dependent stellar feedback}
Once the massive stars are formed, we need to determine their lifetimes, photon production rates, mass-loss rates, and wind velocities based on their initial masses and metallicities. To build our library of these stellar properties, we gathered data on stellar evolutionary tracks, luminosities, mass-loss rates, and wind velocities from various references, as is detailed below, to encompass a range of metallicities from zero to the solar value and an initial mass range of $M\in[8-300]\ \rm M_\odot$. We used the fitting formulae for ZAMS stellar properties in \citet[hereafter \citetalias{2020MNRAS.495.4170T}]{2020MNRAS.495.4170T} to bridge the properties in different literature and convert them into functions of $M$ and $Z$. We found that the metallicity dependence of the ZAMS stellar radius and effective temperature can be well approximated with a power law in the mass range of $8-120\ \rm M_{\odot}$ for $Z\sim 10^{-6}-0.01\ \rm Z_{\odot}$, and that the luminosity is almost independent of $Z$, consistent with \citet{2023AJ....165....2L}. Therefore, we did extrapolation with the power-law metallicity dependence to obtain the ZAMS stellar properties at $Z=\rm Z_{\odot}$, making a new reference model called \citetalias{2020MNRAS.495.4170T} (EXT).

For the convenience of implementation, we used a polynomial formula to fit the stellar properties, ${\cal A}(M,Z')$, where ${\cal A}$ can be the photon production rate in $[\text{s}^{-1}]$, mass-loss rate in $[\Msun\,\text{yr}^{-1}]$, wind velocity in $[\text{km}\,\text{s}^{-1}]$, and main-sequence time in $[\text{yr}]$, as functions of the ZAMS mass at a given metallicity in the solar units $Z'\equiv Z/{\rm Z_\odot}$ and interpolated linearly in $\log_{10}(Z')$ space. In practice, the interpolation over $\log_{10}(Z')$ starts from a nonzero small metallicity $Z_{0}$, and the $Z=Z_{0}$ model is applied to all cases with $Z<Z_{0}$.
We adopted $Z_{0}=10^{-8}\ \rm Z_\odot$, below which stellar evolution is effectively metal-free.\footnote{Our choice of $Z_{0}=10^{-8}\ \rm Z_\odot$ is below or close to the critical metallicity, $Z_{\rm crit,\star}\sim 10^{-10}-10^{-8}$, below which features of metal-free stellar evolution arise \citep{Cassisi1993,Windhorst2018,2023AJ....165....2L}.} The polynomial fitting formula has the form
\begin{align}
    &\log_{10}{\cal A}=\begin{cases}
    b_{0}+b_{1}x+b_{2}x^{2}+b_{3}x^{3}+b_{4}x^{4}\ ,\quad &x\leq x_{0}\ , \\
    {\cal A}_{0}+\frac{d\log_{10} {\cal A}}{dx}\big|_{x=x_{0}}(x-x_{0})\ , & x>x_{0}\ ,
    \end{cases}\label {qfit}\\
    &{\cal A}_{0}= b_{0}+b_{1}x_{0}+b_{2}x_{0}^{2}+b_{3}x_{0}^{3}+b_{4}x_{0}^{4}\ ,\notag\\
&\frac{d\log_{10} {\cal A}}{dx}\Big|_{x=x_{0}} = (b_{1}+2b_{2}x_{0}+3b_{3}x_{0}^{2}+4b_{4}x_{0}^{3})\notag
\end{align}
for $x\equiv \log_{10}(M/\rm M_{\odot})$, 
where $x_{0}$ is a parameter chosen to optimize the fit quality at the high-mass end (which is better described by a linear relation between $\log_{10} {\cal A}$ and $\log M$). The relevant fit parameters, $b_{i}$ ($i=0$, 1, 2, 3, 4) and $x_{0}$, are shown in Appendix~\ref{sec:tables} (see Table~\ref{age:fit}, Table~\ref{uv:fit}, and Table~\ref{wind:fit_v}). In most cases, we fixed $b_{4}=0$ to avoid overfitting.

\subsubsection{Stellar evolution}
\begin{figure*}
	\includegraphics[width=2\columnwidth]{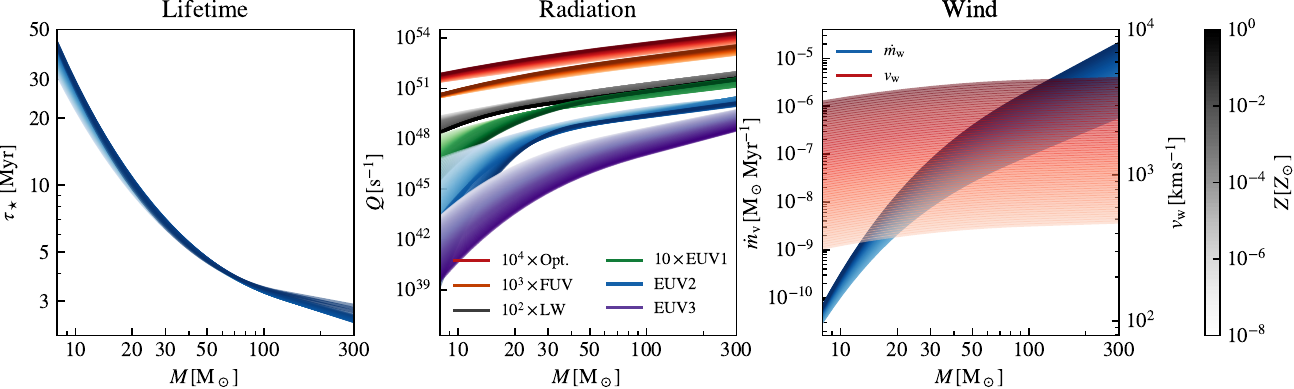}
    \caption{Stellar properties as a function of ZAMS mass and metallicity. Left: The main-sequence lifetime ($\tau_\star$) as a function of ZAMS mass and metallicity. The color of the curves from light to dark indicates the metallicity from low ($10^{-8}\,\Zsun$) to high ($1\,\Zsun$). Middle: Photon production rate ($Q$) of radiation bins from optical to \HeII ionizing band as a function of ZAMS mass and metallicity. Similarly, the color of the curves from light to dark indicates the metal abundance from low ($10^{-8}\,\Zsun$) to high ($1\,\Zsun$). We shifted the curves of the Opt., FUV, LW, and EUV1 bands by multiplying constants of $10^4$, $10^3$, $10^2$, and 10, respectively, to avoid overlapping different bins. Right: Similar to left but for the wind mass-loss rate ($\dot{m}_\text{w}$) and wind velocity ($v_\text{w}$). The color of the curves from light to dark indicates the metal abundance from low ($10^{-6}\,\Zsun$) to high ($1\,\Zsun$). The main-sequence stellar winds from extremely metal-poor stars with $Z<10^{-6}\,\Zsun$ are negligible and not considered in our model.}
    \label{fig:feedbackModel}
\end{figure*}

In our model, we assume that the stellar properties do not evolve during their main-sequence lifetime. The main-sequence lifetimes of stars are determined by the metallicity-dependent function of Pop. II stars from \cite{1998A&A...334..505P} over $Z=0.0004-0.05$ combined with the function of metal-free Pop. III stars ($Z\leq10^{-8}\,\Zsun$) from \citetalias{Schaerer02}. In the left panel of Fig.~\ref{fig:feedbackModel}, we present the main-sequence lifetimes, $\tau_\star$, as a function of the mass and metallicity of the main-sequence of zero age.

During the main-sequence stage, massive stars radiate photons (see Section~\ref{sec:rad} for details) and release mass and metals through stellar winds (Section~\ref{sec:wind}). We assume that massive stars release unenriched gas by main-sequence winds and the yields by post-main-sequence winds and SN explosions are released instantaneously when the stars end their MS lifetimes.

Once stars complete their main-sequence lifetime, they can release mass and metals through three channels: asymptotic giant branch (AGB) winds, core-collapse SNe, and type Ia SNe (SN Ia).
Our AGB yield table covers the mass range of $1-7.5\,\Msun$ over $Z=0.0001 - 0.02$.
This table was mainly adopted from \cite{2010MNRAS.403.1413K} and is supplemented by \cite{2014MNRAS.441..582D} and \cite{2014ApJ...797...44F} (see \citealt{2018MNRAS.473.4077P} for details). 
The yield table for massive stars was adopted from the {\tt set M} of \citet[LC18]{2018ApJS..237...13L}, which covers the mass range of $13-120\,\Msun$. These tables include the final integrated yields and the yields present in the stellar winds. The original \citetalias{2018ApJS..237...13L} tables have four metallicity bins expressed as [Fe/H] = $-3$ to 0. We converted the [Fe / H] values to $Z=\{3.11\times10^{-5},3.11\times10^{-4}, 3.11\times10^{-3}, 0.0128\}$ based on the initial composition of  C, N, O, Ne, Mg, Si, and
Fe provided in their tables. A small fraction of $<8\,\Msun$ low- and intermediate-mass stars will explode as type Ia SNe following a delay-time distribution (see Section~\ref{sec:Ia}). We adopted the `W7' model from \cite{1997NuPhA.621..467N:Nomoto} for the SN Ia yields.

In the rest of this section, we shall introduce our stellar radiation, main-sequence wind, and SN models in detail. We discussed the caveats and limitations of our stellar feedback model in Section~\ref{sec:caveat}.

\subsubsection{Radiation}
\label{sec:rad}

\begin{table*}
\caption{Summary of the radiation energy bins adopted in our simulations.}
\addtolength{\tabcolsep}{2pt}
\renewcommand{\arraystretch}{1.1}
\begin{tabular}{lllll}
\hline
Bin           & Range (eV) & Metallicity bins ($\Zsun$) & Function & Reference \\
\hline
IR & $0.1-1$ & N/A & IR radiation pressure (RP) & \cite{2020MNRAS.499.5732K}\\
Opt.   & $1-5.8$ & $\{10^{-8},10^{-6},10^{-5},10^{-4},0.01,1\}$ & RP & \citetalias{2020MNRAS.495.4170T} (EXT)\\
FUV    & $5.8-11.2$ & $\{10^{-8},10^{-6},10^{-5},10^{-4},0.01,1\}$ & Photoelectric heating (PE), RP &  \citetalias{2020MNRAS.495.4170T} (EXT) \\
LW    & $11.2-13.6$ & $\{10^{-8},10^{-4},0.007\}$ & \ce{H2} dissociation &  \citetalias{2020MNRAS.495.4170T} (EXT), \citetalias{2019MNRAS.482.1304E}, \citetalias{Gessey-Jones2022}\\
& & & \CI ionization, PE, RP &  \\
EUV1    & $13.6-24.6$ & $\{10^{-8},0.001,0.0282,1\}$ & \HI, \ce{H2} ionization, RP & \citetalias{2020MNRAS.495.4170T} (EXT), \citetalias{2019MNRAS.482.1304E}, \citetalias{2003ApJS..146..417L}, \citetalias{Schaerer02}\\
EUV2    & $24.6-54.4$ & $\{10^{-8},0.001,0.0282,1\}$ & \HeI, \HI, \ce{H2} ionization, RP & \citetalias{2020MNRAS.495.4170T} (EXT), \citetalias{2019MNRAS.482.1304E}, \citetalias{2003ApJS..146..417L}, \citetalias{Schaerer02}\\
EUV3    & $54.4-\infty$ & $\{10^{-8},10^{-6},10^{-5},10^{-4},0.01,1\}$ & \HeII, \HeI, \HI, \ce{H2} ionization, RP & \citetalias{2020MNRAS.495.4170T} (EXT) \\
\hline
\end{tabular}
\addtolength{\tabcolsep}{-2pt}
\renewcommand{\arraystretch}{0.9090909090909090909}
\label{tab:RadiationBins}
\end{table*}

We included radiative feedback due to photodissosiation, photoionization, and photoheating of \ce{H2}, \HI, \HeI, and \HeII species and the momentum transfer from photons to gas due to single and multiple scattering (i.e., radiation pressure). In our simulations, the star particles hosting alive $M_\text{fb}>8\,\Msun$ massive stars are sources of local radiation. We assume that the photon production rate, $Q$, remains constant at its ZAMS value throughout the main-sequence lifetime and drops to zero immediately after the main-sequence ends. 

The stellar spectra were discretized into seven bins (IR, Opt., FUV, LW, EUV1, EUV2, and EUV3) corresponding to our radiation bins. The photon production rate, $Q_i$, for each spectral band was determined using polynomial fitting functions derived from the stellar mass-metallicity grid outlined later in this section. The mean energy per photon, mean ionization cross-section, photoheating rate, and the radiation pressure for each of the bins were calculated using a template synthesis spectrum taken from \cite{2003MNRAS.344.1000B} (see again Table~1 in \citealt{2020MNRAS.499.5732K} for details). Similar to \cite{2020MNRAS.499.5732K}, we injected all photons into the nearest gas cell of the star particle to increase the probability of resolving the initial Str\"omgren mass. At time step $\Delta t$, this gas cell receives $Q_i\Delta t$ band $i$ photons with mean energy $e_i$. The direction of the photon flux, ${\bm F}^i_\gamma$, was set to be radially outward from the star particle with a magnitude of $|{\bm F}^i_\gamma| =\tilde{c}e_i Q_i\Delta t/V_\text{cell}$, where $V_\text{cell}$ is the volume of the gas cell receiving the photons.

The photoionization and photon heating rates of species ``$j$'' are given by 
\begin{equation}
    \zeta_{\text{pi},j} = \tilde{c}\sum_i\bar{\sigma}_{ij}N_\gamma^i\,
\end{equation}
and 
\begin{equation}
    \Gamma_{\text{pi},j} = \tilde{c}\sum_iN_\gamma^i\bar{\sigma}_{ij}\mathfrak{h}_{ij}\,,
\end{equation}
where $\bar{\sigma}_{ij}$ and $\mathfrak{h}_{ij}$ are the mean ionization cross-section and photoelectron energy of photons in bin ``$i$'' interacting with species ``$j$'' \citep{2019MNRAS.485..117K,2020MNRAS.499.5732K}.

Radiation can also directly couple momentum to gas by the single scattering. We added a source term in the momentum conservation equation of hydrodynamics and it is given by
\begin{equation}
\left.\frac{\partial \rho v}{\partial t}\right|_\text{rad.} =  \frac{1}{c} \sum_i {\bm F}_\gamma^i \, \left(\sum_j \, n_j \, \bar{\sigma}_{ij} \, p_{ij}  + \kappa_i \, \rho \, e_i\right)\, ,
\label{eq:uvmomentum}
\end{equation}
where $p_{ij}$ is the mean photon momentum and $\kappa_i$ is the opacity due to dust (see Table~1 in \citealt{2020MNRAS.499.5732K} for the values we adopted). The momentum boost due to multiple scattering was also modelled self-consistently by the method described in \cite{2019MNRAS.485..117K}.

In general, the photon production rate, $Q_i$, increases with decreasing metallicity, as more metal-poor stars are more compact and hotter. To capture this trend, we constructed a metallicity grid for each photon energy band of ionizing photons 
with select models that satisfy a quasi-monotonic evolution of the total production rate of ionizing photons ($>13.6\ \rm eV$, combining EUV1, EUV2, and EUV3)
with metallicity. Similarly, we selected models with monotonic evolution with metallicity for the Opt., FUV, and LW bands. For the IR band, we ignored the contribution of stellar IR radiation for simplicity, as it is employed for IR dust-gas coupling \citep{2019MNRAS.485..117K,2020MNRAS.499.5732K}. For the six photon energy bands above 1~eV, the stellar evolution models\footnote{We convert the absolute metallicities $Z$ used in these stellar evolution models to the metallicities in the solar unit $Z'$ using the bulk solar metallicity $\rm Z_\odot=0.0142$ from \citet{2009ARA&A..47..481A}, which is slightly larger than the value $\rm Z_\odot=0.0136$ used in our ISM chemistry and cooling model because not all metal elements are tracked in the ISM (see Sec.~\ref{sec:chemistry}). Here, we assume that the mass fraction of all metal elements in a newly born star is proportional to the mass fraction of the seven tracked metal elements in the natal ISM by a constant coefficient 0.0142/0.0136. } used to construct the metallicity grids are described as follows.
\begin{enumerate}
    \item For the Opt. and FUV bands, since the relevant feedback channels are unimportant for stellar feedback in metal-poor gas clouds, we calculated their production rates with the ZAMS stellar properties at $Z=10^{-8}$, $10^{-6}$, $10^{-5}$, $10^{-4}$, 0.01, and $1\ \rm Z_{\odot}$ in \citetalias{2020MNRAS.495.4170T} (EXT), assuming black-body spectra for simplicity.
    \item  For the LW band, the metallicity dependence is unclear in the literature. 
    For simplicity, we considered three models at $Z=10^{-8}, 10^{-4}$, and $0.007\ \rm Z_\odot$ in which $Q$ decreases monotonically with $Z$. We adopted the results for $Z\geq 0.007\,\Zsun$ from \citetalias{2019MNRAS.482.1304E} (see their Appendix B) based on the OSTAR2002 stellar atmosphere models (\citetalias{2003ApJS..146..417L}) and the ZAMS stellar properties from PARSEC \citep{2012MNRAS.427..127B,2014MNRAS.445.4287T:Tang}, which show minor variations with metallicity at $Z\sim 0.007-1.2\,\Zsun$. At lower metallicities, we derived the photon production rates from the $Z=10^{-4}\,\Zsun$ ZAMS properties in \citetalias{2020MNRAS.495.4170T}, assuming black-body spectra, and used the results for metal-free stars from \citet[GJ22, see their Fig.~5]{Gessey-Jones2022} to cover $Z\le 10^{-8}\ \rm Z_\odot$. 
    \item For the EUV1 and EUV2 bands that dominate the ionization feedback from massive stars, we considered four models at $Z=10^{-8}, 0.001, 0.0282$, and $1\ \rm Z_\odot$. 
    We used the $Z=0$ model in \citetalias{Schaerer02} (see their table.~6) for $Z\le 10^{-8}\ \rm Z_\odot$. For $Z=0.001\ \rm Z_\odot$, we applied the OSTAR2002 grid (\citetalias{2003ApJS..146..417L}) to the stellar parameters in \citetalias{2020MNRAS.495.4170T} and took the average of the rates at ZAMS and the end of the main-sequence. Here, the stellar parameters at $Z=0.001\ \rm\ Z_{\odot}$ were calculated by interpolating between the $10^{-4}$ and $0.01\ \rm Z_{\odot}$ models in \citetalias{2020MNRAS.495.4170T}. For $Z=0.0282\ \rm Z_\odot$, we used the fit in \citetalias{Schaerer02}. For $Z\ge 1\ \rm Z_\odot$, we used the results from \citetalias{2019MNRAS.482.1304E} at solar metallicity. 
    \item For the EUV3 band, as optimistic estimates, we used the ZAMS stellar properties from \citetalias{2020MNRAS.495.4170T} (EXT) to derive $Q$ with black-body spectra for $Z=10^{-8}$, $10^{-6}$, $10^{-5}$, $10^{-4}$, 0.01, and $1\ \rm Z_{\odot}$. 
\end{enumerate}
We summarize the properties of the seven radiation bands in Table~\ref{tab:RadiationBins}. In the middle panel of Fig.~\ref{fig:feedbackModel}, we present the photon production rate, $Q_i$, as a function of ZAMS mass and metallicity for the six radiation bands above 1~eV with metallicity dependence (excluding the IR band).

\subsubsection{Winds of massive stars}
\label{sec:wind}
We modeled the main-sequence winds of massive stars by injecting the mass, $\dot{m}_\text{w}\Delta t$,
and momentum, $\sqrt{2L_\text{w}\dot{m}_\text{w}}\Delta t$, fluxes from a star to its nearest 32 neighboring gas cells weighted by their solid angle opening to the star (see \citealt{2019MNRAS.487..364L} for details), where $\dot{m}_{\rm w}$ is the wind mass-loss rate and $L_{\rm w}=0.5\dot{m}_{\rm w} v_{\rm}^{2}$ is the wind luminosity or power given the (terminal) wind velocity, $v_{\rm w}$. 

Since the massive stars sampled in our simulations are typically hotter than the bi-stability-jump (with
effective temperatures $T_\text{eff}>30\,000$\,K), we used the fit formula of Eq.~3 in \cite{2021MNRAS.504.2051V} to calculate $v_\text{w}$ given the stellar luminosity, $L$, effective temperature, $T_{\rm eff}$, and metallicity, $Z$. The wind velocity was then combined with $L$, $T_{\rm eff}$, $Z$, the stellar mass, $M$, and escape velocity, $v_{\rm esc}=\sqrt{2GM/R_{\star}}$ (given the stellar radius $R_{\star}$), to derive the mass-loss rate, $\dot{m}_{\rm w}$, through Eq.~24 in \cite{2001A&A...369..574V}. Similar to the calculation of photon production rates in the Opt. and FUV bands, we used the ZAMS values of $M$, $L$, $T_{\rm eff}$, and $R_{\star}$ from \citetalias{2020MNRAS.495.4170T} (EXT) to calculate $v_\text{w}$ and $\dot{m}_{\rm w}$ for $Z=10^{-6}$, $10^{-5}$, $10^{-4}$, 0.01, and $1\ \rm Z_{\odot}$. Here, we ignored the weak main-sequence stellar winds from extremely metal-poor stars with $Z<10^{-6}\,\Zsun$ for simplicity. The results were fit with Eq.~(\ref{qfit}), and the fit parameters are presented in Appendix~\ref{sec:tables}.
We present the wind mass-loss rate, $\dot{m}_\text{v}$, and wind velocity, $v_\text{v}$, as a function of ZAMS mass and metallicity in the right panel of Fig.~\ref{fig:feedbackModel}.

\subsubsection{Core-collapse supernovae}

We assume that all stars with initial masses larger than $8\,\Msun$ explode as core-collapse SNe and release $10^{51}$\,erg of thermal energy into their neighboring gas cells at the end of their main-sequence lives. We notice that the SN explosion energy has been modeled as a mass-dependent function in a few works \citep[e.g.][]{2021MNRAS.501.5597G,2023arXiv231011495S}, while its dependence on metallicity is still unclear (see Section~\ref{sec:caveat} for details).

Since we adopted a ``pure thermal'' scheme to inject SN energy, the feedback from SN can be significantly underestimated if we cannot resolve the Sedov–Taylor (ST) phase during which the thermal energy converts to kinetic energy while the total energy is conserved \citep[i.e., the over-cooling issue,][]{1992ApJ...391..502K}. To maximize the possibility of resolving the ST phase, we injected all the explosion energy to the nearest neighboring cell of the SN while distributing the ejected mass and metals to 32 neighbors weighted by the solid angle opening to the SN. 

\cite{2015ApJ...802...99K} show that the resolution is sufficiently high to ensure the momentum convergence if $m_\text{gas,SN}<M_\text{shell}/27$, where $m_\text{gas,SN}$ is the mass of the gas cell that received the SN energy and $M_\text{shell}$ is the total swept-up mass at the end of the energy conserving ST phase (shell formation) given by
\begin{equation}
    M_\text{shell} = 1680\,\Msun\,\left(\frac{E}{10^{51}\text{ erg}}\right)^{0.87}\left(\frac{n}{\text{ cm}^{-3}}\right)^{-0.26}\,.
    \label{equ:shell-formation-mass}
\end{equation}
Following equation~(\ref{equ:shell-formation-mass}), we can estimate the maximum density where the ST phase can be safely resolved, and it is $8\times10^6$\,cm$^{-3}$ for our fiducial $1\,\Msun$ resolution and $10^3$\,cm$^{-3}$ for $10\,\Msun$ resolution. We have verified that the ST phase is resolved in most cases when the gas mass resolution is below $10\,\Msun$ \citep[see Appendix~\ref{sec:SN_res}, also see][for the convergence tests in other numerical models showing similar results]{2019MNRAS.483.3363H,2020MNRAS.495.1035S:Steinwandel}.

\subsubsection{Type Ia supernovae}
\label{sec:Ia}
For type Ia SNe, we adopted a delay-time distribution (DTD) formalism to compute the explosion probability of each star particle as a function of time. In the DTD formalism, the type Ia rate is determined by 
\begin{equation}
    \dot{N}_\text{Ia}(t)=N_\text{Ia}\int_0^{t}\dot{M}_\star(t')\Psi(t-t'){\rm d}t'\,,
\end{equation}
where $N_\text{Ia}$ is an observable, $\dot{M}_\star(t)$ is the star formation rate (SFR), and $\Psi(t)$ is the DTD. For each star particle in our simulations, $\dot{M}_\star(t) = m_\star\delta(t-t_0)$, where $\delta(t)$ is the Dirac Delta function, and $t_0$ is the formation time of the star particle. For simplicity, we denote $t_0=0$. This gives a simplified SN Ia rate for each star particle: $\dot{N}_\text{Ia}(t)=m_\star N_\text{Ia}\Psi(t)$.

In our high-resolution simulations, each star particle should at most have one SN Ia. Similar to our IMF-sampling scheme, we first drew a random number to determine whether the star particle is an SN Ia progenitor. Each star
particle with mass $m_\star$ has a possibility, ${\cal P}_\text{Ia}$, of being an SN Ia progenitor,
\begin{equation}
    {\cal P}_\text{Ia} = m_\star N_\text{Ia}\,.
\end{equation}
If this star particle is an SN Ia progenitor, we drew another random number, $y$, to determine its explosion time from the inverse function of the cumulative function of DTD by $t=F^{-1}(y)$, where
\begin{equation}
    F(t) = \int_0^{t}\Psi(t'){\rm d}t'\,.
\end{equation}

We adopted a power-law DTD model that consists of a power law in time,
\begin{equation}
\Psi(t)=\left\{
\begin{aligned}
    &0\, , & t<\tau_\star(8\Msun,Z)\,\\
    &At^{-s} \,, & t>\tau_\star(8\Msun,Z)\,,
\end{aligned}
\right.
\end{equation}
where $A$ is a normalization factor such that the integration of $\Psi(t)$ over a Hubble time, $t_\text{H}$, equals 1 (i.e.,  $\int_0^{t_\text{H}}\Psi(t'){\rm d}t'=1$) and $\tau_\star(8\Msun,Z)$ is the lifetime of an $8\,\Msun$ massive star with an initial metallicity of $Z$. Specifically, we adopted $s=1.12$ and $N_\text{Ia}=1.3\times10^{-3}$\,SN\,$\Msun^{-1}$ \citep{2012MNRAS.426.3282M:Maoz,2013MNRAS.436.3031V}. While our model accounts for it, we leave the discussion of SN Ia feedback and enrichment for our future cosmological simulations, as it occurs too late to influence the cloud-scale star formation that is the focus of this paper.

\section{Isolated dwarf galaxy simulations}
\label{sec:isolateddwarf}
\subsection{Initial conditions and model parameters}
\label{sec:IC}
To evaluate the performance of the RIGEL model, we ran a suite of simulations of an isolated dwarf galaxy. We generated the isolated disk initial conditions using the {\sc MakeDisk} code developed by \citet{2005MNRAS.361..776S}. We created a disk with parameters identical to the fiducial initial condition used in \citetalias{2017MNRAS.471.2151H}, which is also extensively used in recent isolated dwarf simulations (e.g., \citealt{2021MNRAS.501.5597G,2022MNRAS.509.5938H}; \citetalias{2023MNRAS.522.3092L}). The dark matter halo follows a Hernquist profile with an NFW-equivalent concentration parameter, $c=10$. The virial mass of the halo is $M_\text{vir}=2\times10^{10}\,\Msun$, the virial radius is $R_\text{vir}=44$\,kpc, and the spin parameter is $\lambda =0.03$. The rotation-supported galaxy disk has an initial stellar disk with a mass of $2\times10^7\,\Msun$ and a scale length of $0.73$\,kpc and an initial gas disk with a mass of $4\times10^7\,\Msun$ and a scale length of $1.46$\,kpc. The scale height of both the gas and stellar disks is 0.35\,kpc. The star particles in the initial disk only interact with gas via gravity but do not contribute to stellar feedback or enrichment. 

The gravitational softening lengths for dark matter (DM) and star were determined following \cite{2018MNRAS.480..800H} by $\epsilon_\text{DM} = 1.7\,\text{pc}\,m^{1/3}
_{\text{DM}}(\rho_{\text{DM}}/10^9\Msun\,\text{kpc}^{-3})^{-1/3}$ and $\epsilon_\star = 0.34\,\text{pc}\,m^{1/3}_{\star}(\left \langle n_\text{SF} \right \rangle/10^3\,\text{cm}^{-3})^{-1/3}$, where $\left \langle n_\text{SF} \right \rangle$ is the mean density for star formation. We list the softening lengths used in this work in Table~\ref{tab:simulationSets}. The softening of gas cells was determined adaptively with a minimum length of $0.004$\,pc. We set the reduced speed of light as $\tilde{c}=1000$\,km~s$^{-1}$ to capture the propagation of the ionization front in a typical ISM environment with a reasonable time lapse \citep{2013MNRAS.436.2188R}.                     

To avoid initial artificial starbursts, we relaxed the initial conditions by injecting thermal energy following the method outlined in \citetalias{2017MNRAS.471.2151H} for 500\,Myr to create a turbulent ISM within the gas disc. In production simulations, we adopted a spatially uniform ionizing UV background radiation field from \cite{2009ApJ...703.1416F} with corrections for self-shielding in dense ISM \citep{2013MNRAS.430.2427R}. In addition, we set a background FUV field of $G_{0,\text{UVB}} = 0.00\,324$ for the \CI ionization and photoelectric heating, which is the UVB value adopted from \cite{2012ApJ...746..125H} integrated over the 5.8--13.6 eV energy range. The star formation criteria used for production simulations are $T_\text{th}=100$\,K, $n_\text{th}=10^4$\,cm$^{-3}$ for $1\,\Msun$ resolution and $n_\text{th}=10^3$\,cm$^{-3}$ for $10\,\Msun$ resolution. This choice of this density threshold was motivated by the observational implications \citep[e.g.][]{2011MNRAS.416..783P,2012ApJ...745..190L}. The necessary and sufficient Jeans numbers for star formation were set as $f_\text{J,n} = 0.1$ and $f_\text{J,s} = 0.5$. 

In this paper, we present a total of six simulations with different combinations of feedback channels, initial metallicities, and resolutions. We list the naming conventions and simulation setups in Table~\ref{tab:simulationSets}. We refer to the full RIGEL physics model as the ``full'' model. In this paper, the full RIGEL physics simulation with $1\,\Msun$ baryon resolution initialized with $Z=0.02\,\Zsun$ is our fiducial run, named as ``$0.02\Zsun$/full.'' For comparison, we simulated the same galaxy without local radiative feedback (but with UVB), referred to as ``$0.02\Zsun$/noRT.'' We also simulated a galaxy initialized with $0.2\,\Zsun$ with the ``full'' physics, and it is referred to as the ``$0.2\Zsun$/full'' run. For the convergence test, we also ran each simulation with a lower resolution of $10\,\Msun$ and these runs are referred to as ``low.'' 

We ran all six simulations for $1$\,Gyr, which corresponds to roughly six orbit times at a $1$\,kpc galactocentric radius. As our simulated dwarf galaxies exhibit regular rotation and steady star formation activity, each $\sim200$\,Myr time segment of the simulation can be considered as a realization of different random seeds. We additionally ran the $0.02\Zsun$/low simulation using another random seed, which demonstrates minor variations in the global SFR and ISM structures. Since different disks spend different times to cool down and form the first star from the initial condition, we define the time origin of each simulation as the moment its first star forms.

For each $1\,\Msun$-resolution simulation, $\sim1.3 \times 10^6$ core-hours are required for every $1$\,Gyr of simulated time using the Intel Xeon Platinum 8160 (``Skylake'') CPUs; the $10\,\Msun$-resolution simulations, which have ten times fewer particles and roughly $10^{1/3}$ larger time steps, reduce the computational cost to $\sim3 \times 10^4$ core-hours per $1$\,Gyr of simulation. 

We notice that though the disk properties of our galaxies are the same as those in \citetalias{2017MNRAS.471.2151H} and \citetalias{2023MNRAS.522.3092L}, the initial metallicity of \citetalias{2017MNRAS.471.2151H} is $0.15\,\Zsun$ in our normalization, slightly lower than our $0.2\,\Zsun$ galaxies, while the initial metallicity of \citetalias{2023MNRAS.522.3092L} is $0.015\,\Zsun$, also slightly lower than our $0.02\,\Zsun$ galaxies. Nonetheless, we think our results can be directly compared with them because such differences are minor.

We use cylindrical coordinates $R$ and $z$ to describe our simulations, where $R$ is the galactocentric radius and $z$ is along the rotation axis of the disc. We define the ``ISM region'' as the $R < 2$\,kpc and $|z| < 1$\,kpc region.

\begin{table*}
\caption{Overview of the simulations reported in this work and simulation parameters.}
\addtolength{\tabcolsep}{-0.75pt}
\renewcommand{\arraystretch}{1.1}
\begin{tabular}{lcccccccl}
\hline
Simulation           & $Z_\text{init}$($\Zsun$) & $m_\text{gas}$ ($\Msun$) & $m_\star$ ($\Msun$) & $m_\text{DM}$ ($\Msun$) & $\epsilon_\text{gas,max}$ (pc) & $\epsilon_\star$ (pc) & $\epsilon_\text{DM}$ (pc) & Feedback channels\\
\hline
$0.02\Zsun$/full  & 0.02 & 1 & 1 & $2.5\times10^3$ & 0.05 & 0.05 & 39 & PI+RP+PE+WD+SN\\
$0.02\Zsun$/full/low   & 0.02 & 10 & 10 & $2.5\times10^4$ & 0.3 & 0.3 & 84 & PI+RP+PE+WD+SN\\
$0.02\Zsun$/noRT     & 0.02 & 1 & 1 & $2.5\times10^3$ & 0.05 & 0.05 & 39 & WD+SN \\
$0.02\Zsun$/noRT/low    & 0.02 & 10 & 10 & $2.5\times10^4$ & 0.3 & 0.3 & 84 & WD+SN\\
$0.2\Zsun$/full    &  0.2 & 1 & 1 & $2.5\times10^3$ & 0.05 & 0.05 & 39 & PI+RP+PE+WD+SN\\    
$0.2\Zsun$/full/low     & 0.2 & 10 & 10 & $2.5\times10^4$& 0.3 & 0.3 & 84 & PI+RP+PE+WD+SN\\

\hline
\end{tabular}
\tablefoot{The columns from left to right are the model name (first column), initial metallicity (second column), target mass of gas particles (third column),  mass of star particles (both background and active stars, fourth column), mass of dark matter particles (fifth column), max softening length of gas (sixth column),  softening length of stars (seventh column) and DM (eighth column), and the stellar feedback channels included in the models (ninth column). Abbreviations: PI: photoionization, RP: radiation pressure, PE: photoelectric heating, WD: stellar wind, SN: supernova.}
\addtolength{\tabcolsep}{0.75pt}
\renewcommand{\arraystretch}{0.9090909090909090909}
\label{tab:simulationSets}
\end{table*}

\subsection{Star formation rates and the Kennicutt–Schmidt relation}
\label{sec:starformation}
\begin{figure*}
	\includegraphics[width=2\columnwidth]{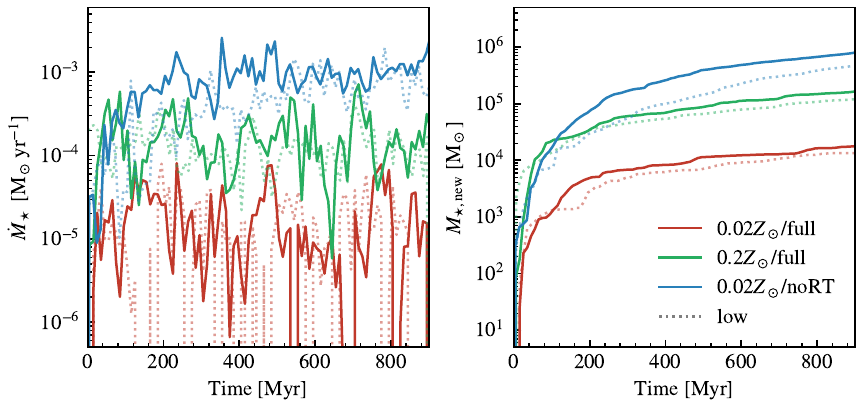}
    \caption{Star formation histories of the simulated dwarf galaxies. Left: SFR as a function of time. The SFR is averaged over a time-scale of 10\,Myr. The solid curves are the results from the high-resolution runs while the dotted curves are from corresponding low-resolution runs. Right: Cumulative mass of newly formed stars in our simulations. The SFRs exhibit good convergence with resolution. The 0.02$\Zsun$/noRT simulations show the most significant difference in cumulative stellar mass between the low- and high-resolution simulations; nonetheless, the difference remains within a factor of two.}
    \label{fig:SFR}
\end{figure*}

\begin{figure*}
	\includegraphics[width=2\columnwidth]{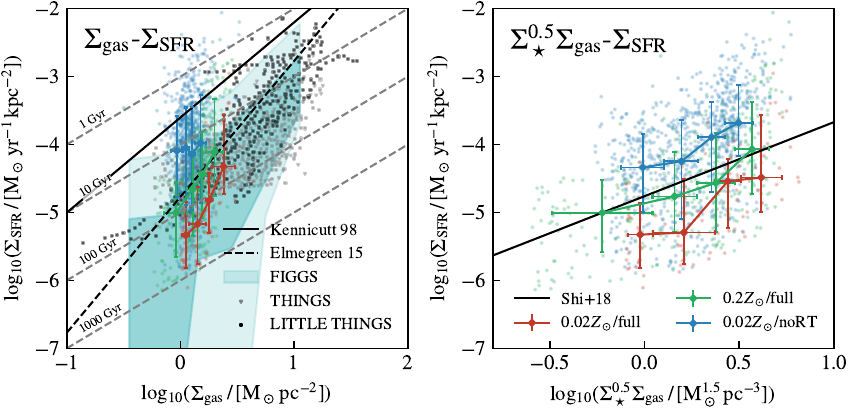}
    \caption{Star formation relations in the simulated dwarf galaxies. Left: Kennicutt--Schmidt relation for the simulated high-resolution galaxies for the $0.02\Zsun$/full (red), $0.2\Zsun$/full (green), and $0.02\Zsun$/noRT (blue) model. The translucent data points are 1\,Myr average SFR within a series of $100$\,pc annuli measured from the simulation time $t=300$\,Myr to $1000$\,Myr. The data with errorbars are the median $\Sigma_\text{SFR}$ of the [0,25), [25,50), [50,75), [75,100] gas surface density percentile bins, and the errorbars of $\Sigma_\text{SFR}$ show the 16 and 84 percentiles of the SFR surface density in each bin. The black solid line shows the standard KS relation from \protect\cite{1998ApJ...498..541K}; the black dashed line shows the power-2 relation for outer disk regions and dwarf irregular galaxies from \protect\cite{2015ApJ...814L..30E}. The observational data collected by \protect\cite{2015ApJ...814L..30E}, including the far outer regions of spirals and dwarf galaxies from the THINGS survey \protect\citep{2010AJ....140.1194B} and the dwarf galaxies from the LITTLE THINGS survey \protect\citep{2015ApJ...805..145E:Elmegreen}, are shown as grey and black dots, respectively. The blue-shaded region is the 5–95 and 16–84 percentile range of the observed dwarf galaxies in the FIGGS survey \protect\citep{2015MNRAS.449.3700R}. Right: extended KS relation \protect\citep[$\Sigma_\text{SFR}\propto(\Sigma_\star^{0.5}\Sigma_\text{gas})^{1.09}$,][]{2011ApJ...733...87S,2018ApJ...853..149S} measured in the same way; the black solid line is the fitting of observed data of star-forming galaxies of various types from \protect\cite{2018ApJ...853..149S}.  
    }
    \label{fig:KSR}
\end{figure*}

We first analyze the star formation properties of our simulations. The left panel in Fig.~\ref{fig:SFR} shows the SFR as a function of simulation time, representing the star formation histories of the simulated galaxies, and the right panel shows the cumulative mass of newly formed stars in our simulations. In each simulation, there is a rapid rise in the SFR initially, followed by the SFR stabilizing at a self-regulated state due to the presence of feedback. Despite the average SFRs remaining relatively stable over long timescales, there are noticeable bursty fluctuations at shorter scales. In particular, the SFRs in the full feedback models exhibit fluctuations exceeding two orders of magnitude. 

The SFRs show an evident dependence on metallicity. The median SFR of $0.02\Zsun$/full is only $\sim6\%$ of that in $0.2\Zsun$/full. This phenomenon is mainly attributed to the low cooling rate, which suppresses the formation of cold, dense gas. Consequently, once the gas is heated to $\gtrsim10^4$\,K, a longer period is required for the metal-poor gas to cool down and convert to a star-forming state. Although the total feedback budget (which scales with the SFR) is relatively limited in low-metallicity  environments, it becomes more difficult for the ISM to revert to its cold phase once heated to its warm-diffuse phase due to ineffective cooling mechanisms and suppression of thermal instability \citep[e.g.][]{2004ApJ...609..667S,2011ApJ...733...47W,2019ApJ...881..160B}.

Turning off radiative feedback has a larger impact on the SFR. The median SFR of $0.02\Zsun$/noRT is 92 times higher than that of $0.02\Zsun$/full and even five times higher than $0.2\Zsun$/full. We will see in Section~\ref{sec:ISM_properties} that this difference is due to the fact that $0.02\Zsun$/noRT contains the highest amount of cold and molecular gas, as there is no ISRF present to heat the diffuse gas and keep it warm and atomic.

We note that the SFRs exhibit good convergence with resolution. In all models, the cumulative stellar masses of the low-resolution simulations are slightly lower, but the difference in stellar mass formed between the low- and high-resolution simulations is within a factor of two. The 0.02$\Zsun$/noRT simulations show the most significant difference in cumulative stellar mass between the low- and high-resolution simulations, because the SFR of the high-resolution run is systematically higher than that of the low-resolution run before $t\approx500$\,Myr. These systematic differences can be attributed to some specific features of initial conditions for high- and low-resolution simulations, as we stochastically injected thermal energy to relax them (Section~\ref{sec:IC}).

To assess the performance of our model in regulating star formation, we analyze the Kennicutt--Schmidt (KS) relation \citep{1998ApJ...498..541K} which establishes a connection between the surface densities of gas ($\Sigma_\text{gas}$) and the SFR ($\Sigma_\text{SFR}$) in our simulations. We also analyze the extended KS relation which substitutes the total gas surface density by 
including the contribution of stars' gravity \citep[$\Sigma_\star$, e.g.][]{2011ApJ...733...87S,2018ApJ...853..149S}. In Fig.~\ref{fig:KSR}, we present the KS relation ($\Sigma_\text{gas}$--$\Sigma_\text{SFR}$, left panel) 
and the extended KS relation ($\Sigma_\star^{0.5}\Sigma_\text{gas}$--$\Sigma_\text{SFR}$, right panel) of the simulated high-resolution galaxies for the $0.02\Zsun$/full (red), $0.2\Zsun$/full (green), $0.02\Zsun$/noRT (blue) model. The SFR surface densities are measured as the average over 1\,Myr within a series of $100$\,pc annuli from the simulation time $t=300$\,Myr to $1000$\,Myr. We also plot the mean $\Sigma_\text{SFR}$ of the [0,25], [25,50], [50,75], and [75,100] $\Sigma_\text{gas}$ ($\Sigma_\star^{0.5}\Sigma_\text{gas}$) percentile bins to show the trend.

On the $\Sigma_\text{gas}$--$\Sigma_\text{SFR}$ plane (Fig.~\ref{fig:KSR} left panel), 
the results from both full feedback simulations are in good agreement with the observed dwarf galaxies in the FIGGS survey measured with 400\,pc resolution \citep[][]{2015MNRAS.449.3700R}, whose 16–84 and 5–95 percentile ranges are presented as the dark blue and light blue shaded regions. For a specific $\Sigma_\text{gas}$ bin, the scatter in $\Sigma_\text{SFR}$ is  $1.0$--$1.4$\,dex, which is comparable to FIGGS's scatter of $1.3$--$1.8$\,dex for the $0.15<\log(\Sigma_\text{gas})<0.75$ range. The medians of $0.2\Zsun$/full data roughly follows the \cite{2015ApJ...814L..30E} star formation relation for dwarf galaxies and outer disks, characterized by a power-law index of $2$ (black dashed line). This power-2 star formation relation suggests that the thickness of the disk is regulated by the balance between self-gravity and vertical pressure, in regions with low stellar surface density \citep[e.g.][]{2010ApJ...721..975O,2018ApJ...854...16E}. The $0.02\Zsun$/full also roughly follows the power-$2$ star formation relation, while the intercept is smaller, implying a longer dynamical time-scale for star formation in the low metallicity environments due to the inefficient cooling. 
 The $0.02\Zsun$/noRT results lie outside the 16-84 percentile range of the FIGGS data, exhibiting higher SFR surface density, but still lower than the classical \cite{1998ApJ...498..541K} relation. The slope of its KS relation is even shallower than the classic slope of 1.4, indicating a more rapid and efficient star formation process in the absence of momentum injection and photoheating from radiative feedback.

On the $\Sigma_\star^{0.5}\Sigma_\text{gas}$--$\Sigma_\text{SFR}$ plane (Fig.~\ref{fig:KSR} middle panel), $0.2\Zsun$/full roughly follows the extended KS relation given by \cite{2018ApJ...853..149S}. All three simulations exhibit a similar slope, indicating the significant role of pre-existing disk stars in regulating star formation in our simulated galaxies. 

The results in the KS and extended KS planes of our full feedback simulations demonstrate the ability of the RIGEL model to capture the self-regulated galactic star formation environments. In the $0.2\Zsun$/full galaxy, both the KS and extended KS relations are consistent with the observed empirical relations, while the results of $0.02\Zsun$/full suggest a longer dynamical time-scale for star formation (a smaller intercept for the relations) in extremely metal-poor environments.

\subsection{Morphologies}
\label{sec:Morph}

\begin{figure*}
	\includegraphics[width=2\columnwidth]{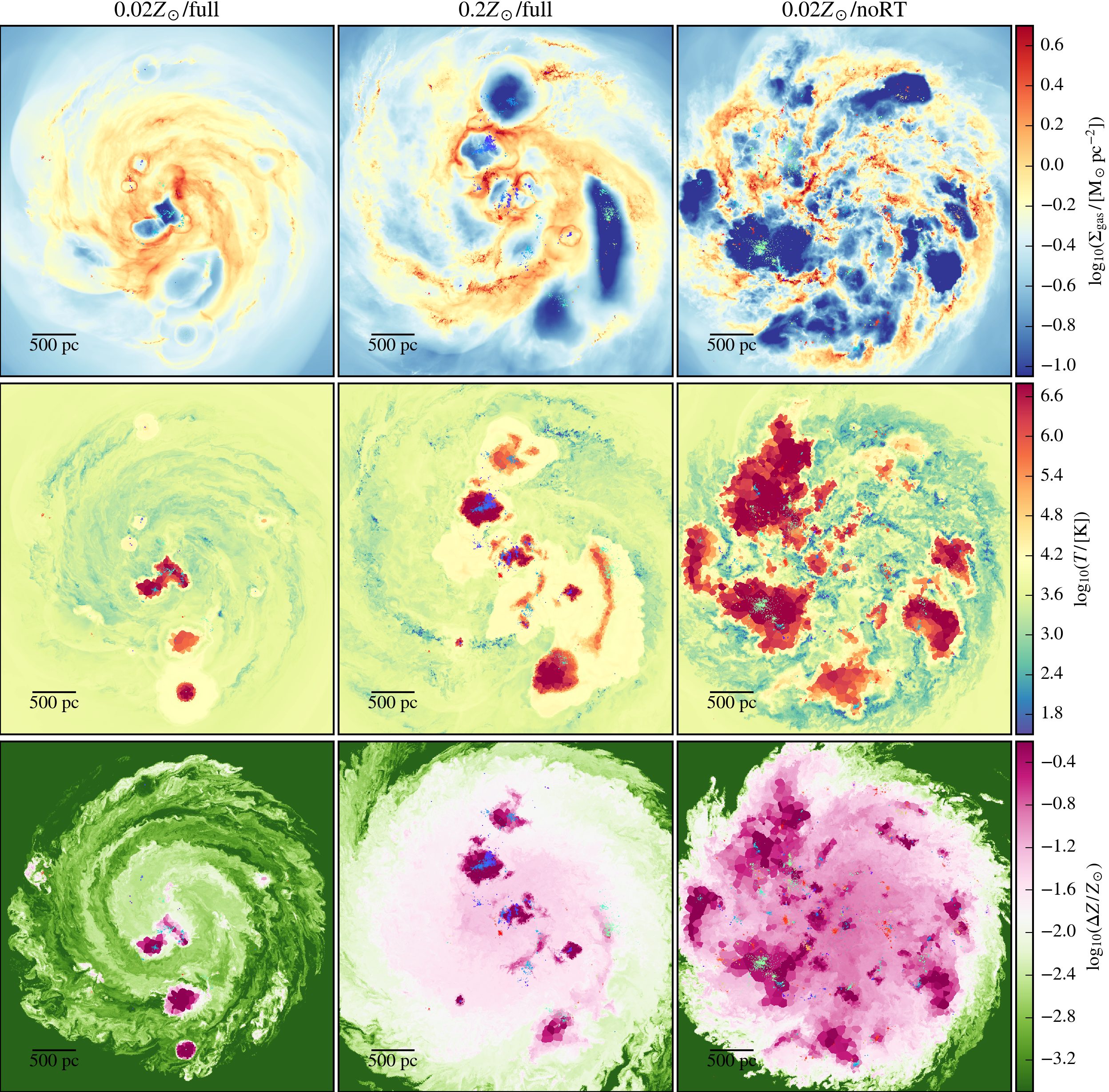}
    \caption{Face-on images of the integrated gas mass surface density ($\Sigma_\text{gas}$; top row), along with mid-plane slices of gas temperature ($T$; middle row), and metallicity increment ($\Delta Z=Z-Z_\text{init}$; bottom row) of the simulated galaxies at 900\,Myr.  The three columns correspond to the different runs, as labeled above the panels. From left to right, they are $0.02\Zsun$/full, $0.2\Zsun$/full, and $0.02\Zsun$/noRT, respectively. The overset points represent young stars ($<100$\,Myr) with ages color-coded from violet (youngest) to red (oldest).}
    \label{fig:face_on_prjection}
\end{figure*}
To provide a panoramic view of the entire galactic disk, Fig.~\ref{fig:face_on_prjection} presents the face-on surface density distribution (the first row) and mid-plane slices of gas temperature (the second row) and metallicity increment ($\Delta Z=Z-Z_\text{init}$, the third row) 
of the simulated galaxies at 900\,Myr for the $0.02\Zsun$/full, $0.2\Zsun$/full, and $0.02\Zsun$/noRT models. Young stars (less than 100\,Myr old) are color-coded, ranging from purple for the youngest (0\,Myr) to red for the oldest (100\,Myr). 

In all three galaxies, the ISM exhibits a multiphase structure. The galactic disks display filamentary and clumpy patterns, with many bubbles amidst the clouds, which illustrate the ongoing stellar feedback processes. Bubbles with moderate density contrast are mainly carved out by early feedback from radiation and stellar winds, while those with high density contrast are primarily the result of SN explosions.

The $0.02\Zsun$/full galaxy presents a relatively smooth gas distribution and less clumpiness. Cold ($<100$\,K) gas is only found in sporadic small clumps, while a large part of the disk remains warm. Hot gas also only appears in relatively isolated bubbles blown by the SN explosion. The smaller warm bubbles are expanding \HII regions of young massive stars. In contrast, in $0.2\Zsun$/full, although warm gas still dominates, the structure is more filamentary and clumpy due to more rapid cooling caused by the higher metal content. Hot SN bubbles are interconnected to form superbubbles on a kiloparsec scale. The ISM in $0.02\Zsun$/noRT is ripped into fragments by intense SN feedback due to its highest SFR (see Section~\ref{sec:starformation}). Interestingly, though a large fraction of the disk is occupied by hot superbubbles, a considerable amount of cold gas coexists with these large hot bubbles (see Section~\ref{sec:ISM_properties} for the phase diagrams and distributions of gas). This coexistence is explained by the fact that winds and SNe primarily heat and disrupt molecular clouds locally, while maintaining the warm state of diffuse gas requires FUV radiation. Furthermore, the absence of radiation prevents efficient evacuation of the dense gas surrounding massive stars, which suppresses the efficiency of SN feedback. Consequently, the cold molecular gas can persist instead of being destroyed by FUV radiation that permeates the entire disc. 

The metallicity distribution in these galaxies shows significant spatial variations, where areas that are enriched beyond normal levels are associated with SN bubbles. Small-scale eddies can be found that trace the turbulent flow of gas. Generally, the central regions exhibit higher metallicity due to the higher SFR, while metallicity decreases gradually as distance from the galactic center increases. In the non-star-forming disk, the metallicity distribution appears relatively uniform and smooth, suggesting that the over-enriched bubbles could blend with the ISM and fade away rapidly.

\begin{figure*}
	\includegraphics[width=2\columnwidth]{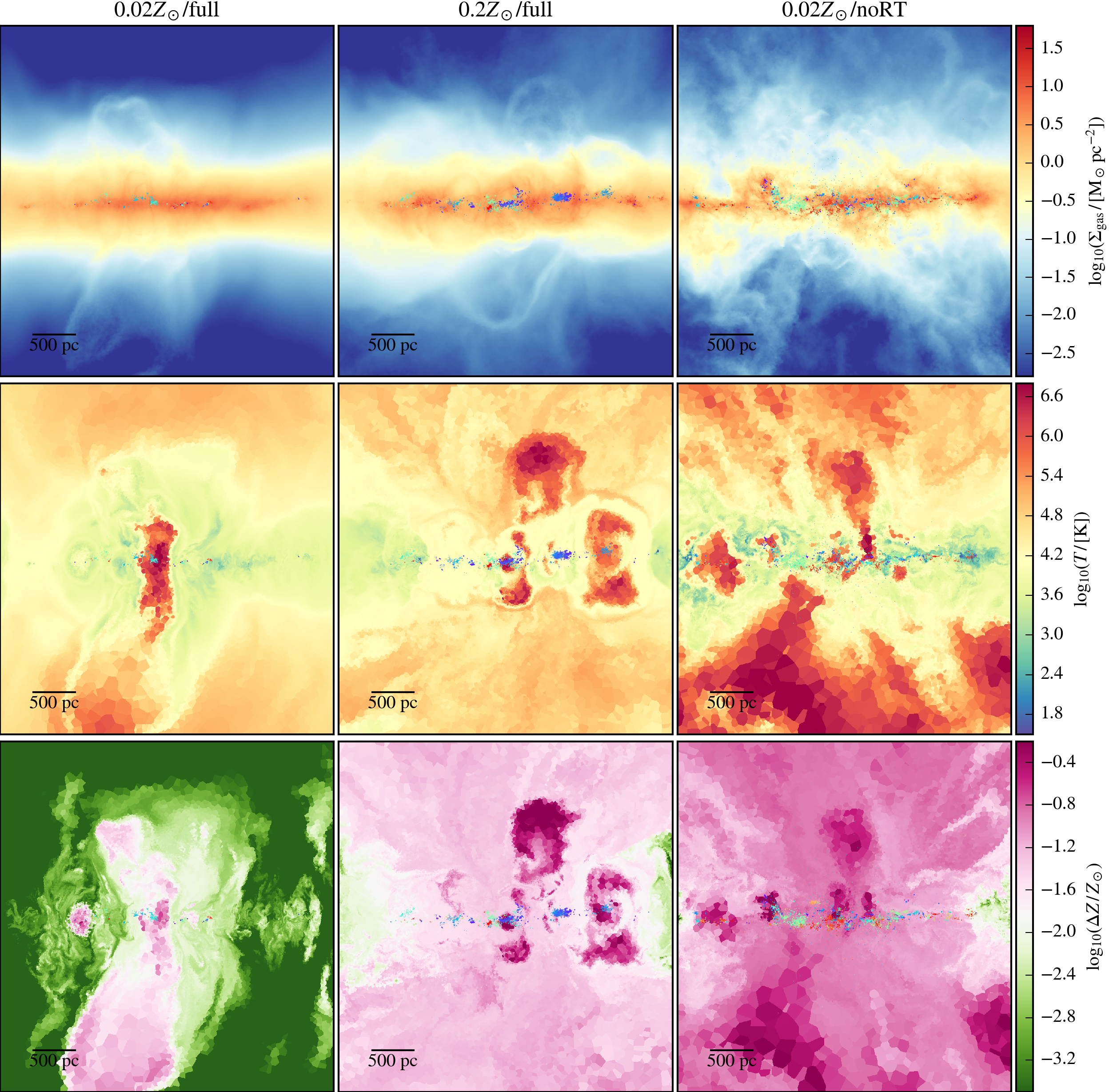}
\caption{Similar to Fig.~\ref{fig:face_on_prjection}, but viewed from an edge-on perspective.}
    \label{fig:edge_on_prjection}
\end{figure*}

Fig.~\ref{fig:edge_on_prjection} presents the same projections and slices but viewed from an edge-on perspective. In the $0.02\Zsun$/full galaxy, the disk exhibits a spindle shape that flares outwards in the outer disk due to pressure support. Enriched gas is confined to specific chimneys within its CGM, while a significant portion of the CGM remains pristine. On the other hand, both the $0.2\Zsun$/full and $0.02\Zsun$/noRT galaxies feature a more enriched CGM, with the enriched gas occupying almost the entire volume of the plotted box. The spindle shape is disrupted by multiple feedback-driven chimneys through which hot outflow gas escapes.
Notably, the CGM of the $0.02\Zsun$/noRT galaxy is hotter and more enriched than that of the $0.2\Zsun$/full galaxy. These results demonstrate that the inclusion of energetic radiative feedback reduces the outflow of gas and metals by suppressing the SFR. Further analysis of the outflow properties of our galaxies will be detailed in Section~\ref{sec:outflow}.

\subsection{Interstellar medium properties and structures}
\label{sec:ISM_properties}
\subsubsection{Mass and volume fractions of different interstellar medium phases}
\begin{figure*}
	\includegraphics[width=2\columnwidth]{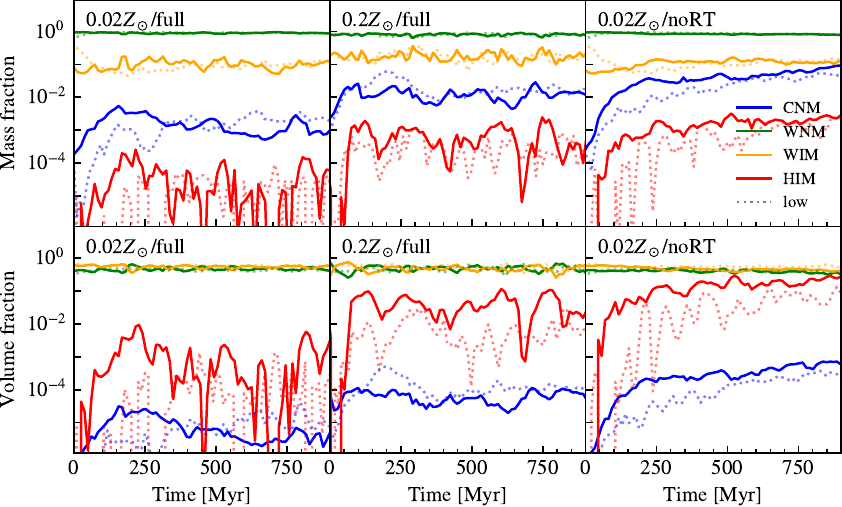}
       \caption{Time evolution of the mass fractions (top row panels) and volume fractions (bottom row panels) of gas in the ISM regions ($R < 2$\,kpc and $|z| < 1$\,kpc) in the cold neutral medium (CNM, $T<100$\,K, blue curves), the warm neutral medium (WNM, $100<T<8000$\,K, green curves), the warm ionized medium (WIM, $8000<T<100\,000$\,K, orange curves), and hot ionized medium (HIM, $T>100\,000$\,K, red curves) phases. The dotted curves are from the low-resolution runs. Fig.~\ref{fig:ISMpdf} shows how these phases are distributed across the mass-weighted ISM probability distribution functions. Both higher metallicity and noRT runs exhibit significantly more CNM and HIM fractions due to the more efficient gas cooling and collapse (induced by higher cooling rate and lower heating rate in $0.2\Zsun$/full and $0.02\Zsun$/noRT, respectively), and the resultant higher SFRs.}
    \label{fig:fractions}
\end{figure*}
Fig.~\ref{fig:fractions} quantifies the time evolution of the mass fractions (top row panels) and volume fractions (bottom row panels) of gas in the ISM region ($R < 2$\,kpc and $|z| < 1$\,kpc) for the cold neutral medium (CNM, $T<100$\,K), the warm neutral medium (WNM, $100$\,K$<T<8000$\,K), the warm ionized medium (WIM, $8000$\,K$<T<10\,000$\,K), and hot ionized medium (HIM, $T>10\,000$\,K) phases. 

In all three simulations, warm gas dominates both the mass and volume fractions. The majority of the mass is found in the WNM phase, while WIM occupies a smaller mass but a similar volume in the ISM as the WNM. The CNM only makes up about $10^{-3}$ of the ISM mass in the $0.02\Zsun$/full galaxy, with its mass fraction increasing by a factor of around 10 in the $0.2\Zsun$/full galaxy due to more effective metal cooling. Notably, in the $0.02\Zsun$/noRT galaxy, the CNM mass fraction is about 40 times higher than in the $0.02\Zsun$/full galaxy in the absence of radiative feedback. As a result, $0.2\Zsun$/full and $0.02\Zsun$/noRT exhibit higher SFRs due to the availability of more cold, dense gas for star formation. The higher SN rates in these two galaxies produce more hot gas as well. In $0.02\Zsun$/full, the volume fraction of HIM is only $\sim10^{-3}$ due to the low SN rate, while it constitutes a significant volume fraction ($\sim0.1$) in $0.2\Zsun$/full and $0.02\Zsun$/noRT. On the other hand, the volume occupied by CNM is always negligibly small in all three galaxies as CNM only appears in dense clumps. The mass and volume fractions of the low-resolution simulations are presented in dotted curves, which show very similar warm and cold gas factions to those of the high-resolution simulations. The HIM volume fractions exhibit the largest systematic differences with resolution, possibly due to the slightly lower SFR and potentially unresolved cooling at low resolution, which results in values up to 0.8~dex lower in the low-resolution simulations.

\subsubsection{Phase diagram and interstellar medium distributions}
\begin{figure*}
	\includegraphics[width=2\columnwidth]{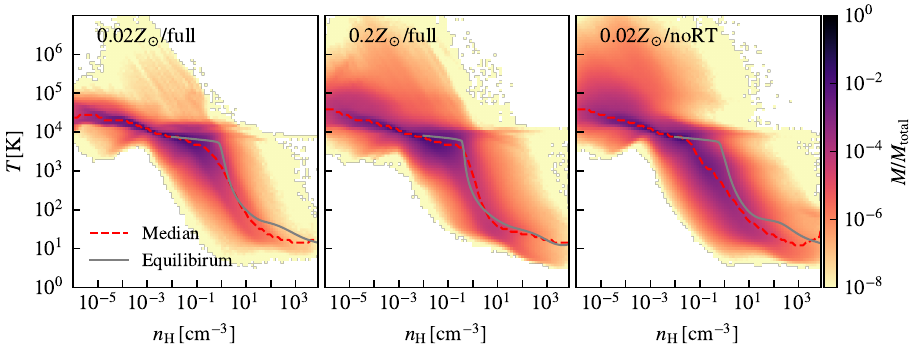}
       \caption{Mass-weighted gas phase diagrams for the high-resolution galaxies averaged over the snapshots from $t = 300$\,Myr to $t = 1000$\,Myr with a time interval of 10\,Myr. The red dashed curves represent the median temperature within each density bin. The grey curves are the equilibrium temperature assuming $G_0=0.01$, $0.1$, and $0.00324$ for $0.02\Zsun$/full, $0.2\Zsun$/full, and $0.02\Zsun$/noRT, respectively.}
    \label{fig:phasediagram}
\end{figure*}

\begin{figure*}
	\includegraphics[width=2\columnwidth]{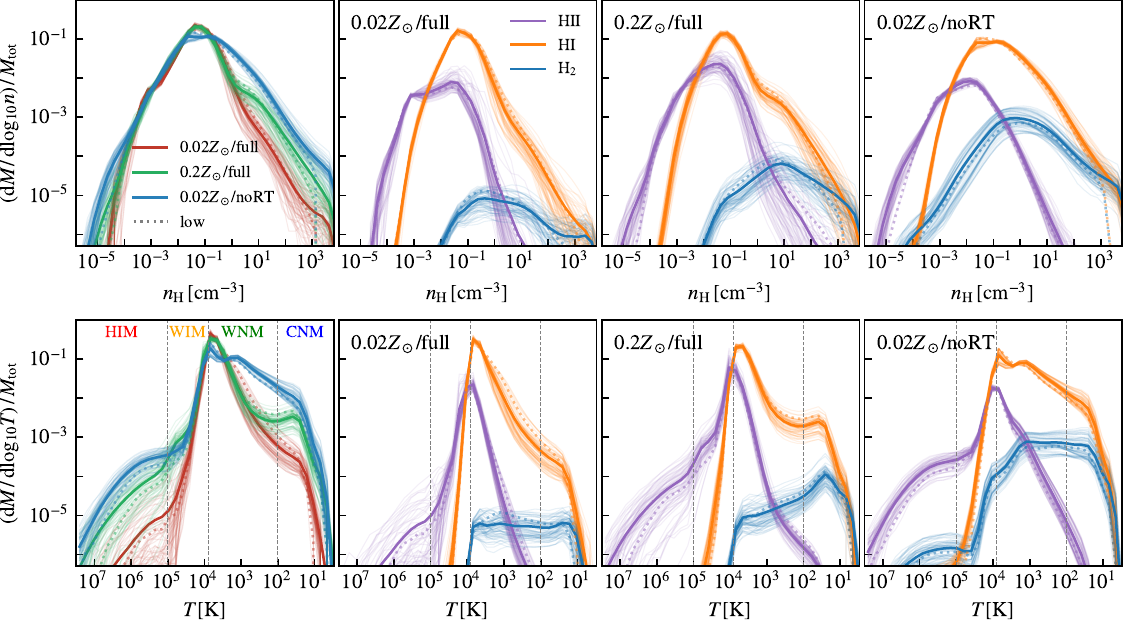}
\caption{Probability Distribution Functions (PDFs) of gas density (top row) and temperature (bottom row) in the ISM region, weighted by mass for a stack of the simulated galaxies from 300\,Myr to 1000\,Myr (solid curves) in the ISM regions. The translucent curves are the PDFs of individual snapshots with a spacing of 10\,Myr. The $x$-axes of the $T$-PDFs are inverted to plot the hot, diffuse gas on the left and the cold, dense gas on the right as the $n$-PDF. The dotted curves show the stacked results for the low-resolution simulation. The first column shows the PDFs of all the gas in the disk, while the second to fourth columns show the mass distribution of gas for different galaxies weighted by \HII (purple), \HI (orange), and \ce{H2} (blue) fractions. The dashed vertical lines on the $T$-PDFs indicate the boundaries of the CNM, WNM, WIM, and HIM phases, whose corresponding mass and volume fractions in the ISM have been presented in Fig.~\ref{fig:fractions}. The ISM in three galaxies with different metallicities and
feedback channels are all dominated by the warm ($\sim10^4$\,K) diffuse ($\sim1$\,cm$^{-3}$) \HI gas. Radiative feedback leads to significantly less buildup of \ce{H2} in metal-poor environments.}
    \label{fig:ISMpdf}
\end{figure*}

To better understand the ISM structures in our simulated galaxies, we present the gas phase diagrams for the high-resolution galaxies averaged over the snapshots from $t = 300$\,Myr to $t = 1000$\,Myr with a time interval of 10\,Myr in Fig.~\ref{fig:phasediagram}. We also show the median temperature (red dashed curves) and the equilibrium temperature (grey curves) as a function of density. The median curves roughly follow the equilibrium curves, though there is a significant dispersion in the gas distribution due to the feedback processes and the turbulence stirring. The gas distribution presents a typical bi-stable structure, featuring stable warm and cold phases connected. The hot gas heated by SN explosions lies in the upper left quadrant of the phase diagrams, and the gas in the diagonal branches of the lower right quadrant represents thermally unstable gas transitioning from its warm to cold phase. In the $0.02\Zsun$/full galaxies, both the upper left and lower right quadrants are less populated due to the lower SN and cooling rates. The photoionized gas within \HII regions manifests as a faint, narrow, $\sim10^4$\,K isothermal line in $0.02\Zsun$/full and $0.2\Zsun$/full galaxies.

In the first column of Fig.~\ref{fig:ISMpdf}, we present the mass-weighted probability distribution functions (PDFs) of gas density (top) and temperature (bottom) in the ISM region of the simulated galaxies. The dotted lines are the results of the low-resolution simulations, which closely resemble the high-resolution ones. 
Across all three galaxies, the predominant gas component is warm ($\sim10^4$\,K) and diffuse ($\sim1$\,cm$^{-3}$). The distributions of warm gas in $0.02\Zsun$/full and $0.2\Zsun$/full exhibit substantial similarity, suggesting that the thermal structure of warm gas is insensitive to metallicity. This is due to the dominance of Ly$\alpha$ cooling over the metallicity-dependent cooling and heating processes in this phase. When the density exceeds $\sim1$\,cm$^{-3}$, the gas in $0.2\Zsun$/full is more susceptible to thermal instability, leading to easier cooling and condensation into cold, dense gas. Thus, $0.2\Zsun$/full displays a distinct two-phase behavior with an additional peak of cold gas at temperatures $\lesssim10^2$\,K. As a result, the $0.2\Zsun$/full galaxy contains significantly more cold, dense gas than $0.02\Zsun$/full. Additionally, the $0.2\Zsun$/full also contains more hot, low-density gas because of its higher SN rate. 

Notably, the ISM phase structure of the $0.02\Zsun$/noRT galaxy differs significantly from that of $0.02\Zsun$/full, even for the warm diffuse gas. As we have noticed from the morphologies, large superbubbles and cold, dense clumps are coexistent in this galaxy. This feature is also reflected in the PDFs, where the fractions of both hot and cold gas are notably elevated compared to the other two galaxies. This discrepancy arises because the absence of heating by the ISRF allows the diffuse ISM to cool more easily, rendering it susceptible to thermal instability. As a result, it contains a higher proportion of cold dense gas compared to $0.02\Zsun$/full, despite its lower metallicity and higher SFR. Furthermore, $0.02\Zsun$/noRT exhibits the highest amount of hot gas. This is not only because its higher SFR leads to more SN explosions, but also because the more clustered star formation in the absence of radiative feedback increases the clustering of SNe \citep[see Section~\ref{sec:cluster} and][]{2021MNRAS.506.3882S}. 

Now, we shift our focus to the distribution of hydrogen species tracked by the nonequilibrium chemical network across its ionized, neutral, and molecular phases (\HII, \HI, and \ce{H2}). In the second to fourth columns of Fig.~\ref{fig:ISMpdf}, we present the mass distribution of gas for different galaxies weighted by the \HII (purple), \HI (orange), and \ce{H2} (blue) fractions. We see that all three galaxies are atomic-dominated. In these galaxies, \HI gas dominates at densities ranging from $n_\text{H}\sim 10^{-2.5}$\,cm$^{-3}$ to at least $10^3$\,cm$^{-3}$, covering a broad range of temperatures of $T\sim10-10^4$\,K. \HI can be ionized to \HII by collisional ionization in warm and hot gas with $T\gtrsim20\,000$\,K, by UVB in unshielded low-density regions ($\lesssim10^{-2}$\,cm$^{-3}$), and by local photoionization in the vicinity of young massive stars. \HII gas dominates the warm diffuse ISM and the hot low-density gas that corresponds to SN-driven bubbles, and it has an extended distribution towards the dense ($\sim10$\,cm$^3$) cold ($\sim100$\,K) gas. In the $0.02\Zsun$/noRT galaxy, the distribution of \HII gas exhibits symmetry in density in the absence of radiation fields, whereas in the full feedback simulations, particularly in $0.2\Zsun$/full, there is a departure from symmetry with a high-density tail comprising the photoionized gas within \HII regions.

The \ce{H2} fractions in the full feedback galaxies are primarily determined by competition between dust-phase formation ($\propto Z_\text{d}$) and photodissociation by LW radiation. On the other hand, in the noRT galaxy, \ce{H2} can only be destroyed slowly by collisional processes and CR ionization. In the $0.02\Zsun$/full galaxy, \ce{H2} remains a minority component even in dense gas with $>10^3$\,cm$^3$ due to its low metallicity. Half of the total \ce{H2} mass exists in $n_\text{H}<1$\,cm$^{-3}$ diffuse gas where the \ce{H2} fraction is low. The situation is similar in the $0.2\Zsun$/full galaxy. Although there is more cold, dense gas where \ce{H2} can form, the \HI-\ce{H2} transition occurs at high densities of $>10^3$\,cm$^{-3}$ with a significant fraction of the \ce{H2} mass existing in \HI-dominated gas. The $0.02\Zsun$/noRT galaxy exhibits a broader distribution of \ce{H2} across large temperature and density ranges. \ce{H2} even exists in warm-hot gas ($T\sim 10^4-10^6$\,K) with densities $n_{\rm H}\sim10^{-4}-10^{-2}\ \rm cm^{-3}$, 
because the gas-phase formation of \ce{H2} can be effective in the high-temperature diffuse gas with a non-negligible abundance of electrons \citep{2019ApJ...881..160B}, which should be suppressed in the presence of the ISRF.

\subsubsection{Interstellar radiation field}
\begin{figure}
	\includegraphics[width=\columnwidth]{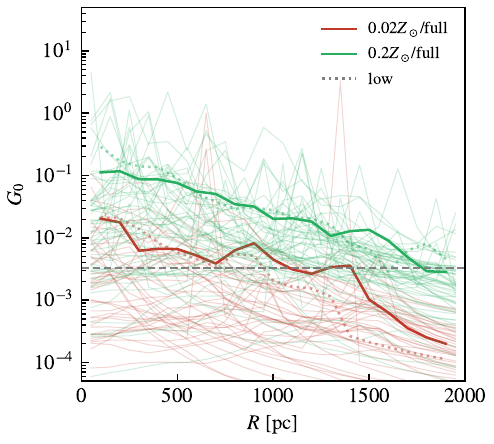}
  \caption{5.8--13.6\,eV ISRF normalized to the \cite{1968BAN....19..421H} unit as a function of $R$ in the ISM regions for the $0.02\Zsun$/full and $0.2\Zsun$/full galaxies. The solid curves are the stacked results from 300\,Myr to 1000\,Myr, while the translucent curves are the results of individual snapshots with a spacing of 20\,Myr. The dotted curves present the stacked results of the corresponding low-resolution galaxies. The grey dashed line shows the value of the background FUV field of $G_{0,\text{UVB}} = 0.00\,324$.}
    \label{fig:ISRF}
\end{figure}

In previous sections, we have seen that the existence of FUV ISRF can significantly change the phase structure of the ISM.
In Fig.~\ref{fig:ISRF}, we present the radial distribution of the FUV ISRF within the ISM regions of the two galaxies with radiative feedback, $0.02\Zsun$/full and $0.2\Zsun$/full. 
The ISFR energy density is described by the conventional parameter $G_0=u_\text{5.8-13.6 eV}/5.29\times10^{-14}$\,erg\,cm$^{-3}$  \citep{1968BAN....19..421H}. The solid curves represent the average ISRF by stacking the snapshots from 300\,Myr to 1000\,Myr, while the faint curves represent individual snapshots taken with intervals of 20\,Myr. In both galaxies, the ISRF decreases gradually from the inner to outer regions, following an approximately exponential decay profile. In the $0.2\Zsun$/full galaxy, the stacked ISRF peaks at around $\sim0.1\,G_0$ in the central area and decreases exponentially to the background level at $R\simeq 2$\,kpc. The ISRF in $0.02\Zsun$/full is roughly one-tenth of that in the $0.2\Zsun$/full galaxy and decreases from the center to the outer regions in a similar manner (for comparison, the SFR is $\sim6\%$ of that in $0.2\Zsun$/full). These results show that the ISRF is proportional to the metallicity of the dwarf galaxy. This correlation arises because the FUV radiation originates from young OB stars, making its intensity proportional to the SFR. As reported in section~\ref{sec:starformation}, the global SFR also shows a positive correlation to the metallicity. However, the increase of ISRF with metalicity is slightly slower than
that of SFR, because of the competition between higher SFR and higher dust opacity, along with the reduced FUV and LW photon production rates of metal-rich stars (see Fig.~\ref{fig:feedbackModel}). The ISRFs of individual snapshots display an exponential decay trend as well, interspersed with multiple spikes indicating local enhancements in FUV intensity near young massive stars and clusters. At any given radius, there can be temporal variations in the ISRF of more than two or even three orders of magnitude across different snapshots. 

The stacked ISRF in the low-resolution galaxies is also plotted with dotted curves, which show minor differences from the high-resolution results for the $0.2\Zsun$ galaxies and the inner part of the $0.02\Zsun$. However, the stacked ISRF in the $0.02\Zsun$/full/low galaxy appears to be two to four times lower than that of $0.02\Zsun$/full in the outer part ($R\gtrsim700$\,pc, roughly half of the scale length of gas disk). This suggests that the low resolution can suppress star formation in the low gas surface density regions. 

\subsubsection{ISM densities around SN explosions}
\label{sec:SN_Density}
\begin{figure}
	\includegraphics[width=\columnwidth]{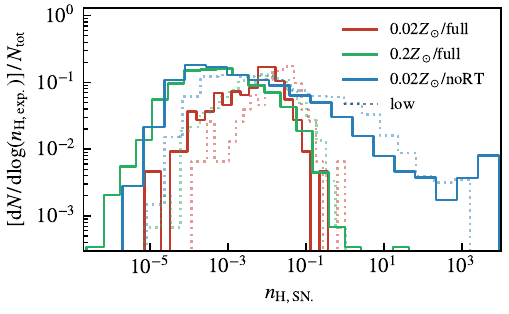}
    \caption{Histogram of the hydrogen number density  of the host gas cell where the SNe injected their explosion energy. The notations of simulations are identical to those in Fig.~\ref{fig:SFR}.}
    \label{fig:SN_density}
\end{figure}

One crucial function of early radiative feedback is to disperse the dense star-forming ISM through radiation pressure and the expansion of \HII regions. This process allows SNe to explode in less dense environments, enhancing the efficiency of coupling the explosion energy to the gas. In Fig.~\ref{fig:SN_density}, we present the distribution of the densities of the gas cells where the SNe released their explosion energy.

In the case of the two full feedback galaxies, most of the SNe exploded in $n_\text{H}<1$\,cm$^{-3}$ diffuse gas because of the preprocessing effect of radiation. The explosion density distributions of these two runs are similar at the high-density end, but for $0.2\Zsun$/full, there are more SNe exploding in $n_\text{H}<0.01$, cm$^{-3}$ low-density gas. This can be understood by the scenario that the radiative feedback in these two simulations rarefies the ISM to a similar degree, so the explosion densities of the first SN in star clusters follow a similar distribution. On the other hand, in $0.2\Zsun$/full, there are more clustered SNe because of its higher SFR, while most star clusters in $0.02\Zsun$/full host only one SN. In a cluster with multiple SNe, the first several SNe will create a hot bubble, allowing the subsequent SNe to explode within this low-density bubble formed by earlier explosions. Consequently, this results in a higher fraction of SNe exploding in low-density environments in $0.2\Zsun$/full.

In the absence of radiative feedback, the explosion density distribution of $0.02\Zsun$/noRT is significantly broader, reaching up to $10^4$\,cm$^{-3}$. Although the main sequence winds can still provide early feedback, their ability to expel cold dense gas is limited by their low luminosities and efficient cooling \citep{2018MNRAS.478.4799H,2021ApJ...914...89L,2021ApJ...914...90L}. Thus, the first SN of a cluster is likely to occur at a high density and be inefficient in providing feedback and driving outflow (see Section~\ref{sec:outflow} for the analysis of outflow properties). 

\subsubsection{Interstellar medium pressure modulated by feedback}
\label{sec:P-SFR}
\begin{figure*}
	\includegraphics[width=2\columnwidth]{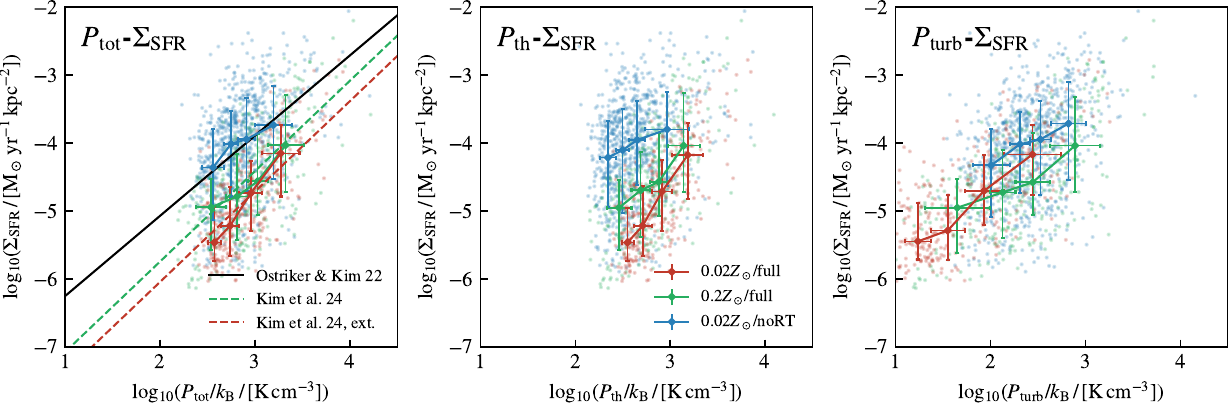}
    \caption{Star formation surface density $\Sigma_\text{SFR}$ as a function of total (left), thermal (middle), and turbulence (right) pressure of the ISM measured at the galactic mid-plane. The $\Sigma_\text{SFR}$ measurement is the same as in Fig.~\ref{fig:KSR}. The black solid line in the left panel presents the relation given by \protect\cite{2022ApJ...936..137O} based on simulations of solar metallicity ISM boxes, while the green and red dashed lines present the metallicity-dependent relation given by \protect\cite{2024arXiv240519227K:Kim}. We noticed that the numerical experiments performed by \protect\cite{2024arXiv240519227K:Kim} only extend to $Z=0.1\,\Zsun$. Therefore, the red dashed line is obtained by extrapolating their fitting function to $Z=0.02\,\Zsun$. The full feedback simulations follows this metallicity-dependent superlinear relation, which exhibit steeper slopes and lower intercepts compared to 0.02$\Zsun$/noRT. }
    \label{fig:P_SFR}
\end{figure*}

Star formation activities and ISM properties interplay with each other. Stellar feedback injects energy and momentum into the ISM, sustaining its turbulent and thermal pressures, whereas the local deficit of pressure due to cooling leads to the gravitational collapse of the ISM and star formation. This concept is encapsulated in the pressure-regulated, feedback-modulated (PRFM) theory of star-forming ISM \citep{2010ApJ...721..975O,2011ApJ...731...41O,2022ApJ...936..137O}. 

In vertical dynamical equilibrium of the galactic disk, the weight of the ISM per unit area, $\mathcal{W}$, is balanced by the total ISM pressure, $P_\text{tot}$, at the galactic mid-plane. The ISM pressure is sustained by star formation activities that injected momentum and energy through feedback. To satisfy the vertical energy and momentum equations simultaneously, $\Sigma_\text{SFR}$ should be proportional to $\mathcal{W}$, and consequently, to $P_\text{tot}$. Therefore, the correlation between ISM pressure and star formation surface density serves as a valuable diagnostic for stellar feedback models.

In Fig.~\ref{fig:P_SFR}, we present the $\Sigma_\text{SFR}$ as a function of the ISM pressure measured at the galactic mid-plane. Here, the total pressure $P_\text{tot} = P_\text{th}+P_\text{turb}$, the thermal pressure $P_\text{th} = \rho c^2_\text{s}/(5/3)$, 
 and the turbulent pressure $P_\text{turb} = \rho \sigma_z^2$ are plotted separately. 
In the two full feedback galaxies, $0.02\Zsun$/full and $0.2\Zsun$/full, the ISM pressure is dominated by the thermal pressure which spans a similar range of $10^{2.5}$\,K\,cm$^{-3}<P/k_\text{B}<10^{3.5}$\,K\,cm$^{-3}$. The turbulent pressure in these galaxies shows a wider variation, with $P_\text{turb}$ being overall lower in $0.02\Zsun$/full compared to $0.2\Zsun$/full. The similarity in thermal pressure is due to the prevalence of the warm diffuse ISM, maintaining nearly identical thermal conditions when $P/k_\text{B}<10^4$\,K\,cm$^{-3}$. The lower turbulent pressure in $0.02\Zsun$/full results from the reduced injection of turbulence attributed to its lower SFR. On the other hand, $0.02\Zsun$/noRT exhibits a slightly lower thermal pressure due to the lack of heating by ISRF, while its turbulent pressure is significantly higher compared to $0.02\Zsun$/full, attributed to its enhanced SFR.

Now we turn our attention to the correlation between pressure and SFR surface density. The left panel of Fig.~\ref{fig:P_SFR} shows that the two full feedback simulations roughly follow the superlinear relation between $P_\text{tot}$ and $\Sigma_\text{SFR}$ given by \cite{2024arXiv240519227K:Kim} in their Fig.~10, which exhibit steeper slopes and lower intercepts compared to $0.02\Zsun$/noRT. The $0.02\Zsun$/noRT simulation roughly follows the relation given by equation~(26a) in \cite{2022ApJ...936..137O}. We noticed that the numerical experiments performed by \cite{2024arXiv240519227K:Kim} only extend to $Z=0.1\,\Zsun$ and we extrapolate their metallicity dependence of $\propto Z^{0.30}$ to $Z=0.02\,\Zsun$ to make a comparison with our 0.02$\Zsun$/full galaxy, showing that their fitting functions might still be applicable at metallicity lower than $0.1\,\Zsun$.
The difference in the intercept between the noRT and full models can be understood qualitatively from two perspectives. Firstly, when considering the same $P_\text{tot}$, or in other words, the ISM weight, the full feedback model exhibits a lower SFR due to the impact of radiative feedback in reducing SFE. On the other hand, with the same $\Sigma_\text{SFR}$, the full feedback model will yield more feedback as a result of the presence of radiative feedback and its preprocessing impact on the ISM. The difference in slopes will be explored in a separate study, which will cover a broader dynamical range of $P_\text{tot}$ and $\Sigma_\text{SFR}$, and will conduct a more detailed quantitative analysis. 

The feedback yields ($P/\Sigma_\text{SFR}$) of the two full feedback galaxies also have differences. The feedback yields for the total pressure are comparable when $\Sigma_\text{SFR}$ is high ($\gtrsim10^{-4.5}\,\Msun$\,yr$^{-1}$\,kpc$^{-2}$), but the $0.2\Zsun$/full yields lower pressure in the low $\Sigma_\text{SFR}$ regions. This is because the total pressure is dominated by the thermal component. In the case of $0.02\Zsun$/full, feedback is expected to yield higher thermal pressure compared to $0.2\Zsun$/full, especially at lower $\Sigma_\text{SFR}$. This phenomenon can be attributed to inefficient cooling in metal-poor environments, particularly in regions with low $\Sigma_\text{SFR}$ where the gas surface density is low, making the gas more susceptible to heating and less prone to cooling. In contrast, feedback in $0.02\Zsun$/full will yield a lower turbulent pressure than in $0.2\Zsun$/full, 
suggesting a less efficient coupling of feedback to ISM in metal-poor environments. This could be due to the fact that high thermal pressure in low-metallicity environments results in a less efficient momentum coupling and turbulence driving from stellar feedback \citep[e.g.][]{2018MNRAS.478.4799H}. The interplay between thermal and turbulent pressure then sets the total ISM pressure which equilibrates the ISM weight and regulates the star formation.

\subsection{The formation and properties of star clusters}
\label{sec:cluster}
To study the properties of star clusters in our simulations, we perform a four-dimensional ‘Friends of Friends’ (4D-FoF) analysis on the newly formed stars in the ${\bm x}_\text{form}$--$t_\text{form}$ space, where ${\bm x}_\text{form}$ and $t_\text{form}$ are the formation coordinates and formation time of star particles recorded in the simulation snapshots. Particles with physical connection are linked as a group using a linking length $l_\text{link}$ and a linking time $t_\text{link}$. We choose $l_\text{link}=5$\,pc and $t_\text{link}=5$\,Myr motivated by the observed size and age spread of star clusters. We found that tuning $l_\text{link}$ or (and) $t_\text{link}$ by a factor of two has minor effects on the cluster mass functions and age spreads, while larger or smaller choices result in either overlinking or underlinking. We only keep the identified groups that have at least 35 members as star clusters \citep{2003ARA&A..41...57L}. In the case of low-resolution simulations, we set a minimum membership of 4 to achieve a similar minimum cluster mass to that of high-resolution simulations.

\subsubsection{Cluster initial mass functions}
\begin{figure}
	\includegraphics[width=\columnwidth]{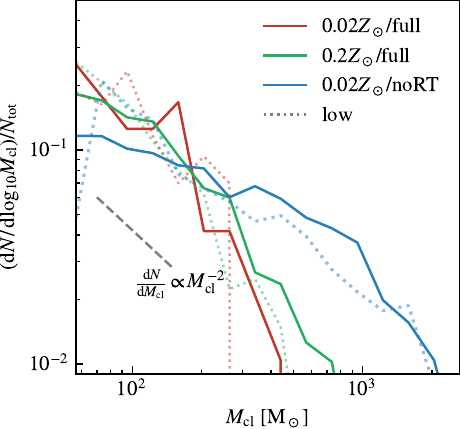}
    \caption{Cluster initial mass functions of the simulated galaxies. The solid curves are the results of the high-resolution simulations, while the dotted curves are the results of the corresponding low-resolution galaxies. The grey dashed line presents the observed power-law relation with a slope of $-2$. Radiative feedback steepens the mass functions by rapidly dispersing the clouds and shutting off the further star formation.}
    \label{fig:CIMF}
\end{figure}

In Fig.~\ref{fig:CIMF}, we present the cluster initial mass functions (CIMFs) of our simulated galaxies. The initial mass of each cluster is defined as the sum of the initial mass of its member star particles. In the $0.2\Zsun$/full galaxies, the CIMF follows a power-law distribution of $\mathrm{d}N/\mathrm{d}M_\text{cl}\propto M_\text{cl}^{-\alpha}$ with a slope $\alpha = 2.07$, which is in agreement with the observed $-2$ slope \citep[e.g.][]{2010ARA&A..48..431P,2019ARA&A..57..227K}. The CIMF of the $0.02\Zsun$/full galaxies has a slightly steeper slope of $-2.40$, although the uncertainty of the slope is large due to low number statistics. Turning off the radiative feedback results in a significantly shallower power-law slope of -1.63, indicating that more massive clusters are formed. Thus, we conclude that radiative feedback reduces the clustering of star formation by decreasing the number of massive clusters and shaping the CIMF steeper. This indicates that radiative feedback disperses clouds quickly and efficiently.

The power-law slopes for the low-resolution simulations are $-2.16$ for $0.2\Zsun$/full/low, $-2.34$ for $0.02\Zsun$/full/low, and $-1.80$ for $0.02\Zsun$/noRT/low, respectively. Despite a minor deviation from the high-resolution values, our qualitative findings regarding the shapes of their mass distributions remain robust to numerical resolution.

\subsubsection{Age spreads and feedback timing}
\label{sec:fbtiming}
\begin{figure}
	\includegraphics[width=\columnwidth]{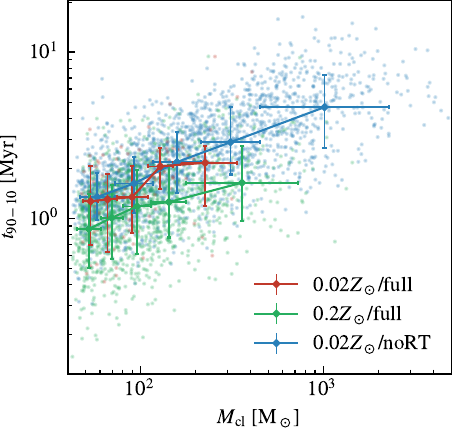}
    \caption{Age spreads of the clusters ($t_{90-10}$) as the function of the cluster initial mass. The data
with errorbars present the median age spreads of the [0,20), [20,40), [40,60), [60,80), [80,100] cluster initial mass percentile bins, and the vertical errorbars show the 16 and 84 percentiles of the age spreads in each bin. More massive clusters exhibit larger age spreads, suggesting that the final mass of the cluster is influenced by the duration of star formation within the cloud when there is no disruptive stellar feedback.}
    \label{fig:Mcl}
\end{figure}

\begin{figure}
	\includegraphics[width=\columnwidth]{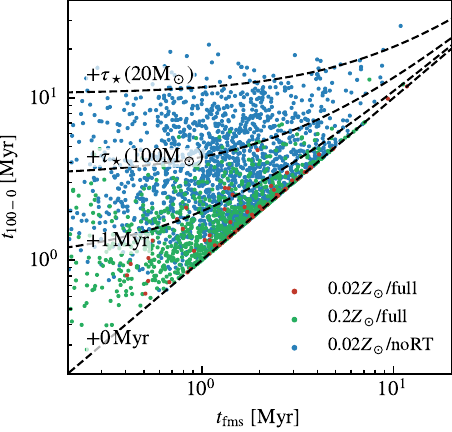}
    \caption{Time between the formation of the first and last member star of a cluster ($t_{100-0}$) as the function of the time of the first massive star in the cluster ($t_\text{fms}$). The black dashed curves present the line of $t_{100-0}=t_\text{fms}+\{0\,\text{Myr},1\,\text{Myr},\tau_\star(100\,\Msun),\tau_\star(20\,\Msun)\}$, where $\tau_\star(100\,\Msun)$ and $\tau_\star(20\,\Msun)$ are the lifetimes for $100\,\Msun$ and $20\,\Msun$ massive stars. The time between $t_{100}$ and $t_\text{fms}$ are smaller than $1$\,Myr for most clusters in the full feedback simulations, much shorter than the time required for the first SN explosion.}
    \label{fig:tfms}
\end{figure}

Now, we study the cluster formation time based on the age spreads of the member stars. We use the age spreads as a proxy to study the time-scale of feedback dispersing the star-forming clouds. We quantify the age spread of a cluster by estimating the duration needed for a specific percentage of member stars to be formed within the cluster. Since the mass of each star particle (inherited from its progenitor gas cell, not $M_\text{fb}$) is almost the same, it describes a mass-weighted cluster formation timescale. In Fig.~\ref{fig:Mcl}, we plot the age spreads of the cluster when 10 per cent to 90 per cent of its final members are formed ($t_{90-10}$) as a function of the cluster initial mass. As can be seen, the age spread shows a clear positive correlation with the cluster initial mass, where the more massive clusters exhibit larger age spreads, especially for the $0.02\Zsun$/noRT. This correlation suggests that the final mass of the cluster is influenced by the duration of star formation within the cloud in the absence of disruptive stellar feedback.

To quantitatively analyze the correlation between feedback and cluster growth, we study the formation time of the first massive star in the cluster ($t_\text{fms}$). In Fig.~\ref{fig:tfms}, we present the duration between the formation of the first and last member star of a cluster ($t_{100-0}$) as a function of $t_\text{fms}$. The timing at which SN feedback occurs is plotted with black dashed curves. Among these curves, the ``$+\tau_\star(100\Msun)$'' curve corresponds to the minimum explosion time of the first potential SN in the cluster, given that $100\,\Msun$ represents the maximum stellar mass in our model. As shown by the results of $0.02\Zsun$/full and $0.2\Zsun$/full, the time differences between $t_{100}$ and $t_\text{fms}$ are smaller than $1$\,Myr for most clusters in the full feedback simulations, much shorter than the time required for the first SN explosion. The clusters cease to grow rapidly after the formation of the first massive stars, suggesting that the clouds are dispersed rapidly by the pre-SN radiative feedback. Without radiative feedback, the clouds can remain in their star-forming phase for longer periods. While the stellar winds can also contribute to early feedback, its dynamic influence is relatively ineffective. Therefore, the clusters grow over long periods and accumulate higher final masses.

In summary, in low-surface density, low-metallicity environments such as our dwarf galaxies, the mass growth period and the final mass of low-mass ($<10^4\,\Msun$) young star clusters are closely correlated with the timing of feedback emergence. With effective radiative feedback, the appearance of the first massive star nearly coincides with the dispersal of its birth molecular cloud, leading to the inhibition of further star formation. SN feedback typically does not contribute to cloud dispersal. The absence of radiative feedback allows the cloud to exist longer and form new stars, and SNe also play a role in disrupting clouds in this case.

\subsection{Outflow properties}
\label{sec:outflow}
\begin{figure*}
	\includegraphics[width=2\columnwidth]{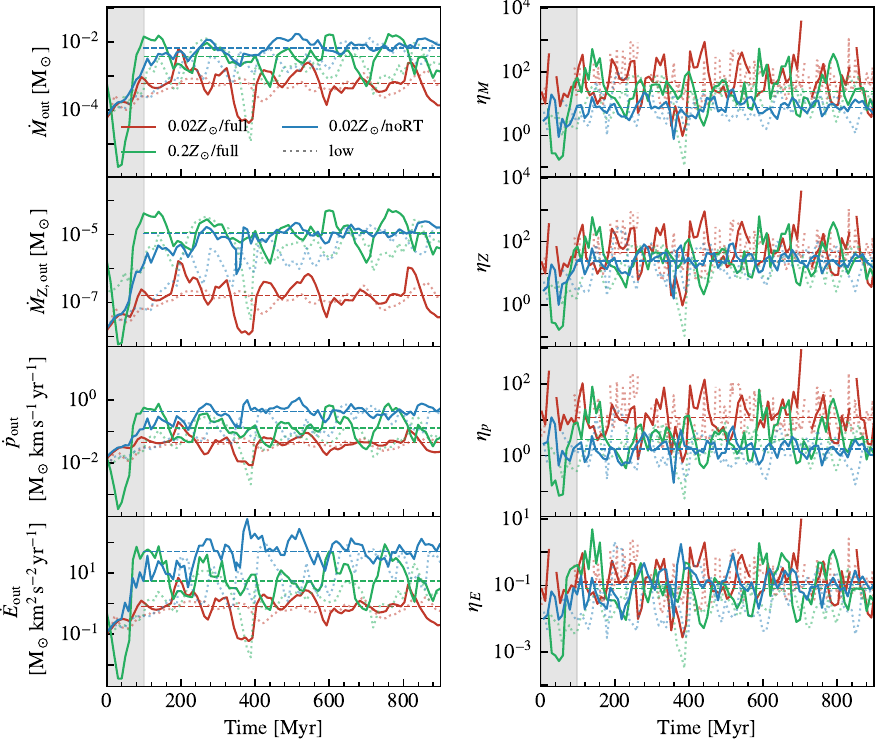}
    \caption{Outflow properties of the simulated dwarf galaxies. Left: the outflow rate of gas mass, metal mass, momentum, and energy measured at $|z|=1$\,kpc. Right: the mass-loading, metal-loading, momentum-loading, and energy-loading
    factors computed as outflow rate per locking rate as defined in equation~(\ref{equ:loading}). Dashed horizontal lines indicate the median values of outflow rates and loading factors in the high-resolution simulations after 100\,Myr (outside grey shaded region).}
    \label{fig:outflow}
\end{figure*}

In previous subsections, we show that radiative feedback and metallicity largely affect star formation activities, the ISM structure, and the properties of star clusters. Now we investigate how they shape the large-scale outflow properties in the disk-halo interface of the simulated dwarf galaxies. We compute the outflow rates of a physical quantity $q$ (i.e., gas mass $\dot{M}_\text{out}$, metal mass $\dot{M}_\text{Z,out}$, momentum $\dot{p}_\text{out}$, energy $\dot{E}_\text{out}$) as the sum of $q_k$ over all gas particles within a slab of thickness $\Delta L$ centered at height $|z|$
\begin{equation}
    \dot{q}_\text{out}=\frac{1}{\Delta L}\sum_k {\bm v}_k\cdot {\bm n} \, q_k\,,\,({\bm v}_k\cdot {\bm n}>0)\,,
\end{equation}
where ${\bm v}_k$ is the velocity of the gas cell $k$ and ${\bm n}$ is the outward norm of the mid-plane. We measure the outflow rates at $|z|=1$\,kpc to capture the outflow properties at the launching interface, and we set $\Delta L=100$\,pc. 

We use loading factors to describe how star formation activities drive the galactic outflow. We define the mass-loading ($\eta_M$), metal-loading ($\eta_Z$), momentum-loading ($\eta_p$), and energy-loading ($\eta_E$) factors as
\begin{equation}
\label{equ:loading}
    \eta_M = \frac{\dot{M}_\text{out}}{\dot{M}_\star}\,,\,
    \eta_Z = \frac{\dot{M}_\text{Z,out}}{Z_\text{ISM}\dot{M}_\star}\,,\,
    \eta_p = \frac{\dot{p}_\text{out}}{p_\text{ej}\dot{N}_\text{SN}}\,,\,
    \eta_E = \frac{\dot{E}_\text{out}}{E_\text{SN}\dot{N}_\text{SN}}\,,
\end{equation}
respectively, where the SFR $\dot{M}_\star$ is time-averaged over 10\,Myr, $Z_\text{ISM}$ is the metallicity of the ISM region, $\dot{N}_\text{SN}=\dot{M}_\star/(100\,\Msun)$ is the SN rate assuming a canonical IMF, $p_\text{ej}=3\times10^4\,\Msun$\,km\,s$^{-1}$ and $E_\text{SN}=10^{51}$\,erg are the momentum carried by ejecta and injected energy of a typical SN.

In the left column of Fig.~\ref{fig:outflow}, we present the outflow rates of gas mass, metal mass, momentum, and energy in 10\,Myr steps for our simulated galaxies. The low-resolution results are presented with dotted lines, which are similar to the high-resolution results. 
The $0.02\Zsun$/noRT run has the highest outflow rates for all quantities due to its highest SFR. 
The outflow rates of the two full feedback galaxies show evident temporal fluctuations, especially the $0.02\Zsun$/full ones, due to their bursty star formation. On the contrary, the $0.02\Zsun$/noRT results present the smallest fluctuations due to its more continuous star formation.

In the right column of Fig.~\ref{fig:outflow}, we present the loading factors defined by equation~(\ref{equ:loading}). The two full feedback galaxies exhibit comparable (less than 1\,dex difference) loading factors for mass, metal, and energy, despite differences in their metallicities and SFRs. The momentum-loading factor of $0.02\Zsun$/full is slightly higher than that of $0.2\Zsun$/full. The similarity in these loading factors is attributed to the stellar feedback mechanisms driving the outflow. The early radiative feedback preprocesses the ISM and reduces the clustering of stars so that sparse SNe explode in less dense gas, though the details of feedback budgets and cooling rates are different in these two galaxies. The mass- and metal-loading factors of $0.02\Zsun$/full and $0.2\Zsun$/full fluctuate rapidly between $10$ and $500$, with a median of $\eta_{M}\approx46$, $\eta_{Z}\approx45$ for $0.02\Zsun$/full and $\eta_{M}\approx22$ and $\eta_{Z}\approx25$ for $0.2\Zsun$/full, respectively. Such high loading factors indicate that the gas and metals expelled from the ISM far exceed the mass consumed in star formation in dwarf galaxies. The energy-loading factors of these two runs are similar as well and vary between $0.01$ and $1$ with a median of $\sim0.1$, which are significantly smaller than the momentum-loading factors that vary mostly between $1$ and $100$. This difference can be explained by the fact that the momentum of the SN ejecta is boosted by converting the SN thermal energy to the kinetic energy of the gas through $P{\rm d}V$ work, while the total energy dissipates during the outflow process.

Remarkably, the mass-loading factor of $0.02\Zsun$/noRT is lower than that of the full feedback simulations by almost an order of magnitude with a median of 7, while the metal- and energy- loading factors are similar to those of the other two runs. 
The reduction of mass and momentum loading factors can be intuitively explained by the scenario in which more SNe explode in dense gas without early radiative feedback, making it more challenging for them to drive outflows (e.g., \citetalias{2017MNRAS.471.2151H}). However, this finding contradicts the recent results of \cite{2021MNRAS.506.3882S}, who suggest that photoionization feedback could suppress galactic outflows by reducing the clustering of SNe. 
Indeed, radiative feedback prohibits the formation of massive clusters so that de-clustering the SNe in our simulations as well (Section~\ref{sec:cluster}). However, this effect is mitigated in our simulations due to the low gas-surface density ($3\,\Msun\,\text{pc}^{-2}$ at the center) of our galaxies, which limits the ability of clustered SNe to drive outflows. The mass-loading factor becomes saturated once the gas has been cleared by multiple SNe, causing subsequent SNe in the same cluster to directly eject their material into the CGM. Moreover, all our galaxies disfavour the formation of large clumps and massive clusters due to the relatively low gas surface density and gas fraction of our initial condition \citep[see][for the impact of gas fraction]{2021MNRAS.508..352R}. Even in $0.02\Zsun$/noRT, only 6 per cent of the clusters are more massive than $1000\,\Msun$ which allows them to host ten SNe. In large clusters, clustered SNe can occur within the low-density bubble created by a previous SN, leading to more effective feedback despite the lack of radiative feedback. In contrast, smaller clusters with a few SNe may struggle to provide substantial feedback. Thus, the ability to form sufficient large clusters also determines the effectiveness of SN explosion in driving galactic outflows in the noRT case.

The loading factors in the low-resolution simulations show a good convergence with the high-resolution ones. Variations in the mass-, metal-, and momentum-loading factors remain below a factor of 1.5, while the energy-loading factors are systematically lower than the high-resolution ones by a factor of up to 2.

\section{Discussion}
\label{sec:discussion}

\subsection{Effects of metallicity and radiative feedback on simulated dwarf galaxies}

In this work, we simulated two galaxies alongside our fiducial simulation, one with ten times higher metallicity ($0.2\Zsun$/full) and another with radiative feedback turned off ($0.02\Zsun$/noRT), to study the effects of metallicity and radiative feedback.

The ten-times-higher metallicity results in a roughly ten-times-higher metal cooling rate in $T>10^5$\,K hot gas and $T<10^4$\,K warm and cold gas. The gas heated by stellar feedback can rapidly cool down, leading to a new cycle of star formation \citep{2017ApJ...845..133S}. As a result, the SFR increases by a factor of 17 due to the availability of more cold gas fuel for star formation, and the intensity of ISRF also increases correspondingly by a factor of 10. The increase of ISRF is slightly lower than that of SFR. This can be explained by the decrease of FUV and LW photon production rates with metallicity and the increase of dust absorption with metallicity. Nevertheless, both the KS and extended KS relations of $0.02\Zsun$/full and $0.2\Zsun$/full are consistent with observed dwarf galaxies \citep[e.g.][]{2015MNRAS.449.3700R} and exhibit superlinear relationships with similar slopes. The results of $0.2\Zsun$/full roughly follow the empirical relations given by \cite{2015ApJ...814L..30E} and \cite{2018ApJ...853..149S}, while those of $0.02\Zsun$/full show a smaller intercept compared with these relations, suggesting a longer dynamical time-scale for star formation in extremely metal-poor environments.
The metal-poor and metal-rich galaxies share several properties in common, including the rapid dispersion of clouds, a power-law CIMF with a steep slope, and the loading factors. This similarity arises from the fact that radiative feedback operates similarly on a cluster scale, and the difference in ionizing photon productions between $0.02\,\Zsun$ and $0.2\,\Zsun$ is within a factor of three. On the other hand, more massive clusters with larger age spreads can be formed in $0.2\Zsun$/full because this galaxy can form larger molecular clouds that require more energetic feedback to disperse.

A key feature of radiative feedback is that it acts immediately following the birth of massive stars, without any delay. On the cloud scale, radiative feedback can rapidly disrupt the star-forming clouds through photoionization and direct radiation pressure, thereby halting further star formation within approximately 1\,Myr after the emergence of massive stars (see Section~\ref{sec:fbtiming}). Thus, a small portion of the star-forming gas is converted into stars before being dispersed by radiative feedback. The growth of cluster masses is also prevented after radiative feedback is initiated, leading to a steep slope of $\lesssim-2$ in the CIMF. However, this situation may not apply in denser high-redshift environments, where FUV radiation from nearby massive stars can even enhance the formation of massive clusters by delaying further star formation and enabling the clouds to accumulate more gas \citep[e.g.][]{2023MNRAS.522.2495G:Garcia,2024ApJ...970...14S:Sugimura}. Future investigations will aim to understand the impact of radiative feedback on star cluster formation in various environments. 

On the galaxy scale, the local and long-range photoheating processes remove the cold fuel for star formation and prevent the cooling of warm ISM. These combined local and global effects of radiative feedback consequently reduce the SFR in our dwarf galaxy by a factor of 92 when compared to the noRT model. The inclusion of radiative feedback naturally regulates the star formation process in consistency with empirical relationships. In terms of the galactic outflows, the outflow rates are suppressed in models with radiative feedback because of the lower SFR. 
On the other hand, radiative feedback preprocesses the environments of SN explosion, causing most SNe explode in $n_\text{H}<1$\,cm$^{-3}$ low-density gas. This effect increases the occurrence of efficient SN explosions, which in turn increases the mass loading factor of our simulated dwarf galaxy.

\subsection{Necessity of early feedback}
\label{sec:necessity}
\begin{figure}
	\includegraphics[width=\columnwidth]{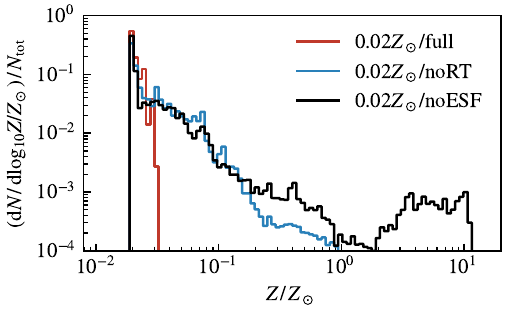}
    \caption{Stellar metallicity distribution of the simulation without any early feedback mechanism (black curve, $0.02\Zsun$/noESF) in comparison with $0.02\Zsun$/noRT and $0.02\Zsun$/full. Stars with extremely high metallicity ($\log{(Z/\Zsun)}>1$) can be formed from the unmixed ejecta in $0.02\Zsun$/noESF due to the absence of early feedback.}
    \label{fig:Zstar}
\end{figure}

We emphasize that early feedback is indispensable for high-resolution multiphase ISM simulations. The maximum gas density of the simulation is constrained by the Jeans criterion ($m_\text{gas}>f_\text{J,s}^3M_\text{J}$, see Section~\ref{sec:SFmodel}), indicating that with higher mass resolutions, stars can be formed in significantly denser gas, which scales with $m_\text{gas}^2$. In the absence of effective early feedback, the explosion density of SNe is linked to the density at which they are formed. For example, if a SN explodes in gas with a density of $10^6$\,cm$^{-3}$, the cool radius and cooling time of its ST phase are only $0.07$\,pc and $22$\,yr \citep[equation~(7) and (8) from][]{2015ApJ...802...99K}. The hot SN remnant can only occupy a very small volume for a very short time before rapidly cooling down to cold gas. Although the SN can still inject a substantial amount of momentum ($p_\text{final}\propto n^{-1/7}$) as long as the ST phase is resolved, it does not heat the gas to the hot and warm phases. Consequently, the SN ejecta may be misidentified as star-forming gas because of its cold and dense nature. Stars can therefore be sampled from the unmixed ejecta and result in the formation of star particles with extremely high metallicities. 

To demonstrate this issue, we run an additional simulation without any early feedback mechanism (i.e., SN only), and present the distribution of stellar metallicity in Fig.~\ref{fig:Zstar}. The SN-only simulation is denoted as $0.02\Zsun$/noESF, in comparison to $0.02\Zsun$/noRT and $0.02\Zsun$/full. Stars with extremely high metallicity ($\log{(Z/\Zsun)}>1$) can be formed from the unmixed ejecta in $0.02\Zsun$/noESF due to the absence of early feedback. Such stars have never been observed so far. In contrast, the stellar metallicity distribution in $0.02\Zsun$/full exhibits a spread of $\sim0.2$\,dex. While $0.02\Zsun$/noRT shows a broad range of stellar metallicities extending to the solar value, the inclusion of wind feedback prevents the formation of excessively metal-rich stars. It is important to note that these phenomena are not a result of numerical over-cooling issues but rather the ``physical'' results of SN exploded in extremely dense gas. Therefore, we must turn at least the wind feedback on to avoid SNe occurring in very dense gas.

\subsection{Comparison to previous work}
\label{sec:comparison}
The initial conditions of our simulations are intentionally set to be similar to that in \citetalias{2017MNRAS.471.2151H} and \citetalias{2023MNRAS.522.3092L} for a direct comparison. The SFRs of $0.2\Zsun$/full and $0.02\Zsun$/full show good agreement with the {\it PE-PI-SN} galaxy in \citetalias{2017MNRAS.471.2151H} and the ``extended'' galaxy in \citetalias{2023MNRAS.522.3092L}. This result is expected since all these simulations incorporate detailed stellar feedback, particularly the pre-SN early feedback, and accurately model the thermodynamics of the multiphase ISM. Consistent with simulations with explicit RT (e.g., \citealt{2018ApJ...865L..22E}; \citetalias{2019MNRAS.482.1304E}; \citealt{2024A&A...681A..28A}) or approximate radiative feedback (e.g., \citetalias{2017MNRAS.471.2151H}; \citealt{2021MNRAS.506.3882S}), our simulations also demonstrate that the inclusion of radiative feedback alongside SN feedback leads to a reduction in the SFR. However, in \cite{2018ApJ...865L..22E}, the radiative feedback only decreased the SFR by about a factor of 5 compared to their ``No RT'' run, whereas in our simulations, this reduction is 92. We note that \cite{2018ApJ...865L..22E} simulated a much smaller system than ours. In their ``No RT'' run, the SFR experiences a significant decline after the initial 50\,Myr and does not reach a self-regulating state. It is unknown if the SFR could further increase when more gas is available, leading to better agreement with our results.

The warm \HI gas dominates the ISM in all three simulated dwarf galaxies, which is consistent with previous simulations of dwarf galaxies (e.g., \citealt{2016MNRAS.458.3528H}; \citetalias{2017MNRAS.471.2151H,2019MNRAS.482.1304E}). Hot ionized gas also fills a non-negligible ISM volume (up to $\sim0.1$) in our $0.2\Zsun$/full galaxies, as reported by \citetalias{2017MNRAS.471.2151H}. In contrast, the volume fraction of HIM is only $\sim 10^{-3}$ in $0.02\Zsun$/full due to its low SFR. In line with the findings of \cite{2021ApJ...920...44H}, we found that there is a substantial fraction of \ce{H2} mass in warm diffuse gas with a low \ce{H2}  fraction in low-metallicity environments.

The ISRF of our $0.2\Zsun$/full presents a similar radial profile as that of {\it PE-PI-SN} in \citetalias{2017MNRAS.471.2151H}. The intensity of our ISRF is slightly lower than that of \citetalias{2017MNRAS.471.2151H} at a given radius. This can be attributed to the slightly higher metallicity of our galaxy, which results in more dust absorption. The significant temporal fluctuations in our ISRF also align with the findings of \citetalias{2019MNRAS.482.1304E}, who similarly reported a variation exceeding two orders of magnitude.

Similar to \cite{2024A&A...681A..28A}, we found that the masses and time spreads of star clusters are sensitive to the timing of the onset of radiative feedback. \cite{2024A&A...681A..28A} also found that radiative feedback results in a steeper power-law slope of the CIMF. However, they report a slope around $-3$ for their ``SNe+winds+radiation'' model. Despite this discrepancy, our work again underscores the importance of incorporating the early feedback in galaxy formation models from the perspective of star cluster formation.

The mass-loading factor of our $0.02\Zsun$/full galaxy varies in the order of $10-500$, which is similar to the results of the ``extended'' galaxy in \citetalias{2023MNRAS.522.3092L}. \cite{2018ApJ...865L..22E} also reported a comparable range for the mass-loading factor in their fiducial model. We get enhanced mass-loading factors with radiative feedback turned on, and we attribute this result to the early radiative feedback that enable SNe to explode in less dense gas with $n_\text{H}<1$\,cm$^{-3}$ so that they are more efficient to drive galactic outflow. This pre-pocessing effect of radiative feedback on the SN explosion site has been noticed in simulations of various environments (e.g., \citetalias{2017MNRAS.471.2151H}, \citealt{Kannan2020a}, \citealt{2021MNRAS.506.3882S}, \citealt{2021MNRAS.504.1039R}). Notably, the ``shortrad'' model in \cite{2018ApJ...865L..22E}, which modelled short-range radiative feedback but lacked the long-range effects, exhibited a lower mass-loading factor ranging from 1 to 10. This result emphasizes the importance of explicit RT as we have integrated into the RIGEL model.

As we mentioned in Section~\ref{sec:outflow}, our enhanced mass-loading factor in the full feedback simulations is opposite to the finding by \cite{2021MNRAS.506.3882S}. We explain this difference with the low gas surface density and (relatively) low gas fraction of our initial condition, leading to star clusters that have only a few SNe, even in the noRT simulation. Simulations of higher gas surface density environments might show different behavior. Consistent with our results, the simulations of a smaller galaxy in \cite{2018ApJ...865L..22E} also reported a significantly reduced mass loading factor for their ``No RT'' run. However, they halted this run after $\sim100$\,Myr evolution so the conclusion may still be unclear. Moreover, \cite{2021MNRAS.506.3882S} only models the short-range effects of photoionization using the Str\"omgren-type approximation, which can reduce the loading factor as reported by \cite{2018ApJ...865L..22E}. We will discuss the dependence of the outflow properties on the gas fraction and clustering properties in an accompanying paper.

\subsection{Limitations and future perspectives}
\label{sec:caveat}

We note that there are a few caveats and limitations of this work. Firstly, since the resolution of our simulations is insufficient to resolve the formation and ongoing accretion of individual protostars, we use a sub-grid model outlined in Section~\ref{sec:SFmodel} to model star formation following the \cite{1959ApJ...129..243S} law. Although our model successfully reproduces the observed star formation relations in dwarf galaxies as shown in Section~\ref{sec:starformation}, it does not provide insights into the star formation properties such as the spatial and temporal preferences of star formation across varying stellar mass ranges and the fundamental physics behind the IMF. As a result, though the global properties like SFR are insensitive \citep[][]{2011MNRAS.417..950H}, the properties within star clusters, such as the environmental densities of star formation, compactness and boundness of star clusters, and cluster formation efficiency, can vary significantly with the underlying sub-grid model \citep{2022MNRAS.509.5938H}. Moreover, the IMF can be more top-heavy than the canonical one in metal-poor dwarf galaxies \citep[e.g.][]{2012MNRAS.422.2246M,2018ApJ...863...38G}, thereby increasing the efficiency of stellar feedback in regulating star formation \citep[e.g.][]{2019MNRAS.482..118G,2022MNRAS.513.2326P}. Although the incorporation of a variable IMF is relatively straightforward, the relationship between the slope of the IMF and physical properties such as metallicity remains unclear. On the other hand, a variety of observational clues also suggest a relation between the highest initial stellar mass and the cluster mass \citep{2003ASPC..287...65L,2023A&A...670A.151Y}. As a result, realistic low-mass clusters may not be able to host very massive stars stochastically sampled from the IMF. More sophisticated star formation models can help us capture this observed relation at the cost of computational efficiency (e.g., \citealt{2021PASJ...73.1036H}; \citetalias{2023MNRAS.522.3092L}).

Recent observations suggest that the classical stellar wind model we adopted can overestimate wind luminosities \citep{2014ARA&A..52..487S}. Although there are still large theoretical and observational uncertainties, we have compiled a table of fitting parameters based on the weak-wind model following \cite{2021MNRAS.506.2199G} for future study. 
We also do not model post-main-sequence winds because i) in our metal-poor dwarfs, only a handful number (order of $10$) of $\gtrsim20\,\Msun$ stars can be formed during $1$\,Gyr, which have a Wolf-Rayet (WR) phase; ii) for these stars, though their mass-loss rates are enhanced by a factor of $\sim10$, they have similar wind luminosities to the main-sequence winds (e.g., \citealt{2005A&A...429..581M}); iii) the duration of the WR phase is short ($\lesssim1$\,Myr, \citetalias{2018ApJS..237...13L}). Thus, we argue that the post-main-sequence winds have negligible effects on the global properties of our simulated galaxies.
We assume that massive stars release unenriched gas by main-sequence winds and release yields by post-main-sequence winds together with SN ejecta. This assumption is valid for the non-rotating massive stars in the \citetalias{2018ApJS..237...13L} model. However, fast-rotating stars with efficient mixing and strong winds can provide continuous yield input throughout their lifetime. Moreover, we do not include binary stars in our model, while feedback from binaries can be very different from that of single stars in terms of production rates of ionizing photons and metal yields \citep[e.g.][]{2017A&A...608A..11G:Gotberg,2020ApJ...901...72S:Secunda,2023ApJ...951...84T:Tsai,2024MNRAS.527.6292Y:Yates}. The enrichment contributions of fast-rotating stars and binary stars are possibly important ingredients in producing multiple populations in clusters \citep[][]{2018ARA&A..56...83B}. One ongoing investigation is implementing the post-main-sequence evolution and yields from fast-rotating stars and binaries to study the multiple populations problem.

We assume that all $8-100\,\Msun$ massive stars have a constant SN energy of $10^{51}$\,erg. However, the ``explodability'' of massive stars and the type of their remnants can be a complicated function of their masses and metallicities \citep{2003ApJ...591..288H}, and the explosion energy can be a spectrum over a wide range \citep[e.g.][]{2016ApJ...818..124E,2016ApJ...821...38S}. \cite{2021MNRAS.501.5597G} modelled the SN explosion with variable energy as a function of mass based on \cite{2016ApJ...821...38S}. They found that their model yielded a factor of $2-3$ lower mass loading factor compared to the fixed SN energy model. The implementation is straightforward for us, but the metallicity dependence of SN energy spectrum across the entire mass range is still unclear.

The metals diffuse passively following mesh advection, and we do not employ any sub-grid diffusion model for the unresolved eddies. In this case, the numerical diffusivity is roughly on the scale of $\sigma_\text{t}\Delta x$, where $\sigma_\text{t}$ is the velocity dispersion and $\Delta x$ is the spatial resolution. Recent work by \cite{2024arXiv240714599S:Steinwandel} has shown that the inclusion of a metal diffusion model has a minor impact on ISM structures and loading factors, while it plays a significant role in the phase structure of the galactic outflows. Investigating metal diffusion in detail will be a significant focus of the RIGEL project.

Dust formation and destruction in the gas are not explicitly accounted for in our current model. We assumed a constant dust-to-metal ratio of 0.5 based on the MW observations. However, \cite{2014A&A...563A..31R} suggest a lower dust-to-metal ratio at low metallicity. To improve the accuracy of our dust physics modeling, future iterations of the model will need to explicitly consider processes such as dust formation and destruction, as well as the evolution of dust size and its interaction with radiation \citep[e.g][]{2018MNRAS.478.2851M,2021MNRAS.502.1344M}.  These advanced features will be integrated into our model in the future.

Lastly, in this work, we focused on studying an idealized dwarf galaxy at $z=0$. Therefore, the conclusions are mostly limited to this type of dwarf galaxy with relatively quiescent star formation. Future works on simulating dwarf galaxies in different environments, such as high-z gas-rich analogs, merging systems, and ultimately the cosmological zoom-in boxes are needed to investigate the stellar feedback and multiphase ISM in dwarf galaxies across cosmic time. Specifically, we assume a uniform $z=0$ UVB in this work, which is appropriate for our non-cosmological isolated dwarf galaxy. However, the spatial and temporal variations of UVB have a significant impact on the low-mass galaxy populations, affecting their abundance, star formation history, physical extent, and metal enrichment  \citep{2023ApJ...959...31K:Kim,2023MNRAS.525.5932B}.
One immediate application of RIGEL model is to explore the cosmological evolution of the low-mass reionization survivors \citep[$M_\text{halo}\gtrsim10^9\,\Msun$,][]{2017ApJ...848...85J,2021MNRAS.502....1J:Jeon,2022ApJ...941..120G,2024MNRAS.527.1257L:Lee} in synergy with the inhomogeneous UVB predicted by the THESAN simulations \citep{2022MNRAS.511.4005K,Smith2022,Garaldi2022}. 

\section{Summary}
\label{sec:summary}

This paper introduces the RIGEL model, a framework for studying self-regulating star formation, multiphase ISM, galactic outflows, and cluster formation in dwarf galaxies and other galactic environments. The primary advancement of RIGEL, in comparison to similar numerical models, lies in its incorporation of explicit RT and a detailed feedback model for massive stars across a broad range of metallicities from zero to solar. We incorporate metallicity-dependent, multi-channel stellar feedback on a star-by-star basis, whereby we explicitly sample individual massive stars from the IMF to account for feedback mechanisms such as radiation, stellar winds, and core-collapse SNe. This treatment of feedback, combined with the efficient M1 RT solver and improved heating-cooling model that consider nonequilibrium primordial chemistry and equilibrium abundances of C/O species, enables us to comprehensively model self-regulating star formation processes and the thermodynamics of all ISM phases, spanning from cold molecular gas to hot ionized gas. In this work, we validate our model by simulating the evolution of an isolated dwarf galaxy at gas mass resolutions of $1\,\Msun$ and $10\,\Msun$.
For our simulated dwarf galaxies, we find that:
\begin{enumerate}
    \item The RIGEL model self-consistently produces a multiphase ISM (Fig.~\ref{fig:face_on_prjection} and \ref{fig:edge_on_prjection}) in which the star formation is regulated to a rate consistent with the observed dwarf galaxies (see Section~\ref{sec:starformation}).
    \item The SFRs of our simulated dwarf galaxies show a strong positive correlation with the metallicity. Radiative feedback can reduce the SFR by almost two orders of magnitude (see Fig.~\ref{fig:SFR} and \ref{fig:KSR}) by removing the cold, dense gas fuel available for star formation (Fig.~\ref{fig:fractions}). Turning off radiative feedback boosts the SFR to an unreasonably high value by about two orders of magnitude.
    \item The ISMs in three galaxies with different metallicities and feedback channels are all dominated by warm ($100$\,K$<T<10^5$\,K) \HI gas in both mass and volume. Specifically, \HI gas dominates at densities from $n_\text{H}\sim 10^{-2.5}$\,cm$^{-3}$ to at least $10^3$\,cm$^{-3}$, covering a broad range of temperatures of $T\sim10-10^4$\,K. \HII gas dominates the hot and photoionized gas. \ce{H2} is always a minority phase in the $0.02\Zsun$/full galaxy, even at a high density of $n_\text{H}>10^3$\,cm$^{-3}$ (see Fig.~\ref{fig:ISMpdf}). With a higher metallicity, $0.2\Zsun$/full is still dominated by \HI below $10^3$\,cm$^{-3}$. Turning off radiative feedback leads to diffuse \ce{H2} formed in warm-hot gas with temperatures and densities extended to $10^6$\,K and $10^{-4}$\,cm$^{-3}$ in the absence of the LW radiation.
    \item The ISRFs of our galaxies have significant temporal variations of two orders of magnitude, but on average they have an exponentially declining radial profile that peaks at the galactic center (see Fig.~\ref{fig:ISRF}). The average ISFR of $0.2\Zsun$/full is roughly ten times higher than $0.02\Zsun$/full because of its higher SFR.
    \item Radiative feedback plays a crucial role in removing the dense gas from the star-forming regions to create rarefied gas cavities where SNe explode (see Fig.~\ref{fig:SN_density}). Although stellar winds may not be very efficient in dispersing dense gas, they still play an important role in preventing SN explosions in highly dense environments. In Section~\ref{sec:necessity}, we show that turning off all the early feedback results in a rapid cooling of (unmixed) SN ejecta within dense gas, leading to the formation of excessively metal-rich stars.
    \item The ISM pressure of our three galaxies is dominated by the thermal component. The total pressure and star formation surface density follow a superlinear relation. The full feedback galaxies have a lower SFR at a given pressure compared to the noRT one, due to the impact of radiative feedback in reducing SFE. On the other hand, with the same star formation surface density, the full feedback galaxies yield more pressure because of the presence of radiative feedback and its preprocessing effect on the ISM (see Section~\ref{sec:P-SFR}).
    \item The emergence of radiative feedback can rapidly disperse the molecular clouds and suppress further star formation. The age spread of low-mass star clusters is therefore tightly related to the timing of the onset of radiative feedback (Fig.~\ref{fig:Mcl}). The efficient early feedback prohibits the formation of massive star clusters, shapes the CIMF slope at the high-mass end (Fig.~\ref{fig:CIMF}), and shrinks the age spread of star clusters to less than 2\,Myr (Fig.~\ref{fig:tfms}).
    \item The mass-loading factors of the galactic outflows in our full feedback galaxies vary between 10 and 500 (Fig.~\ref{fig:outflow}). Turning off the radiative feedback leads to reduced mass- and momentum-loading factors, while having minor effects on the metal- and energy-loading factors. The relationship between outflow properties and the properties of galactic disks will be explored in an accompanying paper.
    \item Finally, the $10\,\Msun$ low-resolution simulations present a satisfactory convergence with the $1\,\Msun$ fiducial simulations concerning various aspects such as the SFRs, the ISM structures, the ISRF distribution, the cluster mass functions, and the outflow properties. This convergence enables us to investigate the evolution of larger systems over cosmic time with a resolution of $10\,\Msun$ with high fidelity. 
    
\end{enumerate}

\begin{acknowledgements}
We thank Volker Springel for giving us access to \arepo. YD is grateful to Yang Ni and Zhiqiang Yan for useful discussions. HL is supported by the National Key R\&D Program of China No. 2023YFB3002502, the National Natural Science Foundation of China under No. 12373006, and the China Manned Space Program through its Space Application System. BL is supported by the Royal Society University Research Fellowship. RK acknowledges support of the Natural Sciences and Engineering Research Council of Canada (NSERC) through a Discovery Grant and a Discovery Launch Supplement, funding reference numbers RGPIN-2024-06222 and DGECR-2024-00144. The simulations of this work were run on the Stampede2 HPC resource at the Texas Advanced Computing Center and the Anvil cluster at Purdue University as part of ACCESS through TG-MCA06N030 and TG-PHY220084. We use \textsc{python} packages {\sc NumPy} \citep{harris2020array}, {\sc SciPy} \citep{2020SciPy-NMeth}, {\sc astropy} \citep{2013A&A...558A..33A,2018AJ....156..123A}, and {\sc matplotlib} \citep{Hunter:2007} to analyze and visualize the simulation data.
\end{acknowledgements}



\bibliographystyle{aa}
\bibliography{DwarfGalaxy} 

\begin{thebibliography}{}
\makeatletter
\relax
\def\mn@urlcharsother{\let\do\@makeother \do\$\do\&\do\#\do\^\do\_\do\%\do\~}
\def\mn@doi{\begingroup\mn@urlcharsother \@ifnextchar [ {\mn@doi@}
  {\mn@doi@[]}}
\def\mn@doi@[#1]#2{\def\@tempa{#1}\ifx\@tempa\@empty \href
  {http://dx.doi.org/#2} {doi:#2}\else \href {http://dx.doi.org/#2} {#1}\fi
  \endgroup}
\def\mn@eprint#1#2{\mn@eprint@#1:#2::\@nil}
\def\mn@eprint@arXiv#1{\href {http://arxiv.org/abs/#1} {{\tt arXiv:#1}}}
\def\mn@eprint@dblp#1{\href {http://dblp.uni-trier.de/rec/bibtex/#1.xml}
  {dblp:#1}}
\def\mn@eprint@#1:#2:#3:#4\@nil{\def\@tempa {#1}\def\@tempb {#2}\def\@tempc
  {#3}\ifx \@tempc \@empty \let \@tempc \@tempb \let \@tempb \@tempa \fi \ifx
  \@tempb \@empty \def\@tempb {arXiv}\fi \@ifundefined
  {mn@eprint@\@tempb}{\@tempb:\@tempc}{\expandafter \expandafter \csname
  mn@eprint@\@tempb\endcsname \expandafter{\@tempc}}}

\bibitem[\protect\citeauthoryear{{Agertz} \& {Kravtsov}}{{Agertz} \&
  {Kravtsov}}{2015}]{2015ApJ...804...18A}
{Agertz} O.,  {Kravtsov} A.~V.,  2015, \mn@doi [\apj]
  {10.1088/0004-637X/804/1/18}, \href
  {https://ui.adsabs.harvard.edu/abs/2015ApJ...804...18A} {804, 18}

\bibitem[\protect\citeauthoryear{{Agertz}, {Kravtsov}, {Leitner}  \&
  {Gnedin}}{{Agertz} et~al.}{2013}]{2013ApJ...770...25A}
{Agertz} O.,  {Kravtsov} A.~V.,  {Leitner} S.~N.,   {Gnedin} N.~Y.,  2013,
  \mn@doi [\apj] {10.1088/0004-637X/770/1/25}, \href
  {https://ui.adsabs.harvard.edu/abs/2013ApJ...770...25A} {770, 25}

\bibitem[\protect\citeauthoryear{{Agertz} et~al.,}{{Agertz}
  et~al.}{2020}]{2020MNRAS.491.1656A}
{Agertz} O.,  et~al., 2020, \mn@doi [\mnras] {10.1093/mnras/stz3053}, \href
  {https://ui.adsabs.harvard.edu/abs/2020MNRAS.491.1656A} {491, 1656}

\bibitem[\protect\citeauthoryear{{Andersson}, {Agertz}, {Renaud}  \&
  {Teyssier}}{{Andersson} et~al.}{2023}]{2023MNRAS.521.2196A}
{Andersson} E.~P.,  {Agertz} O.,  {Renaud} F.,   {Teyssier} R.,  2023, \mn@doi
  [\mnras] {10.1093/mnras/stad692}, \href
  {https://ui.adsabs.harvard.edu/abs/2023MNRAS.521.2196A} {521, 2196}

\bibitem[\protect\citeauthoryear{{Andersson}, {Mac Low}, {Agertz}, {Renaud}  \&
  {Li}}{{Andersson} et~al.}{2024}]{2024A&A...681A..28A}
{Andersson} E.~P.,  {Mac Low} M.-M.,  {Agertz} O.,  {Renaud} F.,   {Li} H.,
  2024, \mn@doi [\aap] {10.1051/0004-6361/202347792}, \href
  {https://ui.adsabs.harvard.edu/abs/2024A&A...681A..28A} {681, A28}

\bibitem[\protect\citeauthoryear{{Applebaum}, {Brooks}, {Quinn}  \&
  {Christensen}}{{Applebaum} et~al.}{2020}]{2020MNRAS.492....8A}
{Applebaum} E.,  {Brooks} A.~M.,  {Quinn} T.~R.,   {Christensen} C.~R.,  2020,
  \mn@doi [\mnras] {10.1093/mnras/stz3331}, \href
  {https://ui.adsabs.harvard.edu/abs/2020MNRAS.492....8A} {492, 8}

\bibitem[\protect\citeauthoryear{{Asplund}, {Grevesse}, {Sauval}  \&
  {Scott}}{{Asplund} et~al.}{2009}]{2009ARA&A..47..481A}
{Asplund} M.,  {Grevesse} N.,  {Sauval} A.~J.,   {Scott} P.,  2009, \mn@doi
  [\araa] {10.1146/annurev.astro.46.060407.145222}, \href
  {https://ui.adsabs.harvard.edu/abs/2009ARA&A..47..481A} {47, 481}

\bibitem[\protect\citeauthoryear{{Astropy Collaboration} et~al.,}{{Astropy
  Collaboration} et~al.}{2013}]{2013A&A...558A..33A}
{Astropy Collaboration} et~al., 2013, \mn@doi [\aap]
  {10.1051/0004-6361/201322068}, \href
  {https://ui.adsabs.harvard.edu/abs/2013A&A...558A..33A} {558, A33}

\bibitem[\protect\citeauthoryear{{Astropy Collaboration} et~al.,}{{Astropy
  Collaboration} et~al.}{2018}]{2018AJ....156..123A}
{Astropy Collaboration} et~al., 2018, \mn@doi [\aj] {10.3847/1538-3881/aabc4f},
  \href {https://ui.adsabs.harvard.edu/abs/2018AJ....156..123A} {156, 123}

\bibitem[\protect\citeauthoryear{{Aubert} \& {Teyssier}}{{Aubert} \&
  {Teyssier}}{2008}]{2008MNRAS.387..295A}
{Aubert} D.,  {Teyssier} R.,  2008, \mn@doi [\mnras]
  {10.1111/j.1365-2966.2008.13223.x}, \href
  {https://ui.adsabs.harvard.edu/abs/2008MNRAS.387..295A} {387, 295}

\bibitem[\protect\citeauthoryear{{Bagla}}{{Bagla}}{2002}]{2002JApA...23..185B}
{Bagla} J.~S.,  2002, \mn@doi [Journal of Astrophysics and Astronomy]
  {10.1007/BF02702282}, \href
  {https://ui.adsabs.harvard.edu/abs/2002JApA...23..185B} {23, 185}

\bibitem[\protect\citeauthoryear{{Barnes} \& {Hut}}{{Barnes} \&
  {Hut}}{1986}]{1986Natur.324..446B}
{Barnes} J.,  {Hut} P.,  1986, \mn@doi [\nat] {10.1038/324446a0}, \href
  {https://ui.adsabs.harvard.edu/abs/1986Natur.324..446B} {324, 446}

\bibitem[\protect\citeauthoryear{{Bastian} \& {Lardo}}{{Bastian} \&
  {Lardo}}{2018}]{2018ARA&A..56...83B}
{Bastian} N.,  {Lardo} C.,  2018, \mn@doi [\araa]
  {10.1146/annurev-astro-081817-051839}, \href
  {https://ui.adsabs.harvard.edu/abs/2018ARA&A..56...83B} {56, 83}

\bibitem[\protect\citeauthoryear{{Bialy} \& {Sternberg}}{{Bialy} \&
  {Sternberg}}{2019}]{2019ApJ...881..160B}
{Bialy} S.,  {Sternberg} A.,  2019, \mn@doi [\apj] {10.3847/1538-4357/ab2fd1},
  \href {https://ui.adsabs.harvard.edu/abs/2019ApJ...881..160B} {881, 160}

\bibitem[\protect\citeauthoryear{{Bigiel}, {Leroy}, {Walter}, {Brinks}, {de
  Blok}, {Madore}  \& {Thornley}}{{Bigiel} et~al.}{2008}]{2008AJ....136.2846B}
{Bigiel} F.,  {Leroy} A.,  {Walter} F.,  {Brinks} E.,  {de Blok} W.~J.~G.,
  {Madore} B.,   {Thornley} M.~D.,  2008, \mn@doi [\aj]
  {10.1088/0004-6256/136/6/2846}, \href
  {https://ui.adsabs.harvard.edu/abs/2008AJ....136.2846B} {136, 2846}

\bibitem[\protect\citeauthoryear{{Bigiel}, {Leroy}, {Walter}, {Blitz},
  {Brinks}, {de Blok}  \& {Madore}}{{Bigiel}
  et~al.}{2010}]{2010AJ....140.1194B}
{Bigiel} F.,  {Leroy} A.,  {Walter} F.,  {Blitz} L.,  {Brinks} E.,  {de Blok}
  W.~J.~G.,   {Madore} B.,  2010, \mn@doi [\aj] {10.1088/0004-6256/140/5/1194},
  \href {https://ui.adsabs.harvard.edu/abs/2010AJ....140.1194B} {140, 1194}

\bibitem[\protect\citeauthoryear{{Bisbas} et~al.,}{{Bisbas}
  et~al.}{2015}]{2015MNRAS.453.1324B}
{Bisbas} T.~G.,  et~al., 2015, \mn@doi [\mnras] {10.1093/mnras/stv1659}, \href
  {https://ui.adsabs.harvard.edu/abs/2015MNRAS.453.1324B} {453, 1324}

\bibitem[\protect\citeauthoryear{{Bode}, {Ostriker}  \& {Xu}}{{Bode}
  et~al.}{2000}]{2000ApJS..128..561B}
{Bode} P.,  {Ostriker} J.~P.,   {Xu} G.,  2000, \mn@doi [\apjs]
  {10.1086/313398}, \href
  {https://ui.adsabs.harvard.edu/abs/2000ApJS..128..561B} {128, 561}

\bibitem[\protect\citeauthoryear{{Borrow}, {Kannan}, {Garaldi}, {Smith},
  {Vogelsberger}, {Pakmor}, {Springel}  \& {Hernquist}}{{Borrow}
  et~al.}{2023}]{2023MNRAS.525.5932B}
{Borrow} J.,  {Kannan} R.,  {Garaldi} E.,  {Smith} A.,  {Vogelsberger} M.,
  {Pakmor} R.,  {Springel} V.,   {Hernquist} L.,  2023, \mn@doi [\mnras]
  {10.1093/mnras/stad2523}, \href
  {https://ui.adsabs.harvard.edu/abs/2023MNRAS.525.5932B} {525, 5932}

\bibitem[\protect\citeauthoryear{{Bressan}, {Marigo}, {Girardi}, {Salasnich},
  {Dal Cero}, {Rubele}  \& {Nanni}}{{Bressan}
  et~al.}{2012}]{2012MNRAS.427..127B}
{Bressan} A.,  {Marigo} P.,  {Girardi} L.,  {Salasnich} B.,  {Dal Cero} C.,
  {Rubele} S.,   {Nanni} A.,  2012, \mn@doi [\mnras]
  {10.1111/j.1365-2966.2012.21948.x}, \href
  {https://ui.adsabs.harvard.edu/abs/2012MNRAS.427..127B} {427, 127}

\bibitem[\protect\citeauthoryear{{Bruzual} \& {Charlot}}{{Bruzual} \&
  {Charlot}}{2003}]{2003MNRAS.344.1000B}
{Bruzual} G.,  {Charlot} S.,  2003, \mn@doi [\mnras]
  {10.1046/j.1365-8711.2003.06897.x}, \href
  {https://ui.adsabs.harvard.edu/abs/2003MNRAS.344.1000B} {344, 1000}

\bibitem[\protect\citeauthoryear{{Burke} \& {Hollenbach}}{{Burke} \&
  {Hollenbach}}{1983}]{1983ApJ...265..223B}
{Burke} J.~R.,  {Hollenbach} D.~J.,  1983, \mn@doi [\apj] {10.1086/160667},
  \href {https://ui.adsabs.harvard.edu/abs/1983ApJ...265..223B} {265, 223}

\bibitem[\protect\citeauthoryear{{Calura} et~al.,}{{Calura}
  et~al.}{2022}]{2022MNRAS.516.5914C:Calura}
{Calura} F.,  et~al., 2022, \mn@doi [\mnras] {10.1093/mnras/stac2387}, \href
  {https://ui.adsabs.harvard.edu/abs/2022MNRAS.516.5914C} {516, 5914}

\bibitem[\protect\citeauthoryear{{Cassisi} \& {Castellani}}{{Cassisi} \&
  {Castellani}}{1993}]{Cassisi1993}
{Cassisi} S.,  {Castellani} V.,  1993, \mn@doi [\apjs] {10.1086/191831}, \href
  {https://ui.adsabs.harvard.edu/abs/1993ApJS...88..509C} {88, 509}

\bibitem[\protect\citeauthoryear{{Cen}}{{Cen}}{1992}]{1992ApJS...78..341C}
{Cen} R.,  1992, \mn@doi [\apjs] {10.1086/191630}, \href
  {https://ui.adsabs.harvard.edu/abs/1992ApJS...78..341C} {78, 341}

\bibitem[\protect\citeauthoryear{{Chabrier}}{{Chabrier}}{2003}]{2003PASP..115..763C}
{Chabrier} G.,  2003, \mn@doi [\pasp] {10.1086/376392}, \href
  {https://ui.adsabs.harvard.edu/abs/2003PASP..115..763C} {115, 763}

\bibitem[\protect\citeauthoryear{{Chan}, {Theuns}, {Bower}  \& {Frenk}}{{Chan}
  et~al.}{2021}]{Chan2021}
{Chan} T.~K.,  {Theuns} T.,  {Bower} R.,   {Frenk} C.,  2021, \mn@doi [\mnras]
  {10.1093/mnras/stab1686}, \href
  {https://ui.adsabs.harvard.edu/abs/2021MNRAS.505.5784C} {505, 5784}

\bibitem[\protect\citeauthoryear{{Chevance} et~al.,}{{Chevance}
  et~al.}{2022}]{2022MNRAS.509..272C}
{Chevance} M.,  et~al., 2022, \mn@doi [\mnras] {10.1093/mnras/stab2938}, \href
  {https://ui.adsabs.harvard.edu/abs/2022MNRAS.509..272C} {509, 272}

\bibitem[\protect\citeauthoryear{{Chevance}, {Krumholz}, {McLeod}, {Ostriker},
  {Rosolowsky}  \& {Sternberg}}{{Chevance} et~al.}{2023}]{2023ASPC..534....1C}
{Chevance} M.,  {Krumholz} M.~R.,  {McLeod} A.~F.,  {Ostriker} E.~C.,
  {Rosolowsky} E.~W.,   {Sternberg} A.,  2023, in {Inutsuka} S.,  {Aikawa} Y.,
  {Muto} T.,  {Tomida} K.,   {Tamura} M.,  eds,  Astronomical Society of the
  Pacific Conference Series Vol. 534, Protostars and Planets VII. p.~1
  (\mn@eprint {arXiv} {2203.09570}), \mn@doi{10.48550/arXiv.2203.09570}

\bibitem[\protect\citeauthoryear{{Conroy} \& {Wechsler}}{{Conroy} \&
  {Wechsler}}{2009}]{2009ApJ...696..620C}
{Conroy} C.,  {Wechsler} R.~H.,  2009, \mn@doi [\apj]
  {10.1088/0004-637X/696/1/620}, \href
  {https://ui.adsabs.harvard.edu/abs/2009ApJ...696..620C} {696, 620}

\bibitem[\protect\citeauthoryear{{Dekel} \& {Silk}}{{Dekel} \&
  {Silk}}{1986}]{1986ApJ...303...39D}
{Dekel} A.,  {Silk} J.,  1986, \mn@doi [\apj] {10.1086/164050}, \href
  {https://ui.adsabs.harvard.edu/abs/1986ApJ...303...39D} {303, 39}

\bibitem[\protect\citeauthoryear{{Deng}, {Li}, {Kannan}, {Smith},
  {Vogelsberger}  \& {Bryan}}{{Deng} et~al.}{2024}]{2024MNRAS.527..478D}
{Deng} Y.,  {Li} H.,  {Kannan} R.,  {Smith} A.,  {Vogelsberger} M.,   {Bryan}
  G.~L.,  2024, \mn@doi [\mnras] {10.1093/mnras/stad3202}, \href
  {https://ui.adsabs.harvard.edu/abs/2024MNRAS.527..478D} {527, 478}

\bibitem[\protect\citeauthoryear{{Dobbs}, {Burkert}  \& {Pringle}}{{Dobbs}
  et~al.}{2011}]{2011MNRAS.417.1318D}
{Dobbs} C.~L.,  {Burkert} A.,   {Pringle} J.~E.,  2011, \mn@doi [\mnras]
  {10.1111/j.1365-2966.2011.19346.x}, \href
  {https://ui.adsabs.harvard.edu/abs/2011MNRAS.417.1318D} {417, 1318}

\bibitem[\protect\citeauthoryear{{Doherty}, {Gil-Pons}, {Lau}, {Lattanzio},
  {Siess}  \& {Campbell}}{{Doherty} et~al.}{2014}]{2014MNRAS.441..582D}
{Doherty} C.~L.,  {Gil-Pons} P.,  {Lau} H. H.~B.,  {Lattanzio} J.~C.,  {Siess}
  L.,   {Campbell} S.~W.,  2014, \mn@doi [\mnras] {10.1093/mnras/stu571}, \href
  {https://ui.adsabs.harvard.edu/abs/2014MNRAS.441..582D} {441, 582}

\bibitem[\protect\citeauthoryear{{Draine}}{{Draine}}{1978}]{1978ApJS...36..595D}
{Draine} B.~T.,  1978, \mn@doi [\apjs] {10.1086/190513}, \href
  {https://ui.adsabs.harvard.edu/abs/1978ApJS...36..595D} {36, 595}

\bibitem[\protect\citeauthoryear{{Draine}}{{Draine}}{2011}]{2011piim.book.....D}
{Draine} B.~T.,  2011, {Physics of the Interstellar and Intergalactic Medium}

\bibitem[\protect\citeauthoryear{{Eftekhari} et~al.,}{{Eftekhari}
  et~al.}{2022}]{2022MNRAS.517.4714E:Eftekhari}
{Eftekhari} F.~S.,  et~al., 2022, \mn@doi [\mnras] {10.1093/mnras/stac2606},
  \href {https://ui.adsabs.harvard.edu/abs/2022MNRAS.517.4714E} {517, 4714}

\bibitem[\protect\citeauthoryear{{Elmegreen}}{{Elmegreen}}{2015}]{2015ApJ...814L..30E}
{Elmegreen} B.~G.,  2015, \mn@doi [\apjl] {10.1088/2041-8205/814/2/L30}, \href
  {https://ui.adsabs.harvard.edu/abs/2015ApJ...814L..30E} {814, L30}

\bibitem[\protect\citeauthoryear{{Elmegreen}}{{Elmegreen}}{2018}]{2018ApJ...854...16E}
{Elmegreen} B.~G.,  2018, \mn@doi [\apj] {10.3847/1538-4357/aaa770}, \href
  {https://ui.adsabs.harvard.edu/abs/2018ApJ...854...16E} {854, 16}

\bibitem[\protect\citeauthoryear{{Elmegreen} \& {Hunter}}{{Elmegreen} \&
  {Hunter}}{2015}]{2015ApJ...805..145E:Elmegreen}
{Elmegreen} B.~G.,  {Hunter} D.~A.,  2015, \mn@doi [\apj]
  {10.1088/0004-637X/805/2/145}, \href
  {https://ui.adsabs.harvard.edu/abs/2015ApJ...805..145E} {805, 145}

\bibitem[\protect\citeauthoryear{{Emerick}, {Bryan}  \& {Mac Low}}{{Emerick}
  et~al.}{2018}]{2018ApJ...865L..22E}
{Emerick} A.,  {Bryan} G.~L.,   {Mac Low} M.-M.,  2018, \mn@doi [\apjl]
  {10.3847/2041-8213/aae315}, \href
  {https://ui.adsabs.harvard.edu/abs/2018ApJ...865L..22E} {865, L22}

\bibitem[\protect\citeauthoryear{{Emerick}, {Bryan}  \& {Mac Low}}{{Emerick}
  et~al.}{2019}]{2019MNRAS.482.1304E}
{Emerick} A.,  {Bryan} G.~L.,   {Mac Low} M.-M.,  2019, \mn@doi [\mnras]
  {10.1093/mnras/sty2689}, \href
  {https://ui.adsabs.harvard.edu/abs/2019MNRAS.482.1304E} {482, 1304}

\bibitem[\protect\citeauthoryear{{Ertl}, {Janka}, {Woosley}, {Sukhbold}  \&
  {Ugliano}}{{Ertl} et~al.}{2016}]{2016ApJ...818..124E}
{Ertl} T.,  {Janka} H.~T.,  {Woosley} S.~E.,  {Sukhbold} T.,   {Ugliano} M.,
  2016, \mn@doi [\apj] {10.3847/0004-637X/818/2/124}, \href
  {https://ui.adsabs.harvard.edu/abs/2016ApJ...818..124E} {818, 124}

\bibitem[\protect\citeauthoryear{{Faucher-Gigu{\`e}re}, {Lidz}, {Zaldarriaga}
  \& {Hernquist}}{{Faucher-Gigu{\`e}re} et~al.}{2009}]{2009ApJ...703.1416F}
{Faucher-Gigu{\`e}re} C.-A.,  {Lidz} A.,  {Zaldarriaga} M.,   {Hernquist} L.,
  2009, \mn@doi [\apj] {10.1088/0004-637X/703/2/1416}, \href
  {https://ui.adsabs.harvard.edu/abs/2009ApJ...703.1416F} {703, 1416}

\bibitem[\protect\citeauthoryear{{Faucher-Gigu{\`e}re}, {Quataert}  \&
  {Hopkins}}{{Faucher-Gigu{\`e}re} et~al.}{2013}]{2013MNRAS.433.1970F}
{Faucher-Gigu{\`e}re} C.-A.,  {Quataert} E.,   {Hopkins} P.~F.,  2013, \mn@doi
  [\mnras] {10.1093/mnras/stt866}, \href
  {https://ui.adsabs.harvard.edu/abs/2013MNRAS.433.1970F} {433, 1970}

\bibitem[\protect\citeauthoryear{{Fishlock}, {Karakas}, {Lugaro}  \&
  {Yong}}{{Fishlock} et~al.}{2014}]{2014ApJ...797...44F}
{Fishlock} C.~K.,  {Karakas} A.~I.,  {Lugaro} M.,   {Yong} D.,  2014, \mn@doi
  [\apj] {10.1088/0004-637X/797/1/44}, \href
  {https://ui.adsabs.harvard.edu/abs/2014ApJ...797...44F} {797, 44}

\bibitem[\protect\citeauthoryear{{Fujimoto}, {Chevance}, {Haydon}, {Krumholz}
  \& {Kruijssen}}{{Fujimoto} et~al.}{2019}]{2019MNRAS.487.1717F}
{Fujimoto} Y.,  {Chevance} M.,  {Haydon} D.~T.,  {Krumholz} M.~R.,
  {Kruijssen} J.~M.~D.,  2019, \mn@doi [\mnras] {10.1093/mnras/stz641}, \href
  {https://ui.adsabs.harvard.edu/abs/2019MNRAS.487.1717F} {487, 1717}

\bibitem[\protect\citeauthoryear{{Garaldi}, {Kannan}, {Smith}, {Springel},
  {Pakmor}, {Vogelsberger}  \& {Hernquist}}{{Garaldi}
  et~al.}{2022}]{Garaldi2022}
{Garaldi} E.,  {Kannan} R.,  {Smith} A.,  {Springel} V.,  {Pakmor} R.,
  {Vogelsberger} M.,   {Hernquist} L.,  2022, \mn@doi [\mnras]
  {10.1093/mnras/stac257}, \href
  {https://ui.adsabs.harvard.edu/abs/2022MNRAS.512.4909G} {512, 4909}

\bibitem[\protect\citeauthoryear{{Garcia}, {Ricotti}, {Sugimura}  \&
  {Park}}{{Garcia} et~al.}{2023}]{2023MNRAS.522.2495G:Garcia}
{Garcia} F. A.~B.,  {Ricotti} M.,  {Sugimura} K.,   {Park} J.,  2023, \mn@doi
  [\mnras] {10.1093/mnras/stad1092}, \href
  {https://ui.adsabs.harvard.edu/abs/2023MNRAS.522.2495G} {522, 2495}

\bibitem[\protect\citeauthoryear{{Geen}, {Rosdahl}, {Blaizot}, {Devriendt}  \&
  {Slyz}}{{Geen} et~al.}{2015}]{2015MNRAS.448.3248G}
{Geen} S.,  {Rosdahl} J.,  {Blaizot} J.,  {Devriendt} J.,   {Slyz} A.,  2015,
  \mn@doi [\mnras] {10.1093/mnras/stv251}, \href
  {https://ui.adsabs.harvard.edu/abs/2015MNRAS.448.3248G} {448, 3248}

\bibitem[\protect\citeauthoryear{{Gennaro} et~al.,}{{Gennaro}
  et~al.}{2018}]{2018ApJ...863...38G}
{Gennaro} M.,  et~al., 2018, \mn@doi [\apj] {10.3847/1538-4357/aaceff}, \href
  {https://ui.adsabs.harvard.edu/abs/2018ApJ...863...38G} {863, 38}

\bibitem[\protect\citeauthoryear{{Gessey-Jones} et~al.,}{{Gessey-Jones}
  et~al.}{2022}]{Gessey-Jones2022}
{Gessey-Jones} T.,  et~al., 2022, \mn@doi [\mnras] {10.1093/mnras/stac2049},
  \href {https://ui.adsabs.harvard.edu/abs/2022MNRAS.516..841G} {516, 841}

\bibitem[\protect\citeauthoryear{{Gnedin} \& {Abel}}{{Gnedin} \&
  {Abel}}{2001}]{2001NewA....6..437G}
{Gnedin} N.~Y.,  {Abel} T.,  2001, \mn@doi [\na]
  {10.1016/S1384-1076(01)00068-9}, \href
  {https://ui.adsabs.harvard.edu/abs/2001NewA....6..437G} {6, 437}

\bibitem[\protect\citeauthoryear{{Goldsmith} \& {Langer}}{{Goldsmith} \&
  {Langer}}{1978}]{1978ApJ...222..881G}
{Goldsmith} P.~F.,  {Langer} W.~D.,  1978, \mn@doi [\apj] {10.1086/156206},
  \href {https://ui.adsabs.harvard.edu/abs/1978ApJ...222..881G} {222, 881}

\bibitem[\protect\citeauthoryear{{Gong}, {Ostriker}  \& {Wolfire}}{{Gong}
  et~al.}{2017}]{2017ApJ...843...38G}
{Gong} M.,  {Ostriker} E.~C.,   {Wolfire} M.~G.,  2017, \mn@doi [\apj]
  {10.3847/1538-4357/aa7561}, \href
  {https://ui.adsabs.harvard.edu/abs/2017ApJ...843...38G} {843, 38}

\bibitem[\protect\citeauthoryear{{G{\"o}tberg}, {de Mink}  \&
  {Groh}}{{G{\"o}tberg} et~al.}{2017}]{2017A&A...608A..11G:Gotberg}
{G{\"o}tberg} Y.,  {de Mink} S.~E.,   {Groh} J.~H.,  2017, \mn@doi [\aap]
  {10.1051/0004-6361/201730472}, \href
  {https://ui.adsabs.harvard.edu/abs/2017A&A...608A..11G} {608, A11}

\bibitem[\protect\citeauthoryear{{Grudi{\'c}}, {Hopkins},
  {Faucher-Gigu{\`e}re}, {Quataert}, {Murray}  \& {Kere{\v{s}}}}{{Grudi{\'c}}
  et~al.}{2018}]{2018MNRAS.475.3511G}
{Grudi{\'c}} M.~Y.,  {Hopkins} P.~F.,  {Faucher-Gigu{\`e}re} C.-A.,  {Quataert}
  E.,  {Murray} N.,   {Kere{\v{s}}} D.,  2018, \mn@doi [\mnras]
  {10.1093/mnras/sty035}, \href
  {https://ui.adsabs.harvard.edu/abs/2018MNRAS.475.3511G} {475, 3511}

\bibitem[\protect\citeauthoryear{{Grudi{\'c}}, {Guszejnov}, {Hopkins}, {Offner}
   \& {Faucher-Gigu{\`e}re}}{{Grudi{\'c}} et~al.}{2021}]{2021MNRAS.506.2199G}
{Grudi{\'c}} M.~Y.,  {Guszejnov} D.,  {Hopkins} P.~F.,  {Offner} S. S.~R.,
  {Faucher-Gigu{\`e}re} C.-A.,  2021, \mn@doi [\mnras]
  {10.1093/mnras/stab1347}, \href
  {https://ui.adsabs.harvard.edu/abs/2021MNRAS.506.2199G} {506, 2199}

\bibitem[\protect\citeauthoryear{{Gutcke} \& {Springel}}{{Gutcke} \&
  {Springel}}{2019}]{2019MNRAS.482..118G}
{Gutcke} T.~A.,  {Springel} V.,  2019, \mn@doi [\mnras]
  {10.1093/mnras/sty2688}, \href
  {https://ui.adsabs.harvard.edu/abs/2019MNRAS.482..118G} {482, 118}

\bibitem[\protect\citeauthoryear{{Gutcke}, {Pakmor}, {Naab}  \&
  {Springel}}{{Gutcke} et~al.}{2021}]{2021MNRAS.501.5597G}
{Gutcke} T.~A.,  {Pakmor} R.,  {Naab} T.,   {Springel} V.,  2021, \mn@doi
  [\mnras] {10.1093/mnras/staa3875}, \href
  {https://ui.adsabs.harvard.edu/abs/2021MNRAS.501.5597G} {501, 5597}

\bibitem[\protect\citeauthoryear{{Gutcke}, {Pfrommer}, {Bryan}, {Pakmor},
  {Springel}  \& {Naab}}{{Gutcke} et~al.}{2022}]{2022ApJ...941..120G}
{Gutcke} T.~A.,  {Pfrommer} C.,  {Bryan} G.~L.,  {Pakmor} R.,  {Springel} V.,
  {Naab} T.,  2022, \mn@doi [\apj] {10.3847/1538-4357/aca1b4}, \href
  {https://ui.adsabs.harvard.edu/abs/2022ApJ...941..120G} {941, 120}

\bibitem[\protect\citeauthoryear{{Haardt} \& {Madau}}{{Haardt} \&
  {Madau}}{2012}]{2012ApJ...746..125H}
{Haardt} F.,  {Madau} P.,  2012, \mn@doi [\apj] {10.1088/0004-637X/746/2/125},
  \href {https://ui.adsabs.harvard.edu/abs/2012ApJ...746..125H} {746, 125}

\bibitem[\protect\citeauthoryear{{Habing}}{{Habing}}{1968}]{1968BAN....19..421H}
{Habing} H.~J.,  1968, \bain, \href
  {https://ui.adsabs.harvard.edu/abs/1968BAN....19..421H} {19, 421}

\bibitem[\protect\citeauthoryear{{Haid}, {Walch}, {Seifried}, {W{\"u}nsch},
  {Dinnbier}  \& {Naab}}{{Haid} et~al.}{2018}]{2018MNRAS.478.4799H}
{Haid} S.,  {Walch} S.,  {Seifried} D.,  {W{\"u}nsch} R.,  {Dinnbier} F.,
  {Naab} T.,  2018, \mn@doi [\mnras] {10.1093/mnras/sty1315}, \href
  {https://ui.adsabs.harvard.edu/abs/2018MNRAS.478.4799H} {478, 4799}

\bibitem[\protect\citeauthoryear{Harris et~al.,}{Harris
  et~al.}{2020}]{harris2020array}
Harris C.~R.,  et~al., 2020, \mn@doi [Nature] {10.1038/s41586-020-2649-2}, 585,
  357

\bibitem[\protect\citeauthoryear{{Hartmann}, {Ballesteros-Paredes}  \&
  {Heitsch}}{{Hartmann} et~al.}{2012}]{2012MNRAS.420.1457H}
{Hartmann} L.,  {Ballesteros-Paredes} J.,   {Heitsch} F.,  2012, \mn@doi
  [\mnras] {10.1111/j.1365-2966.2011.20131.x}, \href
  {https://ui.adsabs.harvard.edu/abs/2012MNRAS.420.1457H} {420, 1457}

\bibitem[\protect\citeauthoryear{{Hayashi}, {Chiba}  \& {Ishiyama}}{{Hayashi}
  et~al.}{2020}]{2020ApJ...904...45H:Hayashi}
{Hayashi} K.,  {Chiba} M.,   {Ishiyama} T.,  2020, \mn@doi [\apj]
  {10.3847/1538-4357/abbe0a}, \href
  {https://ui.adsabs.harvard.edu/abs/2020ApJ...904...45H} {904, 45}

\bibitem[\protect\citeauthoryear{{Hayward} \& {Hopkins}}{{Hayward} \&
  {Hopkins}}{2017}]{2017MNRAS.465.1682H}
{Hayward} C.~C.,  {Hopkins} P.~F.,  2017, \mn@doi [\mnras]
  {10.1093/mnras/stw2888}, \href
  {https://ui.adsabs.harvard.edu/abs/2017MNRAS.465.1682H} {465, 1682}

\bibitem[\protect\citeauthoryear{{Heays}, {Bosman}  \& {van Dishoeck}}{{Heays}
  et~al.}{2017}]{2017A&A...602A.105H}
{Heays} A.~N.,  {Bosman} A.~D.,   {van Dishoeck} E.~F.,  2017, \mn@doi [\aap]
  {10.1051/0004-6361/201628742}, \href
  {https://ui.adsabs.harvard.edu/abs/2017A&A...602A.105H} {602, A105}

\bibitem[\protect\citeauthoryear{{Heger}, {Fryer}, {Woosley}, {Langer}  \&
  {Hartmann}}{{Heger} et~al.}{2003}]{2003ApJ...591..288H}
{Heger} A.,  {Fryer} C.~L.,  {Woosley} S.~E.,  {Langer} N.,   {Hartmann} D.~H.,
   2003, \mn@doi [\apj] {10.1086/375341}, \href
  {https://ui.adsabs.harvard.edu/abs/2003ApJ...591..288H} {591, 288}

\bibitem[\protect\citeauthoryear{Hindmarsh, Brown, Grant, Lee, Serban, Shumaker
   \& Woodward}{Hindmarsh et~al.}{2005}]{hindmarsh2005sundials}
Hindmarsh A.~C.,  Brown P.~N.,  Grant K.~E.,  Lee S.~L.,  Serban R.,  Shumaker
  D.~E.,   Woodward C.~S.,  2005, ACM Transactions on Mathematical Software
  (TOMS), 31, 363

\bibitem[\protect\citeauthoryear{{Hirai}, {Fujii}  \& {Saitoh}}{{Hirai}
  et~al.}{2021}]{2021PASJ...73.1036H}
{Hirai} Y.,  {Fujii} M.~S.,   {Saitoh} T.~R.,  2021, \mn@doi [\pasj]
  {10.1093/pasj/psab038}, \href
  {https://ui.adsabs.harvard.edu/abs/2021PASJ...73.1036H} {73, 1036}

\bibitem[\protect\citeauthoryear{{Hislop}, {Naab}, {Steinwandel}, {Lah{\'e}n},
  {Irodotou}, {Johansson}  \& {Walch}}{{Hislop}
  et~al.}{2022}]{2022MNRAS.509.5938H}
{Hislop} J.~M.,  {Naab} T.,  {Steinwandel} U.~P.,  {Lah{\'e}n} N.,  {Irodotou}
  D.,  {Johansson} P.~H.,   {Walch} S.,  2022, \mn@doi [\mnras]
  {10.1093/mnras/stab3347}, \href
  {https://ui.adsabs.harvard.edu/abs/2022MNRAS.509.5938H} {509, 5938}

\bibitem[\protect\citeauthoryear{{Hollenbach} \& {McKee}}{{Hollenbach} \&
  {McKee}}{1979}]{1979ApJS...41..555H}
{Hollenbach} D.,  {McKee} C.~F.,  1979, \mn@doi [\apjs] {10.1086/190631}, \href
  {https://ui.adsabs.harvard.edu/abs/1979ApJS...41..555H} {41, 555}

\bibitem[\protect\citeauthoryear{{Hollyhead}, {Bastian}, {Adamo},
  {Silva-Villa}, {Dale}, {Ryon}  \& {Gazak}}{{Hollyhead}
  et~al.}{2015}]{2015MNRAS.449.1106H}
{Hollyhead} K.,  {Bastian} N.,  {Adamo} A.,  {Silva-Villa} E.,  {Dale} J.,
  {Ryon} J.~E.,   {Gazak} Z.,  2015, \mn@doi [\mnras] {10.1093/mnras/stv331},
  \href {https://ui.adsabs.harvard.edu/abs/2015MNRAS.449.1106H} {449, 1106}

\bibitem[\protect\citeauthoryear{{Hopkins}, {Quataert}  \& {Murray}}{{Hopkins}
  et~al.}{2011}]{2011MNRAS.417..950H}
{Hopkins} P.~F.,  {Quataert} E.,   {Murray} N.,  2011, \mn@doi [\mnras]
  {10.1111/j.1365-2966.2011.19306.x}, \href
  {https://ui.adsabs.harvard.edu/abs/2011MNRAS.417..950H} {417, 950}

\bibitem[\protect\citeauthoryear{{Hopkins}, {Quataert}  \& {Murray}}{{Hopkins}
  et~al.}{2012a}]{2012MNRAS.421.3488H}
{Hopkins} P.~F.,  {Quataert} E.,   {Murray} N.,  2012a, \mn@doi [\mnras]
  {10.1111/j.1365-2966.2012.20578.x}, \href
  {https://ui.adsabs.harvard.edu/abs/2012MNRAS.421.3488H} {421, 3488}

\bibitem[\protect\citeauthoryear{{Hopkins}, {Quataert}  \& {Murray}}{{Hopkins}
  et~al.}{2012b}]{2012MNRAS.421.3522H}
{Hopkins} P.~F.,  {Quataert} E.,   {Murray} N.,  2012b, \mn@doi [\mnras]
  {10.1111/j.1365-2966.2012.20593.x}, \href
  {https://ui.adsabs.harvard.edu/abs/2012MNRAS.421.3522H} {421, 3522}

\bibitem[\protect\citeauthoryear{{Hopkins}, {Narayanan}  \& {Murray}}{{Hopkins}
  et~al.}{2013}]{2013MNRAS.432.2647H}
{Hopkins} P.~F.,  {Narayanan} D.,   {Murray} N.,  2013, \mn@doi [\mnras]
  {10.1093/mnras/stt723}, \href
  {https://ui.adsabs.harvard.edu/abs/2013MNRAS.432.2647H} {432, 2647}

\bibitem[\protect\citeauthoryear{{Hopkins}, {Kere{\v{s}}}, {O{\~n}orbe},
  {Faucher-Gigu{\`e}re}, {Quataert}, {Murray}  \& {Bullock}}{{Hopkins}
  et~al.}{2014}]{2014MNRAS.445..581H}
{Hopkins} P.~F.,  {Kere{\v{s}}} D.,  {O{\~n}orbe} J.,  {Faucher-Gigu{\`e}re}
  C.-A.,  {Quataert} E.,  {Murray} N.,   {Bullock} J.~S.,  2014, \mn@doi
  [\mnras] {10.1093/mnras/stu1738}, \href
  {https://ui.adsabs.harvard.edu/abs/2014MNRAS.445..581H} {445, 581}

\bibitem[\protect\citeauthoryear{{Hopkins} et~al.,}{{Hopkins}
  et~al.}{2018}]{2018MNRAS.480..800H}
{Hopkins} P.~F.,  et~al., 2018, \mn@doi [\mnras] {10.1093/mnras/sty1690}, \href
  {https://ui.adsabs.harvard.edu/abs/2018MNRAS.480..800H} {480, 800}

\bibitem[\protect\citeauthoryear{{Hopkins}, {Grudi{\'c}}, {Wetzel},
  {Kere{\v{s}}}, {Faucher-Gigu{\`e}re}, {Ma}, {Murray}  \& {Butcher}}{{Hopkins}
  et~al.}{2020}]{2020MNRAS.491.3702H}
{Hopkins} P.~F.,  {Grudi{\'c}} M.~Y.,  {Wetzel} A.,  {Kere{\v{s}}} D.,
  {Faucher-Gigu{\`e}re} C.-A.,  {Ma} X.,  {Murray} N.,   {Butcher} N.,  2020,
  \mn@doi [\mnras] {10.1093/mnras/stz3129}, \href
  {https://ui.adsabs.harvard.edu/abs/2020MNRAS.491.3702H} {491, 3702}

\bibitem[\protect\citeauthoryear{{Hopkins} et~al.,}{{Hopkins}
  et~al.}{2023}]{2023MNRAS.519.3154H}
{Hopkins} P.~F.,  et~al., 2023, \mn@doi [\mnras] {10.1093/mnras/stac3489},
  \href {https://ui.adsabs.harvard.edu/abs/2023MNRAS.519.3154H} {519, 3154}

\bibitem[\protect\citeauthoryear{{Hu}}{{Hu}}{2019}]{2019MNRAS.483.3363H}
{Hu} C.-Y.,  2019, \mn@doi [\mnras] {10.1093/mnras/sty3252}, \href
  {https://ui.adsabs.harvard.edu/abs/2019MNRAS.483.3363H} {483, 3363}

\bibitem[\protect\citeauthoryear{{Hu}, {Naab}, {Walch}, {Glover}  \&
  {Clark}}{{Hu} et~al.}{2016}]{2016MNRAS.458.3528H}
{Hu} C.-Y.,  {Naab} T.,  {Walch} S.,  {Glover} S. C.~O.,   {Clark} P.~C.,
  2016, \mn@doi [\mnras] {10.1093/mnras/stw544}, \href
  {https://ui.adsabs.harvard.edu/abs/2016MNRAS.458.3528H} {458, 3528}

\bibitem[\protect\citeauthoryear{{Hu}, {Naab}, {Glover}, {Walch}  \&
  {Clark}}{{Hu} et~al.}{2017}]{2017MNRAS.471.2151H}
{Hu} C.-Y.,  {Naab} T.,  {Glover} S. C.~O.,  {Walch} S.,   {Clark} P.~C.,
  2017, \mn@doi [\mnras] {10.1093/mnras/stx1773}, \href
  {https://ui.adsabs.harvard.edu/abs/2017MNRAS.471.2151H} {471, 2151}

\bibitem[\protect\citeauthoryear{{Hu}, {Sternberg}  \& {van Dishoeck}}{{Hu}
  et~al.}{2021}]{2021ApJ...920...44H}
{Hu} C.-Y.,  {Sternberg} A.,   {van Dishoeck} E.~F.,  2021, \mn@doi [\apj]
  {10.3847/1538-4357/ac0dbd}, \href
  {https://ui.adsabs.harvard.edu/abs/2021ApJ...920...44H} {920, 44}

\bibitem[\protect\citeauthoryear{Hunter}{Hunter}{2007}]{Hunter:2007}
Hunter J.~D.,  2007, \mn@doi [Computing in Science \& Engineering]
  {10.1109/MCSE.2007.55}, 9, 90

\bibitem[\protect\citeauthoryear{{Indriolo} \& {McCall}}{{Indriolo} \&
  {McCall}}{2012}]{2012ApJ...745...91I}
{Indriolo} N.,  {McCall} B.~J.,  2012, \mn@doi [\apj]
  {10.1088/0004-637X/745/1/91}, \href
  {https://ui.adsabs.harvard.edu/abs/2012ApJ...745...91I} {745, 91}

\bibitem[\protect\citeauthoryear{{Jaura}, {Glover}, {Klessen}  \&
  {Paardekooper}}{{Jaura} et~al.}{2018}]{2018MNRAS.475.2822J}
{Jaura} O.,  {Glover} S.~C.~O.,  {Klessen} R.~S.,   {Paardekooper} J.~P.,
  2018, \mn@doi [\mnras] {10.1093/mnras/stx3356}, \href
  {https://ui.adsabs.harvard.edu/abs/2018MNRAS.475.2822J} {475, 2822}

\bibitem[\protect\citeauthoryear{{Jeffreson}, {Krumholz}, {Fujimoto},
  {Armillotta}, {Keller}, {Chevance}  \& {Kruijssen}}{{Jeffreson}
  et~al.}{2021}]{2021MNRAS.505.3470J}
{Jeffreson} S. M.~R.,  {Krumholz} M.~R.,  {Fujimoto} Y.,  {Armillotta} L.,
  {Keller} B.~W.,  {Chevance} M.,   {Kruijssen} J.~M.~D.,  2021, \mn@doi
  [\mnras] {10.1093/mnras/stab1536}, \href
  {https://ui.adsabs.harvard.edu/abs/2021MNRAS.505.3470J} {505, 3470}

\bibitem[\protect\citeauthoryear{{Jeon}, {Besla}  \& {Bromm}}{{Jeon}
  et~al.}{2017}]{2017ApJ...848...85J}
{Jeon} M.,  {Besla} G.,   {Bromm} V.,  2017, \mn@doi [\apj]
  {10.3847/1538-4357/aa8c80}, \href
  {https://ui.adsabs.harvard.edu/abs/2017ApJ...848...85J} {848, 85}

\bibitem[\protect\citeauthoryear{{Jeon}, {Bromm}, {Besla}, {Yoon}  \&
  {Choi}}{{Jeon} et~al.}{2021}]{2021MNRAS.502....1J:Jeon}
{Jeon} M.,  {Bromm} V.,  {Besla} G.,  {Yoon} J.,   {Choi} Y.,  2021, \mn@doi
  [\mnras] {10.1093/mnras/staa4017}, \href
  {https://ui.adsabs.harvard.edu/abs/2021MNRAS.502....1J} {502, 1}

\bibitem[\protect\citeauthoryear{{Kannan}, {Stinson}, {Macci{\`o}}, {Brook},
  {Weinmann}, {Wadsley}  \& {Couchman}}{{Kannan}
  et~al.}{2014}]{2014MNRAS.437.3529K}
{Kannan} R.,  {Stinson} G.~S.,  {Macci{\`o}} A.~V.,  {Brook} C.,  {Weinmann}
  S.~M.,  {Wadsley} J.,   {Couchman} H. M.~P.,  2014, \mn@doi [\mnras]
  {10.1093/mnras/stt2144}, \href
  {https://ui.adsabs.harvard.edu/abs/2014MNRAS.437.3529K} {437, 3529}

\bibitem[\protect\citeauthoryear{{Kannan}, {Vogelsberger}, {Marinacci},
  {McKinnon}, {Pakmor}  \& {Springel}}{{Kannan}
  et~al.}{2019}]{2019MNRAS.485..117K}
{Kannan} R.,  {Vogelsberger} M.,  {Marinacci} F.,  {McKinnon} R.,  {Pakmor} R.,
    {Springel} V.,  2019, \mn@doi [\mnras] {10.1093/mnras/stz287}, \href
  {https://ui.adsabs.harvard.edu/abs/2019MNRAS.485..117K} {485, 117}

\bibitem[\protect\citeauthoryear{{Kannan}, {Marinacci}, {Simpson}, {Glover}  \&
  {Hernquist}}{{Kannan} et~al.}{2020a}]{Kannan2020a}
{Kannan} R.,  {Marinacci} F.,  {Simpson} C.~M.,  {Glover} S. C.~O.,
  {Hernquist} L.,  2020a, \mn@doi [\mnras] {10.1093/mnras/stz3078}, \href
  {https://ui.adsabs.harvard.edu/abs/2020MNRAS.491.2088K} {491, 2088}

\bibitem[\protect\citeauthoryear{{Kannan}, {Marinacci}, {Vogelsberger},
  {Sales}, {Torrey}, {Springel}  \& {Hernquist}}{{Kannan}
  et~al.}{2020b}]{2020MNRAS.499.5732K}
{Kannan} R.,  {Marinacci} F.,  {Vogelsberger} M.,  {Sales} L.~V.,  {Torrey} P.,
   {Springel} V.,   {Hernquist} L.,  2020b, \mn@doi [\mnras]
  {10.1093/mnras/staa3249}, \href
  {https://ui.adsabs.harvard.edu/abs/2020MNRAS.499.5732K} {499, 5732}

\bibitem[\protect\citeauthoryear{{Kannan}, {Garaldi}, {Smith}, {Pakmor},
  {Springel}, {Vogelsberger}  \& {Hernquist}}{{Kannan}
  et~al.}{2022}]{2022MNRAS.511.4005K}
{Kannan} R.,  {Garaldi} E.,  {Smith} A.,  {Pakmor} R.,  {Springel} V.,
  {Vogelsberger} M.,   {Hernquist} L.,  2022, \mn@doi [\mnras]
  {10.1093/mnras/stab3710}, \href
  {https://ui.adsabs.harvard.edu/abs/2022MNRAS.511.4005K} {511, 4005}

\bibitem[\protect\citeauthoryear{{Kannan} et~al.,}{{Kannan}
  et~al.}{2023}]{2023MNRAS.524.2594K}
{Kannan} R.,  et~al., 2023, \mn@doi [\mnras] {10.1093/mnras/stac3743}, \href
  {https://ui.adsabs.harvard.edu/abs/2023MNRAS.524.2594K} {524, 2594}

\bibitem[\protect\citeauthoryear{{Karakas}}{{Karakas}}{2010}]{2010MNRAS.403.1413K}
{Karakas} A.~I.,  2010, \mn@doi [\mnras] {10.1111/j.1365-2966.2009.16198.x},
  \href {https://ui.adsabs.harvard.edu/abs/2010MNRAS.403.1413K} {403, 1413}

\bibitem[\protect\citeauthoryear{{Katz}}{{Katz}}{1992}]{1992ApJ...391..502K}
{Katz} N.,  1992, \mn@doi [\apj] {10.1086/171366}, \href
  {https://ui.adsabs.harvard.edu/abs/1992ApJ...391..502K} {391, 502}

\bibitem[\protect\citeauthoryear{{Katz} et~al.,}{{Katz}
  et~al.}{2022}]{2022arXiv221104626K}
{Katz} H.,  et~al., 2022, \mn@doi [MNRAS submitted]
  {10.48550/arXiv.2211.04626}, \href
  {https://ui.adsabs.harvard.edu/abs/2022arXiv221104626K} {p. arXiv:2211.04626}

\bibitem[\protect\citeauthoryear{{Kennicutt}}{{Kennicutt}}{1998}]{1998ApJ...498..541K}
{Kennicutt} Robert~C. J.,  1998, \mn@doi [\apj] {10.1086/305588}, \href
  {https://ui.adsabs.harvard.edu/abs/1998ApJ...498..541K} {498, 541}

\bibitem[\protect\citeauthoryear{{Kim} \& {Ostriker}}{{Kim} \&
  {Ostriker}}{2015}]{2015ApJ...802...99K}
{Kim} C.-G.,  {Ostriker} E.~C.,  2015, \mn@doi [\apj]
  {10.1088/0004-637X/802/2/99}, \href
  {https://ui.adsabs.harvard.edu/abs/2015ApJ...802...99K} {802, 99}

\bibitem[\protect\citeauthoryear{{Kim} et~al.,}{{Kim}
  et~al.}{2021a}]{2021MNRAS.504..487K}
{Kim} J.,  et~al., 2021a, \mn@doi [\mnras] {10.1093/mnras/stab878}, \href
  {https://ui.adsabs.harvard.edu/abs/2021MNRAS.504..487K} {504, 487}

\bibitem[\protect\citeauthoryear{{Kim}, {Ostriker}  \& {Filippova}}{{Kim}
  et~al.}{2021b}]{2021ApJ...911..128K}
{Kim} J.-G.,  {Ostriker} E.~C.,   {Filippova} N.,  2021b, \mn@doi [\apj]
  {10.3847/1538-4357/abe934}, \href
  {https://ui.adsabs.harvard.edu/abs/2021ApJ...911..128K} {911, 128}

\bibitem[\protect\citeauthoryear{{Kim}, {Gong}, {Kim}  \& {Ostriker}}{{Kim}
  et~al.}{2023a}]{2023ApJS..264...10K}
{Kim} J.-G.,  {Gong} M.,  {Kim} C.-G.,   {Ostriker} E.~C.,  2023a, \mn@doi
  [\apjs] {10.3847/1538-4365/ac9b1d}, \href
  {https://ui.adsabs.harvard.edu/abs/2023ApJS..264...10K} {264, 10}

\bibitem[\protect\citeauthoryear{{Kim}, {Jeon}, {Choi}, {Richstein}, {Sacchi}
  \& {Kallivayalil}}{{Kim} et~al.}{2023b}]{2023ApJ...959...31K:Kim}
{Kim} J.,  {Jeon} M.,  {Choi} Y.,  {Richstein} H.,  {Sacchi} E.,
  {Kallivayalil} N.,  2023b, \mn@doi [\apj] {10.3847/1538-4357/acfe08}, \href
  {https://ui.adsabs.harvard.edu/abs/2023ApJ...959...31K} {959, 31}

\bibitem[\protect\citeauthoryear{{Kim} et~al.,}{{Kim}
  et~al.}{2024}]{2024arXiv240519227K:Kim}
{Kim} C.-G.,  et~al., 2024, \mn@doi [ApJ in press] {10.48550/arXiv.2405.19227},
  \href {https://ui.adsabs.harvard.edu/abs/2024arXiv240519227K} {p.
  arXiv:2405.19227}

\bibitem[\protect\citeauthoryear{{Kroupa}, {Weidner}, {Pflamm-Altenburg},
  {Thies}, {Dabringhausen}, {Marks}  \& {Maschberger}}{{Kroupa}
  et~al.}{2013}]{2013pss5.book..115K}
{Kroupa} P.,  {Weidner} C.,  {Pflamm-Altenburg} J.,  {Thies} I.,
  {Dabringhausen} J.,  {Marks} M.,   {Maschberger} T.,  2013, in {Oswalt}
  T.~D.,  {Gilmore} G.,  eds, , Vol.~5, Planets, Stars and Stellar Systems.
  Volume 5: Galactic Structure and Stellar Populations.
p.~115, \mn@doi{10.1007/978-94-007-5612-0_4}

\bibitem[\protect\citeauthoryear{{Kruijssen} et~al.,}{{Kruijssen}
  et~al.}{2019}]{2019Natur.569..519K}
{Kruijssen} J.~M.~D.,  et~al., 2019, \mn@doi [\nat]
  {10.1038/s41586-019-1194-3}, \href
  {https://ui.adsabs.harvard.edu/abs/2019Natur.569..519K} {569, 519}

\bibitem[\protect\citeauthoryear{{Krumholz}, {McKee}  \&
  {Bland-Hawthorn}}{{Krumholz} et~al.}{2019}]{2019ARA&A..57..227K}
{Krumholz} M.~R.,  {McKee} C.~F.,   {Bland-Hawthorn} J.,  2019, \mn@doi [\araa]
  {10.1146/annurev-astro-091918-104430}, \href
  {https://ui.adsabs.harvard.edu/abs/2019ARA&A..57..227K} {57, 227}

\bibitem[\protect\citeauthoryear{{Lada} \& {Lada}}{{Lada} \&
  {Lada}}{2003}]{2003ARA&A..41...57L}
{Lada} C.~J.,  {Lada} E.~A.,  2003, \mn@doi [\araa]
  {10.1146/annurev.astro.41.011802.094844}, \href
  {https://ui.adsabs.harvard.edu/abs/2003ARA&A..41...57L} {41, 57}

\bibitem[\protect\citeauthoryear{{Lada}, {Forbrich}, {Lombardi}  \&
  {Alves}}{{Lada} et~al.}{2012}]{2012ApJ...745..190L}
{Lada} C.~J.,  {Forbrich} J.,  {Lombardi} M.,   {Alves} J.~F.,  2012, \mn@doi
  [\apj] {10.1088/0004-637X/745/2/190}, \href
  {https://ui.adsabs.harvard.edu/abs/2012ApJ...745..190L} {745, 190}

\bibitem[\protect\citeauthoryear{{Lah{\'e}n}, {Naab}, {Johansson}, {Elmegreen},
  {Hu}  \& {Walch}}{{Lah{\'e}n} et~al.}{2019}]{2019ApJ...879L..18L}
{Lah{\'e}n} N.,  {Naab} T.,  {Johansson} P.~H.,  {Elmegreen} B.,  {Hu} C.-Y.,
  {Walch} S.,  2019, \mn@doi [\apjl] {10.3847/2041-8213/ab2a13}, \href
  {https://ui.adsabs.harvard.edu/abs/2019ApJ...879L..18L} {879, L18}

\bibitem[\protect\citeauthoryear{{Lah{\'e}n}, {Naab}, {Johansson}, {Elmegreen},
  {Hu}, {Walch}, {Steinwandel}  \& {Moster}}{{Lah{\'e}n}
  et~al.}{2020}]{2020ApJ...891....2L}
{Lah{\'e}n} N.,  {Naab} T.,  {Johansson} P.~H.,  {Elmegreen} B.,  {Hu} C.-Y.,
  {Walch} S.,  {Steinwandel} U.~P.,   {Moster} B.~P.,  2020, \mn@doi [\apj]
  {10.3847/1538-4357/ab7190}, \href
  {https://ui.adsabs.harvard.edu/abs/2020ApJ...891....2L} {891, 2}

\bibitem[\protect\citeauthoryear{{Lah{\'e}n} et~al.,}{{Lah{\'e}n}
  et~al.}{2023}]{2023MNRAS.522.3092L}
{Lah{\'e}n} N.,  et~al., 2023, \mn@doi [\mnras] {10.1093/mnras/stad1147}, \href
  {https://ui.adsabs.harvard.edu/abs/2023MNRAS.522.3092L} {522, 3092}

\bibitem[\protect\citeauthoryear{{Lancaster}, {Ostriker}, {Kim}  \&
  {Kim}}{{Lancaster} et~al.}{2021a}]{2021ApJ...914...89L}
{Lancaster} L.,  {Ostriker} E.~C.,  {Kim} J.-G.,   {Kim} C.-G.,  2021a, \mn@doi
  [\apj] {10.3847/1538-4357/abf8ab}, \href
  {https://ui.adsabs.harvard.edu/abs/2021ApJ...914...89L} {914, 89}

\bibitem[\protect\citeauthoryear{{Lancaster}, {Ostriker}, {Kim}  \&
  {Kim}}{{Lancaster} et~al.}{2021b}]{2021ApJ...914...90L}
{Lancaster} L.,  {Ostriker} E.~C.,  {Kim} J.-G.,   {Kim} C.-G.,  2021b, \mn@doi
  [\apj] {10.3847/1538-4357/abf8ac}, \href
  {https://ui.adsabs.harvard.edu/abs/2021ApJ...914...90L} {914, 90}

\bibitem[\protect\citeauthoryear{{Lanz} \& {Hubeny}}{{Lanz} \&
  {Hubeny}}{2003}]{2003ApJS..146..417L}
{Lanz} T.,  {Hubeny} I.,  2003, \mn@doi [\apjs] {10.1086/374373}, \href
  {https://ui.adsabs.harvard.edu/abs/2003ApJS..146..417L} {146, 417}

\bibitem[\protect\citeauthoryear{{Larkin}, {Gerasimov}  \&
  {Burgasser}}{{Larkin} et~al.}{2023}]{2023AJ....165....2L}
{Larkin} M.~M.,  {Gerasimov} R.,   {Burgasser} A.~J.,  2023, \mn@doi [\aj]
  {10.3847/1538-3881/ac9b43}, \href
  {https://ui.adsabs.harvard.edu/abs/2023AJ....165....2L} {165, 2}

\bibitem[\protect\citeauthoryear{{Larson}}{{Larson}}{2003}]{2003ASPC..287...65L}
{Larson} R.~B.,  2003, in {De Buizer} J.~M.,  {van der Bliek} N.~S.,  eds,
  Astronomical Society of the Pacific Conference Series Vol. 287, Galactic Star
  Formation Across the Stellar Mass Spectrum. pp 65--80 (\mn@eprint {arXiv}
  {astro-ph/0205466}), \mn@doi{10.48550/arXiv.astro-ph/0205466}

\bibitem[\protect\citeauthoryear{{Lee}, {Jeon}  \& {Bromm}}{{Lee}
  et~al.}{2024}]{2024MNRAS.527.1257L:Lee}
{Lee} T.,  {Jeon} M.,   {Bromm} V.,  2024, \mn@doi [\mnras]
  {10.1093/mnras/stad3198}, \href
  {https://ui.adsabs.harvard.edu/abs/2024MNRAS.527.1257L} {527, 1257}

\bibitem[\protect\citeauthoryear{{Leroy}, {Walter}, {Brinks}, {Bigiel}, {de
  Blok}, {Madore}  \& {Thornley}}{{Leroy} et~al.}{2008}]{2008AJ....136.2782L}
{Leroy} A.~K.,  {Walter} F.,  {Brinks} E.,  {Bigiel} F.,  {de Blok} W.~J.~G.,
  {Madore} B.,   {Thornley} M.~D.,  2008, \mn@doi [\aj]
  {10.1088/0004-6256/136/6/2782}, \href
  {https://ui.adsabs.harvard.edu/abs/2008AJ....136.2782L} {136, 2782}

\bibitem[\protect\citeauthoryear{{Levermore}}{{Levermore}}{1984}]{1984JQSRT..31..149L}
{Levermore} C.~D.,  1984, \mn@doi [\jqsrt] {10.1016/0022-4073(84)90112-2},
  \href {https://ui.adsabs.harvard.edu/abs/1984JQSRT..31..149L} {31, 149}

\bibitem[\protect\citeauthoryear{{Li}, {Vogelsberger}, {Marinacci}  \&
  {Gnedin}}{{Li} et~al.}{2019}]{2019MNRAS.487..364L}
{Li} H.,  {Vogelsberger} M.,  {Marinacci} F.,   {Gnedin} O.~Y.,  2019, \mn@doi
  [\mnras] {10.1093/mnras/stz1271}, \href
  {https://ui.adsabs.harvard.edu/abs/2019MNRAS.487..364L} {487, 364}

\bibitem[\protect\citeauthoryear{{Limongi} \& {Chieffi}}{{Limongi} \&
  {Chieffi}}{2018}]{2018ApJS..237...13L}
{Limongi} M.,  {Chieffi} A.,  2018, \mn@doi [\apjs] {10.3847/1538-4365/aacb24},
  \href {https://ui.adsabs.harvard.edu/abs/2018ApJS..237...13L} {237, 13}

\bibitem[\protect\citeauthoryear{{Lodders}}{{Lodders}}{2019}]{2019arXiv191200844L:Lodders}
{Lodders} K.,  2019, \mn@doi [arXiv e-prints] {10.48550/arXiv.1912.00844},
  \href {https://ui.adsabs.harvard.edu/abs/2019arXiv191200844L} {p.
  arXiv:1912.00844}

\bibitem[\protect\citeauthoryear{{Maoz}, {Mannucci}  \& {Brandt}}{{Maoz}
  et~al.}{2012}]{2012MNRAS.426.3282M:Maoz}
{Maoz} D.,  {Mannucci} F.,   {Brandt} T.~D.,  2012, \mn@doi [\mnras]
  {10.1111/j.1365-2966.2012.21871.x}, \href
  {https://ui.adsabs.harvard.edu/abs/2012MNRAS.426.3282M} {426, 3282}

\bibitem[\protect\citeauthoryear{{Marinacci}, {Sales}, {Vogelsberger}, {Torrey}
   \& {Springel}}{{Marinacci} et~al.}{2019}]{2019MNRAS.489.4233M}
{Marinacci} F.,  {Sales} L.~V.,  {Vogelsberger} M.,  {Torrey} P.,   {Springel}
  V.,  2019, \mn@doi [\mnras] {10.1093/mnras/stz2391}, \href
  {https://ui.adsabs.harvard.edu/abs/2019MNRAS.489.4233M} {489, 4233}

\bibitem[\protect\citeauthoryear{{Marks}, {Kroupa}, {Dabringhausen}  \&
  {Pawlowski}}{{Marks} et~al.}{2012}]{2012MNRAS.422.2246M}
{Marks} M.,  {Kroupa} P.,  {Dabringhausen} J.,   {Pawlowski} M.~S.,  2012,
  \mn@doi [\mnras] {10.1111/j.1365-2966.2012.20767.x}, \href
  {https://ui.adsabs.harvard.edu/abs/2012MNRAS.422.2246M} {422, 2246}

\bibitem[\protect\citeauthoryear{{Martin-Alvarez}, {Sijacki}, {Haehnelt},
  {Farcy}, {Dubois}, {Belokurov}, {Rosdahl}  \&
  {Lopez-Rodriguez}}{{Martin-Alvarez} et~al.}{2023}]{2023MNRAS.525.3806M}
{Martin-Alvarez} S.,  {Sijacki} D.,  {Haehnelt} M.~G.,  {Farcy} M.,  {Dubois}
  Y.,  {Belokurov} V.,  {Rosdahl} J.,   {Lopez-Rodriguez} E.,  2023, \mn@doi
  [\mnras] {10.1093/mnras/stad2559}, \href
  {https://ui.adsabs.harvard.edu/abs/2023MNRAS.525.3806M} {525, 3806}

\bibitem[\protect\citeauthoryear{{Mateo}}{{Mateo}}{1998}]{1998ARA&A..36..435M}
{Mateo} M.~L.,  1998, \mn@doi [\araa] {10.1146/annurev.astro.36.1.435}, \href
  {https://ui.adsabs.harvard.edu/abs/1998ARA&A..36..435M} {36, 435}

\bibitem[\protect\citeauthoryear{{McKee} \& {Ostriker}}{{McKee} \&
  {Ostriker}}{1977}]{1977ApJ...218..148M}
{McKee} C.~F.,  {Ostriker} J.~P.,  1977, \mn@doi [\apj] {10.1086/155667}, \href
  {https://ui.adsabs.harvard.edu/abs/1977ApJ...218..148M} {218, 148}

\bibitem[\protect\citeauthoryear{{McKinnon}, {Vogelsberger}, {Torrey},
  {Marinacci}  \& {Kannan}}{{McKinnon} et~al.}{2018}]{2018MNRAS.478.2851M}
{McKinnon} R.,  {Vogelsberger} M.,  {Torrey} P.,  {Marinacci} F.,   {Kannan}
  R.,  2018, \mn@doi [\mnras] {10.1093/mnras/sty1248}, \href
  {https://ui.adsabs.harvard.edu/abs/2018MNRAS.478.2851M} {478, 2851}

\bibitem[\protect\citeauthoryear{{McKinnon}, {Kannan}, {Vogelsberger},
  {O'Neil}, {Torrey}  \& {Li}}{{McKinnon} et~al.}{2021}]{2021MNRAS.502.1344M}
{McKinnon} R.,  {Kannan} R.,  {Vogelsberger} M.,  {O'Neil} S.,  {Torrey} P.,
  {Li} H.,  2021, \mn@doi [\mnras] {10.1093/mnras/stab021}, \href
  {https://ui.adsabs.harvard.edu/abs/2021MNRAS.502.1344M} {502, 1344}

\bibitem[\protect\citeauthoryear{{Meynet} \& {Maeder}}{{Meynet} \&
  {Maeder}}{2005}]{2005A&A...429..581M}
{Meynet} G.,  {Maeder} A.,  2005, \mn@doi [\aap] {10.1051/0004-6361:20047106},
  \href {https://ui.adsabs.harvard.edu/abs/2005A&A...429..581M} {429, 581}

\bibitem[\protect\citeauthoryear{{Moster}, {Naab}  \& {White}}{{Moster}
  et~al.}{2013}]{2013MNRAS.428.3121M}
{Moster} B.~P.,  {Naab} T.,   {White} S. D.~M.,  2013, \mn@doi [\mnras]
  {10.1093/mnras/sts261}, \href
  {https://ui.adsabs.harvard.edu/abs/2013MNRAS.428.3121M} {428, 3121}

\bibitem[\protect\citeauthoryear{{Nomoto}, {Iwamoto}, {Nakasato}, {Thielemann},
  {Brachwitz}, {Tsujimoto}, {Kubo}  \& {Kishimoto}}{{Nomoto}
  et~al.}{1997}]{1997NuPhA.621..467N:Nomoto}
{Nomoto} K.,  {Iwamoto} K.,  {Nakasato} N.,  {Thielemann} F.~K.,  {Brachwitz}
  F.,  {Tsujimoto} T.,  {Kubo} Y.,   {Kishimoto} N.,  1997, \mn@doi [\nphysa]
  {10.1016/S0375-9474(97)00291-1}, \href
  {https://ui.adsabs.harvard.edu/abs/1997NuPhA.621..467N} {621, 467}

\bibitem[\protect\citeauthoryear{{Ostriker} \& {Kim}}{{Ostriker} \&
  {Kim}}{2022}]{2022ApJ...936..137O}
{Ostriker} E.~C.,  {Kim} C.-G.,  2022, \mn@doi [\apj]
  {10.3847/1538-4357/ac7de2}, \href
  {https://ui.adsabs.harvard.edu/abs/2022ApJ...936..137O} {936, 137}

\bibitem[\protect\citeauthoryear{{Ostriker} \& {Shetty}}{{Ostriker} \&
  {Shetty}}{2011}]{2011ApJ...731...41O}
{Ostriker} E.~C.,  {Shetty} R.,  2011, \mn@doi [\apj]
  {10.1088/0004-637X/731/1/41}, \href
  {https://ui.adsabs.harvard.edu/abs/2011ApJ...731...41O} {731, 41}

\bibitem[\protect\citeauthoryear{{Ostriker}, {McKee}  \& {Leroy}}{{Ostriker}
  et~al.}{2010}]{2010ApJ...721..975O}
{Ostriker} E.~C.,  {McKee} C.~F.,   {Leroy} A.~K.,  2010, \mn@doi [\apj]
  {10.1088/0004-637X/721/2/975}, \href
  {https://ui.adsabs.harvard.edu/abs/2010ApJ...721..975O} {721, 975}

\bibitem[\protect\citeauthoryear{{Pakmor}, {Bauer}  \& {Springel}}{{Pakmor}
  et~al.}{2011}]{2011MNRAS.418.1392P:Pakmor}
{Pakmor} R.,  {Bauer} A.,   {Springel} V.,  2011, \mn@doi [\mnras]
  {10.1111/j.1365-2966.2011.19591.x}, \href
  {https://ui.adsabs.harvard.edu/abs/2011MNRAS.418.1392P} {418, 1392}

\bibitem[\protect\citeauthoryear{{Pakmor}, {Springel}, {Bauer}, {Mocz},
  {Munoz}, {Ohlmann}, {Schaal}  \& {Zhu}}{{Pakmor}
  et~al.}{2016}]{2016MNRAS.455.1134P}
{Pakmor} R.,  {Springel} V.,  {Bauer} A.,  {Mocz} P.,  {Munoz} D.~J.,
  {Ohlmann} S.~T.,  {Schaal} K.,   {Zhu} C.,  2016, \mn@doi [\mnras]
  {10.1093/mnras/stv2380}, \href
  {https://ui.adsabs.harvard.edu/abs/2016MNRAS.455.1134P} {455, 1134}

\bibitem[\protect\citeauthoryear{{Parmentier}, {Kauffmann}, {Pillai}  \&
  {Menten}}{{Parmentier} et~al.}{2011}]{2011MNRAS.416..783P}
{Parmentier} G.,  {Kauffmann} J.,  {Pillai} T.,   {Menten} K.~M.,  2011,
  \mn@doi [\mnras] {10.1111/j.1365-2966.2011.19096.x}, \href
  {https://ui.adsabs.harvard.edu/abs/2011MNRAS.416..783P} {416, 783}

\bibitem[\protect\citeauthoryear{{Peter}, {Klessen}, {Kanschat}, {Glover}  \&
  {Bastian}}{{Peter} et~al.}{2023}]{Peter2023}
{Peter} T.,  {Klessen} R.~S.,  {Kanschat} G.,  {Glover} S. C.~O.,   {Bastian}
  P.,  2023, \mn@doi [\mnras] {10.1093/mnras/stac3034}, \href
  {https://ui.adsabs.harvard.edu/abs/2023MNRAS.519.4263P} {519, 4263}

\bibitem[\protect\citeauthoryear{{Petkova} \& {Springel}}{{Petkova} \&
  {Springel}}{2009}]{2009MNRAS.396.1383P}
{Petkova} M.,  {Springel} V.,  2009, \mn@doi [\mnras]
  {10.1111/j.1365-2966.2009.14843.x}, \href
  {https://ui.adsabs.harvard.edu/abs/2009MNRAS.396.1383P} {396, 1383}

\bibitem[\protect\citeauthoryear{{Pillepich} et~al.,}{{Pillepich}
  et~al.}{2018}]{2018MNRAS.473.4077P}
{Pillepich} A.,  et~al., 2018, \mn@doi [\mnras] {10.1093/mnras/stx2656}, \href
  {https://ui.adsabs.harvard.edu/abs/2018MNRAS.473.4077P} {473, 4077}

\bibitem[\protect\citeauthoryear{{Ploeckinger} \& {Schaye}}{{Ploeckinger} \&
  {Schaye}}{2020}]{2020MNRAS.497.4857P}
{Ploeckinger} S.,  {Schaye} J.,  2020, \mn@doi [\mnras]
  {10.1093/mnras/staa2172}, \href
  {https://ui.adsabs.harvard.edu/abs/2020MNRAS.497.4857P} {497, 4857}

\bibitem[\protect\citeauthoryear{{Portegies Zwart}, {McMillan}  \&
  {Gieles}}{{Portegies Zwart} et~al.}{2010}]{2010ARA&A..48..431P}
{Portegies Zwart} S.~F.,  {McMillan} S. L.~W.,   {Gieles} M.,  2010, \mn@doi
  [\araa] {10.1146/annurev-astro-081309-130834}, \href
  {https://ui.adsabs.harvard.edu/abs/2010ARA&A..48..431P} {48, 431}

\bibitem[\protect\citeauthoryear{{Portinari}, {Chiosi}  \&
  {Bressan}}{{Portinari} et~al.}{1998}]{1998A&A...334..505P}
{Portinari} L.,  {Chiosi} C.,   {Bressan} A.,  1998, \mn@doi [\aap]
  {10.48550/arXiv.astro-ph/9711337}, \href
  {https://ui.adsabs.harvard.edu/abs/1998A&A...334..505P} {334, 505}

\bibitem[\protect\citeauthoryear{{Prgomet}, {Rey}, {Andersson}, {Segovia
  Otero}, {Agertz}, {Renaud}, {Pontzen}  \& {Read}}{{Prgomet}
  et~al.}{2022}]{2022MNRAS.513.2326P}
{Prgomet} M.,  {Rey} M.~P.,  {Andersson} E.~P.,  {Segovia Otero} A.,  {Agertz}
  O.,  {Renaud} F.,  {Pontzen} A.,   {Read} J.~I.,  2022, \mn@doi [\mnras]
  {10.1093/mnras/stac1074}, \href
  {https://ui.adsabs.harvard.edu/abs/2022MNRAS.513.2326P} {513, 2326}

\bibitem[\protect\citeauthoryear{{Rahmati}, {Pawlik}, {Rai{\v{c}}evi{\'c}}  \&
  {Schaye}}{{Rahmati} et~al.}{2013}]{2013MNRAS.430.2427R}
{Rahmati} A.,  {Pawlik} A.~H.,  {Rai{\v{c}}evi{\'c}} M.,   {Schaye} J.,  2013,
  \mn@doi [\mnras] {10.1093/mnras/stt066}, \href
  {https://ui.adsabs.harvard.edu/abs/2013MNRAS.430.2427R} {430, 2427}

\bibitem[\protect\citeauthoryear{{Rathjen} et~al.,}{{Rathjen}
  et~al.}{2021}]{2021MNRAS.504.1039R}
{Rathjen} T.-E.,  et~al., 2021, \mn@doi [\mnras] {10.1093/mnras/stab900}, \href
  {https://ui.adsabs.harvard.edu/abs/2021MNRAS.504.1039R} {504, 1039}

\bibitem[\protect\citeauthoryear{{R{\'e}my-Ruyer} et~al.,}{{R{\'e}my-Ruyer}
  et~al.}{2014}]{2014A&A...563A..31R}
{R{\'e}my-Ruyer} A.,  et~al., 2014, \mn@doi [\aap]
  {10.1051/0004-6361/201322803}, \href
  {https://ui.adsabs.harvard.edu/abs/2014A&A...563A..31R} {563, A31}

\bibitem[\protect\citeauthoryear{{Renaud}, {Romeo}  \& {Agertz}}{{Renaud}
  et~al.}{2021}]{2021MNRAS.508..352R}
{Renaud} F.,  {Romeo} A.~B.,   {Agertz} O.,  2021, \mn@doi [\mnras]
  {10.1093/mnras/stab2604}, \href
  {https://ui.adsabs.harvard.edu/abs/2021MNRAS.508..352R} {508, 352}

\bibitem[\protect\citeauthoryear{{Revaz}, {Arnaudon}, {Nichols}, {Bonvin}  \&
  {Jablonka}}{{Revaz} et~al.}{2016}]{2016A&A...588A..21R:Revaz}
{Revaz} Y.,  {Arnaudon} A.,  {Nichols} M.,  {Bonvin} V.,   {Jablonka} P.,
  2016, \mn@doi [\aap] {10.1051/0004-6361/201526438}, \href
  {https://ui.adsabs.harvard.edu/abs/2016A&A...588A..21R} {588, A21}

\bibitem[\protect\citeauthoryear{{Richings} \& {Schaye}}{{Richings} \&
  {Schaye}}{2016}]{2016MNRAS.458..270R}
{Richings} A.~J.,  {Schaye} J.,  2016, \mn@doi [\mnras] {10.1093/mnras/stw327},
  \href {https://ui.adsabs.harvard.edu/abs/2016MNRAS.458..270R} {458, 270}

\bibitem[\protect\citeauthoryear{{Rosdahl}, {Blaizot}, {Aubert}, {Stranex}  \&
  {Teyssier}}{{Rosdahl} et~al.}{2013}]{2013MNRAS.436.2188R}
{Rosdahl} J.,  {Blaizot} J.,  {Aubert} D.,  {Stranex} T.,   {Teyssier} R.,
  2013, \mn@doi [\mnras] {10.1093/mnras/stt1722}, \href
  {https://ui.adsabs.harvard.edu/abs/2013MNRAS.436.2188R} {436, 2188}

\bibitem[\protect\citeauthoryear{{Rosdahl}, {Schaye}, {Teyssier}  \&
  {Agertz}}{{Rosdahl} et~al.}{2015}]{2015MNRAS.451...34R}
{Rosdahl} J.,  {Schaye} J.,  {Teyssier} R.,   {Agertz} O.,  2015, \mn@doi
  [\mnras] {10.1093/mnras/stv937}, \href
  {https://ui.adsabs.harvard.edu/abs/2015MNRAS.451...34R} {451, 34}

\bibitem[\protect\citeauthoryear{{Rosen} \& {Bregman}}{{Rosen} \&
  {Bregman}}{1995}]{1995ApJ...440..634R}
{Rosen} A.,  {Bregman} J.~N.,  1995, \mn@doi [\apj] {10.1086/175303}, \href
  {https://ui.adsabs.harvard.edu/abs/1995ApJ...440..634R} {440, 634}

\bibitem[\protect\citeauthoryear{{Roychowdhury}, {Huang}, {Kauffmann}, {Wang}
  \& {Chengalur}}{{Roychowdhury} et~al.}{2015}]{2015MNRAS.449.3700R}
{Roychowdhury} S.,  {Huang} M.-L.,  {Kauffmann} G.,  {Wang} J.,   {Chengalur}
  J.~N.,  2015, \mn@doi [\mnras] {10.1093/mnras/stv515}, \href
  {https://ui.adsabs.harvard.edu/abs/2015MNRAS.449.3700R} {449, 3700}

\bibitem[\protect\citeauthoryear{{Schaerer}}{{Schaerer}}{2002}]{Schaerer02}
{Schaerer} D.,  2002, \mn@doi [\aap] {10.1051/0004-6361:20011619}, \href
  {https://ui.adsabs.harvard.edu/abs/2002A&A...382...28S} {382, 28}

\bibitem[\protect\citeauthoryear{{Schaye}}{{Schaye}}{2004}]{2004ApJ...609..667S}
{Schaye} J.,  2004, \mn@doi [\apj] {10.1086/421232}, \href
  {https://ui.adsabs.harvard.edu/abs/2004ApJ...609..667S} {609, 667}

\bibitem[\protect\citeauthoryear{{Schaye} et~al.,}{{Schaye}
  et~al.}{2015}]{2015MNRAS.446..521S}
{Schaye} J.,  et~al., 2015, \mn@doi [\mnras] {10.1093/mnras/stu2058}, \href
  {https://ui.adsabs.harvard.edu/abs/2015MNRAS.446..521S} {446, 521}

\bibitem[\protect\citeauthoryear{{Schinnerer} \& {Leroy}}{{Schinnerer} \&
  {Leroy}}{2024}]{2024arXiv240319843S:Schinnerer}
{Schinnerer} E.,  {Leroy} A.~K.,  2024, \mn@doi [ARA\&A in press]
  {10.48550/arXiv.2403.19843}, \href
  {https://ui.adsabs.harvard.edu/abs/2024arXiv240319843S} {p. arXiv:2403.19843}

\bibitem[\protect\citeauthoryear{{Schmidt}}{{Schmidt}}{1959}]{1959ApJ...129..243S}
{Schmidt} M.,  1959, \mn@doi [\apj] {10.1086/146614}, \href
  {https://ui.adsabs.harvard.edu/abs/1959ApJ...129..243S} {129, 243}

\bibitem[\protect\citeauthoryear{{Secunda}, {Cen}, {Kimm}, {G{\"o}tberg}  \&
  {de Mink}}{{Secunda} et~al.}{2020}]{2020ApJ...901...72S:Secunda}
{Secunda} A.,  {Cen} R.,  {Kimm} T.,  {G{\"o}tberg} Y.,   {de Mink} S.~E.,
  2020, \mn@doi [\apj] {10.3847/1538-4357/abaefa}, \href
  {https://ui.adsabs.harvard.edu/abs/2020ApJ...901...72S} {901, 72}

\bibitem[\protect\citeauthoryear{{Semenov}, {Kravtsov}  \& {Gnedin}}{{Semenov}
  et~al.}{2017}]{2017ApJ...845..133S}
{Semenov} V.~A.,  {Kravtsov} A.~V.,   {Gnedin} N.~Y.,  2017, \mn@doi [\apj]
  {10.3847/1538-4357/aa8096}, \href
  {https://ui.adsabs.harvard.edu/abs/2017ApJ...845..133S} {845, 133}

\bibitem[\protect\citeauthoryear{{Shi}, {Helou}, {Yan}, {Armus}, {Wu},
  {Papovich}  \& {Stierwalt}}{{Shi} et~al.}{2011}]{2011ApJ...733...87S}
{Shi} Y.,  {Helou} G.,  {Yan} L.,  {Armus} L.,  {Wu} Y.,  {Papovich} C.,
  {Stierwalt} S.,  2011, \mn@doi [\apj] {10.1088/0004-637X/733/2/87}, \href
  {https://ui.adsabs.harvard.edu/abs/2011ApJ...733...87S} {733, 87}

\bibitem[\protect\citeauthoryear{{Shi} et~al.,}{{Shi}
  et~al.}{2018}]{2018ApJ...853..149S}
{Shi} Y.,  et~al., 2018, \mn@doi [\apj] {10.3847/1538-4357/aaa3e6}, \href
  {https://ui.adsabs.harvard.edu/abs/2018ApJ...853..149S} {853, 149}

\bibitem[\protect\citeauthoryear{{Skinner} \& {Ostriker}}{{Skinner} \&
  {Ostriker}}{2013}]{2013ApJS..206...21S}
{Skinner} M.~A.,  {Ostriker} E.~C.,  2013, \mn@doi [\apjs]
  {10.1088/0067-0049/206/2/21}, \href
  {https://ui.adsabs.harvard.edu/abs/2013ApJS..206...21S} {206, 21}

\bibitem[\protect\citeauthoryear{{Smith}}{{Smith}}{2014}]{2014ARA&A..52..487S}
{Smith} N.,  2014, \mn@doi [\araa] {10.1146/annurev-astro-081913-040025}, \href
  {https://ui.adsabs.harvard.edu/abs/2014ARA&A..52..487S} {52, 487}

\bibitem[\protect\citeauthoryear{{Smith}}{{Smith}}{2021}]{2021MNRAS.502.5417S}
{Smith} M.~C.,  2021, \mn@doi [\mnras] {10.1093/mnras/stab291}, \href
  {https://ui.adsabs.harvard.edu/abs/2021MNRAS.502.5417S} {502, 5417}

\bibitem[\protect\citeauthoryear{{Smith} et~al.,}{{Smith}
  et~al.}{2017}]{2017MNRAS.466.2217S}
{Smith} B.~D.,  et~al., 2017, \mn@doi [\mnras] {10.1093/mnras/stw3291}, \href
  {https://ui.adsabs.harvard.edu/abs/2017MNRAS.466.2217S} {466, 2217}

\bibitem[\protect\citeauthoryear{{Smith}, {Kannan}, {Tsang}, {Vogelsberger}  \&
  {Pakmor}}{{Smith} et~al.}{2020}]{2020ApJ...905...27S}
{Smith} A.,  {Kannan} R.,  {Tsang} B. T.~H.,  {Vogelsberger} M.,   {Pakmor} R.,
   2020, \mn@doi [\apj] {10.3847/1538-4357/abc47e}, \href
  {https://ui.adsabs.harvard.edu/abs/2020ApJ...905...27S} {905, 27}

\bibitem[\protect\citeauthoryear{{Smith}, {Bryan}, {Somerville}, {Hu},
  {Teyssier}, {Burkhart}  \& {Hernquist}}{{Smith}
  et~al.}{2021}]{2021MNRAS.506.3882S}
{Smith} M.~C.,  {Bryan} G.~L.,  {Somerville} R.~S.,  {Hu} C.-Y.,  {Teyssier}
  R.,  {Burkhart} B.,   {Hernquist} L.,  2021, \mn@doi [\mnras]
  {10.1093/mnras/stab1896}, \href
  {https://ui.adsabs.harvard.edu/abs/2021MNRAS.506.3882S} {506, 3882}

\bibitem[\protect\citeauthoryear{{Smith}, {Kannan}, {Garaldi}, {Vogelsberger},
  {Pakmor}, {Springel}  \& {Hernquist}}{{Smith} et~al.}{2022}]{Smith2022}
{Smith} A.,  {Kannan} R.,  {Garaldi} E.,  {Vogelsberger} M.,  {Pakmor} R.,
  {Springel} V.,   {Hernquist} L.,  2022, \mn@doi [\mnras]
  {10.1093/mnras/stac713}, \href
  {https://ui.adsabs.harvard.edu/abs/2022MNRAS.512.3243S} {512, 3243}

\bibitem[\protect\citeauthoryear{{Springel}}{{Springel}}{2010}]{2010MNRAS.401..791S}
{Springel} V.,  2010, \mn@doi [\mnras] {10.1111/j.1365-2966.2009.15715.x},
  \href {https://ui.adsabs.harvard.edu/abs/2010MNRAS.401..791S} {401, 791}

\bibitem[\protect\citeauthoryear{{Springel}, {Di Matteo}  \&
  {Hernquist}}{{Springel} et~al.}{2005}]{2005MNRAS.361..776S}
{Springel} V.,  {Di Matteo} T.,   {Hernquist} L.,  2005, \mn@doi [\mnras]
  {10.1111/j.1365-2966.2005.09238.x}, \href
  {https://ui.adsabs.harvard.edu/abs/2005MNRAS.361..776S} {361, 776}

\bibitem[\protect\citeauthoryear{{Steinwandel} \& {Goldberg}}{{Steinwandel} \&
  {Goldberg}}{2023}]{2023arXiv231011495S}
{Steinwandel} U.~P.,  {Goldberg} J.~A.,  2023, \mn@doi [ApJ submitted]
  {10.48550/arXiv.2310.11495}, \href
  {https://ui.adsabs.harvard.edu/abs/2023arXiv231011495S} {p. arXiv:2310.11495}

\bibitem[\protect\citeauthoryear{{Steinwandel}, {Moster}, {Naab}, {Hu}  \&
  {Walch}}{{Steinwandel} et~al.}{2020}]{2020MNRAS.495.1035S:Steinwandel}
{Steinwandel} U.~P.,  {Moster} B.~P.,  {Naab} T.,  {Hu} C.-Y.,   {Walch} S.,
  2020, \mn@doi [\mnras] {10.1093/mnras/staa821}, \href
  {https://ui.adsabs.harvard.edu/abs/2020MNRAS.495.1035S} {495, 1035}

\bibitem[\protect\citeauthoryear{{Steinwandel}, {Bryan}, {Somerville},
  {Hayward}  \& {Burkhart}}{{Steinwandel} et~al.}{2023}]{2023MNRAS.526.1408S}
{Steinwandel} U.~P.,  {Bryan} G.~L.,  {Somerville} R.~S.,  {Hayward} C.~C.,
  {Burkhart} B.,  2023, \mn@doi [\mnras] {10.1093/mnras/stad2744}, \href
  {https://ui.adsabs.harvard.edu/abs/2023MNRAS.526.1408S} {526, 1408}

\bibitem[\protect\citeauthoryear{{Steinwandel}, {Rennehan}, {Orr}, {Fielding}
  \& {Kim}}{{Steinwandel} et~al.}{2024}]{2024arXiv240714599S:Steinwandel}
{Steinwandel} U.~P.,  {Rennehan} D.,  {Orr} M.~E.,  {Fielding} D.~B.,   {Kim}
  C.-G.,  2024, \mn@doi [ApJ submitted] {10.48550/arXiv.2407.14599}, \href
  {https://ui.adsabs.harvard.edu/abs/2024arXiv240714599S} {p. arXiv:2407.14599}

\bibitem[\protect\citeauthoryear{{Su} et~al.,}{{Su}
  et~al.}{2018}]{2018MNRAS.480.1666S}
{Su} K.-Y.,  et~al., 2018, \mn@doi [\mnras] {10.1093/mnras/sty1928}, \href
  {https://ui.adsabs.harvard.edu/abs/2018MNRAS.480.1666S} {480, 1666}

\bibitem[\protect\citeauthoryear{{Sugimura}, {Ricotti}, {Park}, {Garcia}  \&
  {Yajima}}{{Sugimura} et~al.}{2024}]{2024ApJ...970...14S:Sugimura}
{Sugimura} K.,  {Ricotti} M.,  {Park} J.,  {Garcia} F. A.~B.,   {Yajima} H.,
  2024, \mn@doi [\apj] {10.3847/1538-4357/ad499a}, \href
  {https://ui.adsabs.harvard.edu/abs/2024ApJ...970...14S} {970, 14}

\bibitem[\protect\citeauthoryear{{Sukhbold}, {Ertl}, {Woosley}, {Brown}  \&
  {Janka}}{{Sukhbold} et~al.}{2016}]{2016ApJ...821...38S}
{Sukhbold} T.,  {Ertl} T.,  {Woosley} S.~E.,  {Brown} J.~M.,   {Janka} H.~T.,
  2016, \mn@doi [\apj] {10.3847/0004-637X/821/1/38}, \href
  {https://ui.adsabs.harvard.edu/abs/2016ApJ...821...38S} {821, 38}

\bibitem[\protect\citeauthoryear{{Sun} et~al.,}{{Sun}
  et~al.}{2023}]{2023ApJ...945L..19S}
{Sun} J.,  et~al., 2023, \mn@doi [\apjl] {10.3847/2041-8213/acbd9c}, \href
  {https://ui.adsabs.harvard.edu/abs/2023ApJ...945L..19S} {945, L19}

\bibitem[\protect\citeauthoryear{{Tang}, {Bressan}, {Rosenfield}, {Slemer},
  {Marigo}, {Girardi}  \& {Bianchi}}{{Tang}
  et~al.}{2014}]{2014MNRAS.445.4287T:Tang}
{Tang} J.,  {Bressan} A.,  {Rosenfield} P.,  {Slemer} A.,  {Marigo} P.,
  {Girardi} L.,   {Bianchi} L.,  2014, \mn@doi [\mnras]
  {10.1093/mnras/stu2029}, \href
  {https://ui.adsabs.harvard.edu/abs/2014MNRAS.445.4287T} {445, 4287}

\bibitem[\protect\citeauthoryear{{Tanikawa}, {Yoshida}, {Kinugawa}, {Takahashi}
   \& {Umeda}}{{Tanikawa} et~al.}{2020}]{2020MNRAS.495.4170T}
{Tanikawa} A.,  {Yoshida} T.,  {Kinugawa} T.,  {Takahashi} K.,   {Umeda} H.,
  2020, \mn@doi [\mnras] {10.1093/mnras/staa1417}, \href
  {https://ui.adsabs.harvard.edu/abs/2020MNRAS.495.4170T} {495, 4170}

\bibitem[\protect\citeauthoryear{{Tielens} \& {Hollenbach}}{{Tielens} \&
  {Hollenbach}}{1985}]{1985ApJ...291..747T}
{Tielens} A.~G.~G.~M.,  {Hollenbach} D.,  1985, \mn@doi [\apj]
  {10.1086/163112}, \href
  {https://ui.adsabs.harvard.edu/abs/1985ApJ...291..747T} {291, 747}

\bibitem[\protect\citeauthoryear{{Truelove}, {Klein}, {McKee}, {Holliman},
  {Howell}  \& {Greenough}}{{Truelove} et~al.}{1997}]{1997ApJ...489L.179T}
{Truelove} J.~K.,  {Klein} R.~I.,  {McKee} C.~F.,  {Holliman} John~H. I.,
  {Howell} L.~H.,   {Greenough} J.~A.,  1997, \mn@doi [\apjl] {10.1086/310975},
  \href {https://ui.adsabs.harvard.edu/abs/1997ApJ...489L.179T} {489, L179}

\bibitem[\protect\citeauthoryear{{Tsai}, {Chen}, {Whalen}, {Ou}  \&
  {Woods}}{{Tsai} et~al.}{2023}]{2023ApJ...951...84T:Tsai}
{Tsai} S.-H.,  {Chen} K.-J.,  {Whalen} D.,  {Ou} P.-S.,   {Woods} T.~E.,  2023,
  \mn@doi [\apj] {10.3847/1538-4357/acd936}, \href
  {https://ui.adsabs.harvard.edu/abs/2023ApJ...951...84T} {951, 84}

\bibitem[\protect\citeauthoryear{{Vink} \& {Sander}}{{Vink} \&
  {Sander}}{2021}]{2021MNRAS.504.2051V}
{Vink} J.~S.,  {Sander} A. A.~C.,  2021, \mn@doi [\mnras]
  {10.1093/mnras/stab902}, \href
  {https://ui.adsabs.harvard.edu/abs/2021MNRAS.504.2051V} {504, 2051}

\bibitem[\protect\citeauthoryear{{Vink}, {de Koter}  \& {Lamers}}{{Vink}
  et~al.}{2001}]{2001A&A...369..574V}
{Vink} J.~S.,  {de Koter} A.,   {Lamers} H.~J.~G.~L.~M.,  2001, \mn@doi [\aap]
  {10.1051/0004-6361:20010127}, \href
  {https://ui.adsabs.harvard.edu/abs/2001A&A...369..574V} {369, 574}

\bibitem[\protect\citeauthoryear{Virtanen et~al.,}{Virtanen
  et~al.}{2020}]{2020SciPy-NMeth}
Virtanen P.,  et~al., 2020, \mn@doi [Nature Methods]
  {10.1038/s41592-019-0686-2}, \href {https://rdcu.be/b08Wh} {17, 261}

\bibitem[\protect\citeauthoryear{{Vogelsberger}, {Genel}, {Sijacki}, {Torrey},
  {Springel}  \& {Hernquist}}{{Vogelsberger}
  et~al.}{2013}]{2013MNRAS.436.3031V}
{Vogelsberger} M.,  {Genel} S.,  {Sijacki} D.,  {Torrey} P.,  {Springel} V.,
  {Hernquist} L.,  2013, \mn@doi [\mnras] {10.1093/mnras/stt1789}, \href
  {https://ui.adsabs.harvard.edu/abs/2013MNRAS.436.3031V} {436, 3031}

\bibitem[\protect\citeauthoryear{{Vogelsberger} et~al.,}{{Vogelsberger}
  et~al.}{2014}]{2014MNRAS.444.1518V}
{Vogelsberger} M.,  et~al., 2014, \mn@doi [\mnras] {10.1093/mnras/stu1536},
  \href {https://ui.adsabs.harvard.edu/abs/2014MNRAS.444.1518V} {444, 1518}

\bibitem[\protect\citeauthoryear{{Vogelsberger}, {Marinacci}, {Torrey}  \&
  {Puchwein}}{{Vogelsberger} et~al.}{2020}]{2020NatRP...2...42V}
{Vogelsberger} M.,  {Marinacci} F.,  {Torrey} P.,   {Puchwein} E.,  2020,
  \mn@doi [Nature Reviews Physics] {10.1038/s42254-019-0127-2}, \href
  {https://ui.adsabs.harvard.edu/abs/2020NatRP...2...42V} {2, 42}

\bibitem[\protect\citeauthoryear{{Walch}, {W{\"u}nsch}, {Burkert}, {Glover}  \&
  {Whitworth}}{{Walch} et~al.}{2011}]{2011ApJ...733...47W}
{Walch} S.,  {W{\"u}nsch} R.,  {Burkert} A.,  {Glover} S.,   {Whitworth} A.,
  2011, \mn@doi [\apj] {10.1088/0004-637X/733/1/47}, \href
  {https://ui.adsabs.harvard.edu/abs/2011ApJ...733...47W} {733, 47}

\bibitem[\protect\citeauthoryear{{Wang}, {Dutton}, {Stinson}, {Macci{\`o}},
  {Penzo}, {Kang}, {Keller}  \& {Wadsley}}{{Wang}
  et~al.}{2015}]{2015MNRAS.454...83W}
{Wang} L.,  {Dutton} A.~A.,  {Stinson} G.~S.,  {Macci{\`o}} A.~V.,  {Penzo} C.,
   {Kang} X.,  {Keller} B.~W.,   {Wadsley} J.,  2015, \mn@doi [\mnras]
  {10.1093/mnras/stv1937}, \href
  {https://ui.adsabs.harvard.edu/abs/2015MNRAS.454...83W} {454, 83}

\bibitem[\protect\citeauthoryear{{Wechsler} \& {Tinker}}{{Wechsler} \&
  {Tinker}}{2018}]{2018ARA&A..56..435W}
{Wechsler} R.~H.,  {Tinker} J.~L.,  2018, \mn@doi [\araa]
  {10.1146/annurev-astro-081817-051756}, \href
  {https://ui.adsabs.harvard.edu/abs/2018ARA&A..56..435W} {56, 435}

\bibitem[\protect\citeauthoryear{{Weinberger}, {Springel}  \&
  {Pakmor}}{{Weinberger} et~al.}{2020}]{2020ApJS..248...32W:Weinberger}
{Weinberger} R.,  {Springel} V.,   {Pakmor} R.,  2020, \mn@doi [\apjs]
  {10.3847/1538-4365/ab908c}, \href
  {https://ui.adsabs.harvard.edu/abs/2020ApJS..248...32W} {248, 32}

\bibitem[\protect\citeauthoryear{{Whitworth} \& {Jaffa}}{{Whitworth} \&
  {Jaffa}}{2018}]{2018A&A...611A..20W}
{Whitworth} A.~P.,  {Jaffa} S.~E.,  2018, \mn@doi [\aap]
  {10.1051/0004-6361/201731871}, \href
  {https://ui.adsabs.harvard.edu/abs/2018A&A...611A..20W} {611, A20}

\bibitem[\protect\citeauthoryear{{Windhorst} et~al.,}{{Windhorst}
  et~al.}{2018}]{Windhorst2018}
{Windhorst} R.~A.,  et~al., 2018, \mn@doi [\apjs] {10.3847/1538-4365/aaa760},
  \href {https://ui.adsabs.harvard.edu/abs/2018ApJS..234...41W} {234, 41}

\bibitem[\protect\citeauthoryear{{Wise} \& {Abel}}{{Wise} \&
  {Abel}}{2011}]{2011MNRAS.414.3458W}
{Wise} J.~H.,  {Abel} T.,  2011, \mn@doi [\mnras]
  {10.1111/j.1365-2966.2011.18646.x}, \href
  {https://ui.adsabs.harvard.edu/abs/2011MNRAS.414.3458W} {414, 3458}

\bibitem[\protect\citeauthoryear{{Wolfire}, {Hollenbach}, {McKee}, {Tielens}
  \& {Bakes}}{{Wolfire} et~al.}{1995}]{1995ApJ...443..152W}
{Wolfire} M.~G.,  {Hollenbach} D.,  {McKee} C.~F.,  {Tielens} A.~G.~G.~M.,
  {Bakes} E.~L.~O.,  1995, \mn@doi [\apj] {10.1086/175510}, \href
  {https://ui.adsabs.harvard.edu/abs/1995ApJ...443..152W} {443, 152}

\bibitem[\protect\citeauthoryear{{Wolfire}, {McKee}, {Hollenbach}  \&
  {Tielens}}{{Wolfire} et~al.}{2003}]{2003ApJ...587..278W}
{Wolfire} M.~G.,  {McKee} C.~F.,  {Hollenbach} D.,   {Tielens} A.~G.~G.~M.,
  2003, \mn@doi [\apj] {10.1086/368016}, \href
  {https://ui.adsabs.harvard.edu/abs/2003ApJ...587..278W} {587, 278}

\bibitem[\protect\citeauthoryear{{Xu}}{{Xu}}{1995}]{1995ApJS...98..355X}
{Xu} G.,  1995, \mn@doi [\apjs] {10.1086/192166}, \href
  {https://ui.adsabs.harvard.edu/abs/1995ApJS...98..355X} {98, 355}

\bibitem[\protect\citeauthoryear{{Yan}, {Jerabkova}  \& {Kroupa}}{{Yan}
  et~al.}{2023}]{2023A&A...670A.151Y}
{Yan} Z.,  {Jerabkova} T.,   {Kroupa} P.,  2023, \mn@doi [\aap]
  {10.1051/0004-6361/202244919}, \href
  {https://ui.adsabs.harvard.edu/abs/2023A&A...670A.151Y} {670, A151}

\bibitem[\protect\citeauthoryear{{Yang}, {Mo}  \& {van den Bosch}}{{Yang}
  et~al.}{2003}]{2003MNRAS.339.1057Y}
{Yang} X.,  {Mo} H.~J.,   {van den Bosch} F.~C.,  2003, \mn@doi [\mnras]
  {10.1046/j.1365-8711.2003.06254.x}, \href
  {https://ui.adsabs.harvard.edu/abs/2003MNRAS.339.1057Y} {339, 1057}

\bibitem[\protect\citeauthoryear{{Yates}, {Hendriks}, {Vijayan}, {Izzard},
  {Thomas}  \& {Das}}{{Yates} et~al.}{2024}]{2024MNRAS.527.6292Y:Yates}
{Yates} R.~M.,  {Hendriks} D.,  {Vijayan} A.~P.,  {Izzard} R.~G.,  {Thomas}
  P.~A.,   {Das} P.,  2024, \mn@doi [\mnras] {10.1093/mnras/stad3419}, \href
  {https://ui.adsabs.harvard.edu/abs/2024MNRAS.527.6292Y} {527, 6292}

\bibitem[\protect\citeauthoryear{{de Blok} \& {McGaugh}}{{de Blok} \&
  {McGaugh}}{1997}]{1997MNRAS.290..533D}
{de Blok} W.~J.~G.,  {McGaugh} S.~S.,  1997, \mn@doi [\mnras]
  {10.1093/mnras/290.3.533}, \href
  {https://ui.adsabs.harvard.edu/abs/1997MNRAS.290..533D} {290, 533}

\makeatother
\end{thebibliography}

\begin{appendix}
\section{Treatments of cooling channels}
\label{sec:cooling_app}
In this appendix, we provide a detailed description of the various cooling channels in our model.
\subsection{Primordial cooling}

The term ``primordial cooling'' incorporates photoionization, recombination, free–free (bremsstrahlung),
bound–bound and bound–free collisions of the H and He species tracked in the nonequilibrium network. In this work, we use the primordial thermochemistry and cooling equations described by equations~(49)--(51) of \cite{2019MNRAS.485..117K}
and equations~(2)--(5) of \cite{2020MNRAS.499.5732K}. We adopt the rate coefficients accounting for cooling/heating of atomic hydrogen and helium from \cite{1992ApJS...78..341C}. We model the cooling due to molecular hydrogen by 
\begin{equation}
n^2\Lambda_\text{\ce{H2}}=\Lambda_{\text{\HI\ce{H2}},n\rightarrow0}(T)n_\text{\HI}n_\text{\ce{H2}}+\Lambda_{\text{\ce{H2}\ce{H2}},n\rightarrow0}(T)n_\text{\ce{H2}}n_\text{\ce{H2}} \, ,
\end{equation}
where $\Lambda_{\text{\HI\ce{H2}},n\rightarrow0}$ and $\Lambda_{\text{\ce{H2}\ce{H2}},n\rightarrow0}$ are the \HI–\ce{H2} and \ce{H2}–\ce{H2} collisional cooling coefficients in the low-density limit $(n\rightarrow0)$ from \cite{1979ApJS...41..555H}. We temporarily ignore UV pumping heating and \ce{H2} formation heating as they are subdominant in our current framework.

\subsection{Metal collisional excitation lines in hot gas}
\label{sec:CIE}
The metal-line cooling in $\gtrsim10^5$\,K hot gas is modelled assuming ionization equilibrium in a UVB radiation field given by \cite{2009ApJ...703.1416F} with a density-dependent self-shielding factor from \cite{2013MNRAS.430.2427R}. Given the density and temperature of a gas cell, we look up this cooling rate from a pre-calculated {\sc cloudy} table (see \citealt{2013MNRAS.436.3031V} for more details). 

\subsection{Nebular lines in photoionized gas}
In photoionized gas with sub-solar abundance ($Z>0.1\Zsun$), the collisionally excited forbidden lines of metal ions become important coolants. We adopt the fitting formula given by equation~(47) of \cite{2023ApJS..264...10K} for the nebular lines except those from \ce{C+} assuming a fixed ionization state such that 80\% (20\%) of O, N, and Ne are singly (doubly) ionized and 50\% (50\%) of S is singly (doubly) ionized:
\begin{align}
    n^2\Lambda_\text{neb}&=3.86\times10^{-23}\,\text{erg cm}^{-3}\,\text{s}^{-1} \,\notag\\
    &\times Z_\text{g}'n_en_\text{\HII}\frac{e^{-3.86/T_4}f_\text{neb}(T)}{T_4^{1/2}(1+0.12n^{0.38-0.12\ln{T_4}}_{e,2})}\,,
\end{align}
where  $Z_\text{g}'$ is the gas metallicity normalized to the solar value, $T_4=T/10^4$\,K, $n_{e,2}=n_e/10^2$\,cm$^{-3}$, and $\log_{10}{f_\text{neb}}(T)=\sum^6_{i=0}a_i
(\ln{T_4})^i$ with $\{a_0,a_1,a_2,a_3,a_4,a_5\}=\{0.692,-0.586,0.816,-0.505,0.118,7.66\times10^{-3},-5.08\times10^{-3}\}$. This function captures the temperature dependence of the collisional excitation rate of the dominant nebular coolant [\OII]3729\,\AA\,line and the critical densities of multiple lines for their collisional deexcitation in dense gas \citep[for more details, see][]{2023ApJS..264...10K}.

\subsection{Metal fine structure lines cooling in warm gas}
 The main metal coolants of warm atomic gas are fine-structure transitions of \ce{C+} and \ce{O} with minor contributions from C and other species \citep[e.g.][]{1995ApJ...443..152W,2003ApJ...587..278W}.
 As noticed by \cite{2023ApJS..264...10K},  the low-temperature ($<10^4$\,K) metal cooling models widely used by the galaxy formation community have significant discrepancies with those developed with detailed atomic/molecular physics by the ISM community. To accurately model such low-temperature metal cooling, we calculate the cooling rates of the predominant C/O fine structure lines including \CII~$158$\,$\mu$m, \CI$^*$ 610\,$\mu$m, \CI$^*$ 370\,$\mu$m, \OI$^*$ 63\,$\mu$m, and \OI$^*$ 146\,$\mu$m lines. To obtain the cooling rate of a given emission line, we calculate the equilibrium abundances of C, \ce{C+}, CO, O, and \ce{O+} under the radiation field and nonequilibrium H and He abundances tracked by \areport (see Section~\ref{sec:chemistry}), then we solve the two-/three-level population of C/O atoms/ions with the rate coefficients given by \cite{2017ApJ...843...38G}. For example, the cooling rate for the \CII~$158$\,$\mu$m two-level system is 
 \begin{equation}
      n^2\Lambda_\text{\CII} = n_\text{\CII}A_\text{ul}E_\text{ul}\frac{n_jk_\text{lu}}{n_jk_\text{lu}+n_jk_\text{ul}+A_\text{ul}}\,,
 \end{equation}
where $A_\text{ul}$ is the Einstein $A$ coefficient, $E_\text{ul}$ is the energy difference, $k_\text{ul}$ is the collisional rate coefficien and $k_\text{lu}=g_\text{u}/g_\text{l}k_\text{ul}\exp{(-E_\text{ul}/k_\text{B}T)}$, and $n_j$ is the number density of collisional partners including \ce{H2}, \HI, and free electron. The cooling rates for the three-level systems are calculated in the same way and the solution for the three-level population can be found in Section 17.5 of \cite{2011piim.book.....D}.

\begin{figure}
	\includegraphics[width=\columnwidth]{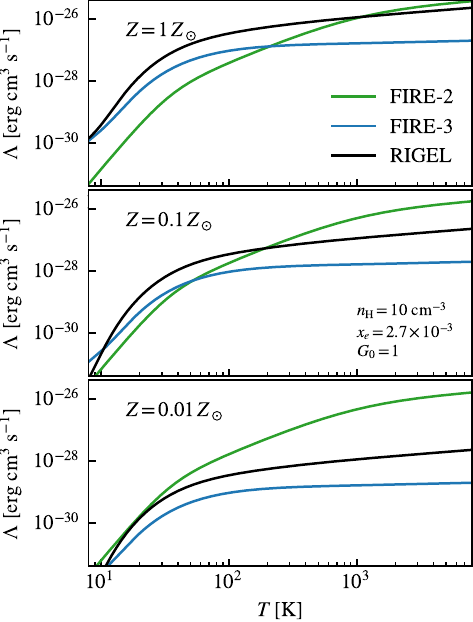}
    \caption{Cooling functions in predominantly neutral gas with $n_\text{H}=10$\,cm$^{-3}$, $x_e=2.7\times10^{-3}$, $G_0=1$  for different metallicities of $Z=1\,\Zsun$ (top panel), $0.1\,\Zsun$ (middle panel), and $0.01\,\Zsun$ (bottom panel) obtained with the FIRE-2 (green), FIRE-3 (blue), and RIGEL (black) model.}
    \label{fig:coolingcurve}
\end{figure}

In Fig.~\ref{fig:coolingcurve}, we compare the cooling functions in predominantly neutral gas with $n_\text{H}=10$\,cm$^{-3}$, $x_e=2.7\times10^{-3}$, $G_0=1$ obtained from our RIGEL model with those from the FIRE-2 \citep{2018MNRAS.480..800H} and FIRE-3 \citep{2023MNRAS.519.3154H} models as an example to show the difference. The cooling curves in different metallicities of $1\,\Zsun$, $0.1\,\Zsun$, and $0.01\,\Zsun$ are shown in the panels from top to  bottom. Both RIGEL and FIRE-3 show a strong dependence on metallicity in their cooling rates, while this dependence in the FIRE-2 model is notably weak. Consequently, the FIRE-2 model can significantly overestimate the cooling rate in metal-poor environments. This weak metallicity dependency in FIRE-2 can be attributed to the inclusion of terms such as $[1+(Z/\Zsun)]/(1+0.001\,43n_\text{H})$ in their fitting function.  

Compared to FIRE-3, RIGEL demonstrates overall higher cooling rates. This variation occurs due to two main factors. First, RIGEL incorporates the cooling effects from the \OI metal fine structure lines alongside the \CI and \CII lines present in the FIRE-3 model. Second, the species abundance and collisional rate coefficients from all these lines display intricate dependencies on temperature, the density of the collision partners, and the radiation field, while FIRE-3 simplified these dependencies to obtain a simple fitting function.

\subsection{CO cooling in cold gas}
In dense $(n\gtrsim10^3\,$\,cm$^{-3}$) molecular gas, the rotational lines of CO and other molecules become the most important coolants. We model the cooling by CO rotational lines following the approach outlined by \cite{2018A&A...611A..20W}, 
\begin{equation}
    n^2\Lambda_\text{CO} = n_\text{CO} x_\text{\ce{H2}}\left(\Lambda_\text{CO,LO}^{-1/\beta}+\Lambda_\text{CO,HI}^{-1/\beta}\right)^{-\beta},
\end{equation}
where $\Lambda_\text{CO,LO}$, $\Lambda_\text{CO,HI}$, and $\beta$ are given by
\begin{align}
    \Lambda_\text{CO,LO}&=2.16\times10^{-27}\,\text{erg s}^{-1}\,\left(\frac{n_\text{H2}}{\text{cm}^{-3}}\right)\left(\frac{T}{K}\right)^{3/2}\,,\notag\\
    \Lambda_\text{CO,HI}&=2.21\times10^{-28}\,\text{erg s}^{-1}\notag\\
    &\times\left[\frac{(x_\text{CO}/x_\text{\ce{H2}})\text{ km s}^{-1}\text{ pc}^{-1}}{|\nabla\cdot {\bm v}|}\right]^{-1}
    \left(\frac{n_\text{\ce{H2}}}{\text{cm}^{-3}}\right)^{-1}\left(\frac{T}{K}\right)^{4}\,,\notag\\
    \beta &= 1.23\left(\frac{n_\text{\ce{H2}}}{\text{cm}^{-3}}\right)^{0.0533}\left(\frac{T}{K}\right)^{0.164}\,.
\end{align}
which reproduces the detailed numerical results of \cite{1978ApJ...222..881G} for $T\in\left[10,100\right]$\,K, $n_\text{\ce{H2}}\in\left[3\times10^2,3\times10^5\right]$\,cm$^{-3}$, and $|\nabla\cdot {\bm v}|[x_\text{CO}/x_{\ce{H2}}/(3\times10^{-4})]^{-1}\in\left[10,10^6\right]$\,km\,s$^{-1}$\,pc$^{-1}$.

\subsection{Dust cooling in cold gas}
In even denser molecular gas $(n\gtrsim10^6$\,cm$^{-3}$) where molecules reach their critical densities, the dust–gas energy exchange can be an important coolant. The dust cooling function is given by \cite{1983ApJ...265..223B} and rearranged by \cite{2020MNRAS.499.5732K}:
\begin{align}
    n^2\Lambda_\text{gd} &= 1.356\times10^{-33}\,\text{erg cm}^{-3}\,\text{s}^{-1} \,\notag\\
    &\times\left( \frac{Z_\text
{d}}{0.01}\right) \left( \frac{0.1 \mu m}{a}\right) \, \sqrt{\frac{T}{\text{K}}} \, \left(\frac{T - T_\text{d}}{\text{K}}\right) \, \left(\frac{n_\text{H}}{\text{cm}^{-3}}\right)^2 \,.
\end{align}
where $T_\text{d}$ is the dust temperature which is self-consistently calculated considering the dust-radiation coupling. The size of the dust grain $a$ is fixed at $0.1$\,$\mu$m. We calculate $T_\text{d}$ by solving the instantaneous equilibrium equation of $\Lambda_\text{gd}+\Gamma_\text{dr}=0$ using Newton’s root-finding method.
The dust heating rate due to the coupling with IR radiation field is given by
\begin{equation}
    n^2\Gamma_\text{dr} = \kappa_\text{P}\rho (4\sigma T^4_\text{d}-cE_\text{IR})\,,
\end{equation}
where $\kappa_\text{P}$ is the Planck mean opacity, $\sigma$ is the Stefan-Boltzmann constant, and $E_\text{IR}$ is the energy density of IR photons.

\subsection{Photoelectric heating}
FUV photons absorbed by small dust grains and polycyclic aromatic hydrocarbons (PAHs) can excite electrons to escape and then heat the ISM, which is referred to as photoelectric heating.
As in \cite{2020MNRAS.499.5732K}, we adopt the photoelectric heating rate given by \cite{2003ApJ...587..278W} 
\begin{equation}
    n^2\Gamma_\text{PE}=1.3\times10^{-24}\,\text{erg s}^{-1}\,\text{cm}^{-3}\left(\frac{Z_\text{d}}{0.01}\right)\epsilon_\text{ff}G_0n_\text{H}\,,
\end{equation}
where $G_0$ is the FUV radiation energy density relative to \cite{1968BAN....19..421H}’s estimate ($u_\text{5.8-13.6 eV}=5.29\times10^{-14}$\,erg\,cm$^{-3}$) and
\begin{equation}
\begin{split}
\epsilon_\mathrm{ff}  &= \frac{0.0487}{1+0.004[G_0 \sqrt{T}/(0.5 n_e)]^{0.73}}\\ &+ \frac{0.0365 (T/10^4)^{0.7}}{1+2\times10^{-4} \times G_0 \sqrt{T}/(0.5 n_e)} \, .
\end{split} 
\end{equation} 
Here, the unit of temperature $T$ is [K], and the unit of electron number density $n_{e}$ is $[\rm cm^{-3}]$.

\subsection{Cosmic ray heating}
We assume constant ionization and heating rates with 0.01 times of the MW value, where the MW values \citep{2012ApJ...745...91I} are
 \begin{align}
 \label{equ:CR}
     \zeta_\text{cr,\ce{H2}}&=3.5\times10^{-16}\,\text{s}^{-1}\,\notag\\
     \zeta_\text{cr,\HI}&=1.78\times10^{-16}\,\text{s}^{-1}\,\notag\\
     \zeta_\text{cr,\HeI}&=1.1\zeta_\text{cr,\HI}\,\notag\\
     \zeta_\text{cr,\CI}&=3.85\zeta_\text{cr,\HI}\,\notag\\
     n^2\Gamma_{\text{cr},j}&=1.6022\times10^{-11}\,\text{erg}\cdot\zeta_{\text{cr},j}n_j\,,
 \end{align}
where $j\in{\text{\ce{H2}},\text{\HI},\text{\HeI}}$.

\clearpage

\section{Tables of fitting parameters}
\label{sec:tables}
In Table~\ref{age:fit}, Table~\ref{uv:fit}, and Table~\ref{wind:fit_v}, we listed the fitting parameters $x_0$ and $b_i(i=0,1,2,3,4)$ the for main-sequence lifetime (in $[\text{yr}]$), UV photon production rates in different bands (in $[\text{s}^{-1}]$), and wind mass-loss rate (in $[\Msun\,\text{yr}^{-1}]$) and velocity (in $[\text{km}\,\text{s}^{-1}$]), defined in Eq.~(\ref{qfit}), respectively. The machine-readable tables with full precision are downloadable online\footnote{\url{https://bitbucket.org/ftldengyw/rigel-tables/src/main/}}.

\begin{table}[h]
    \caption{Fit parameters for main-sequence lifetime of stars.}
    \centering
    \addtolength{\tabcolsep}{2pt}
    \renewcommand{\arraystretch}{1.1}
    \begin{tabular}{ccccccc}
        \hline
        $Z\ [\rm Z_{\odot}]$ & $b_{0}$ & $b_{1}$ & $b_{2}$ & $b_{3}$ & $x_{\rm 0}$ \\
        \hline
        $\le 10^{-8}$ & $9.88$ & $-3.78$ & $1.41$ & $-0.19$ & $2.70$\\
        $0.028$ & $10.10$ & $-3.78$ & $1.26$ & $-0.14$ & $2.08$ \\
        $0.28$ & $10.34$ & $-4.20$ & $1.49$ & $-0.17$ & $2.08$ \\
        $0.56$ & $10.39$ & $-4.28$ & $1.51$ & $-0.17$ & $2.08$ \\
        $1.41$ & $11.00$ & $-5.72$ & $2.52$ & $-0.39$ & $2.08$ \\
        $\ge 3.52$ & $11.19$ & $-6.51$ & $3.11$ & $-0.51$ & $2.08$ \\
        \hline
    \end{tabular}
    \tablefoot{We adopt the \cite{1998A&A...334..505P} model for Pop. II stars and the \citetalias{Schaerer02} model for Pop. III ($Z\leq10^{-8}\,\Zsun$) stars. We use the fit formula of Eq.~\ref{qfit} fixing $b_{4}=0$. The unit of main-sequence lifetime is $[\text{yr}]$.}
    \addtolength{\tabcolsep}{-2pt}
    \renewcommand{\arraystretch}{0.9090909090909090909}
    \label{age:fit}
\end{table}

\begin{table}
    \caption{Fit parameters for photon production rates in different bands (see Table~\ref{tab:RadiationBins}).
    }
    \centering
    \addtolength{\tabcolsep}{-0.5pt}
    \renewcommand{\arraystretch}{1.1}
    \begin{tabular}{cccccccc}
        \hline
         $Z\ [\rm Z_{\odot}]$ & $b_{0}$ & $b_{1}$ & $b_{2}$ & $b_{3}$ & $b_{4}$ & $x_{0}$ \\
        \hline
         Opt. \\
        \hline
        $\le 10^{-8}$ & $45.50$ & $2.41$ & $-0.18$ & $-0.03$ & $0$ & $2$ \\
        $10^{-6}$ & $44.63$ & $4.14$ & $-1.19$ & $0.16$ & $0$ & $2$ \\
        $10^{-5}$ & $44.45$ & $4.49$ & $-1.35$ & $-0.19$ & $0$ & $2$ \\
        $10^{-4}$ & $44.29$ & $4.81$ & $-1.51$ & $-0.21$ & $0$ & $2$ \\
        $10^{-2}$ & $44.02$ & $5.31$ & $-1.73$ & $-0.25$ & $0$ & $2$ \\
        $\ge 1$ & $43.75$ & $5.82$ & $-1.94$ & $-0.28$ & $0$ & $2$ \\
        \hline
         FUV \\
        \hline
        $\le 10^{-8}$ & $46.46$ & $0.52$ & $0.80$ & $-0.20$ & $0$ & $2$ \\
        $10^{-6}$ & $45.42$ & $2.70$ & $-0.53$ & $0.06$ & $0$ & $2$\\
        $10^{-5}$ & $45.27$ & $3.07$ & $-0.72$ & $0.09$ & $0$ & $2$\\
        $10^{-4}$ & $45.19$ & $3.32$ & $-0.85$ & $0.11$ & $0$ & $2$\\
        $10^{-2}$ & $45.15$ & $3.59$ & $-0.97$ & $0.13$ & $0$ & $2$\\
        $\ge 1$ & $45.18$ & $3.76$ & $-1.05$ & $0.14$ & $0$ & $2$\\
        \hline
        LW \\
        \hline
        $\le 10^{-8}$ & $44.79$ & $3.39$ & $-0.75$ & $0.08$ & $0$ & $2.70$\\
        $10^{-4}$ & $42.75$ & $6.38$ & $-2.22$ & $0.33$ & $0$ & $2$\\
        $\ge 0.007$ & $36.20$ & $19.20$ & $-10.73$ & $2.16$ & $0$ & $1.81$\\
        \hline
         EUV1 \\
        \hline
        $\le 10^{-8}$ & $43.50$ & $5.08$ & $-1.16$ & $0.09$ & $0$ & $2.70$ \\
        $0.001$ & $10.41$ & $83.08$ & $-68.83$ & $25.73$ & $-3.60$ & $2$ \\
        $0.0282$ & $27.43$ & $31.73$ & $-15.69$ & $2.72$ & $0$ & $2$ \\
        $\ge 1$ & $15.03$ & $62.92$ & $-46.07$ & $15.93$ & $-2.14$ & $1.93$\\
        \hline
         EUV2 \\
        \hline
        $\le 10^{-8}$ & $42.01$ & $6.62$ & $-1.54$ & $0.09$ & $0$ & $2.70$ \\
        $0.001$ & $-35.92$ & $181.9$ & $-148.7$ & $54.40$ & $-7.43$ & $1.70$ \\
        $0.0282$ & $17.71$ & $45.79$ & $-22.81$ & $3.91$ & $0$ & $2$\\
        $\ge 1$ & $-24.13$ & $120.0$ & $-66.22$ & $12.31$ & $0$ & $1.70$\\
        \hline
         EUV3 \\
        \hline
        $\le 10^{-8}$ & $26.51$ & $29.07$ & $-13.65$ & $2.32$ & $0$ & $2$ \\
        $10^{-6}$ & $28.71$ & $23.19$ & $-9.56$ & $1.45$ & $0$ & $2$\\
        $10^{-5}$ & $28.30$ & $22.96$ & $-9.20$ & $1.36$ & $0$ & $2$\\
        $10^{-4}$ & $27.25$ & $23.83$ & $-9.48$ & $1.39$ & $0$ & $2$\\
        $10^{-2}$ & $24.36$ & $26.68$ & $-10.60$ & $1.54$ & $0$ & $2$\\
        $\ge 1$ & $20.49$ & $30.77$ & $-12.31$ & $1.79$ & $0$ & $2$\\
        \hline
    \end{tabular}
    \tablefoot{Opt. ($1-5.8$~eV), FUV ($5.8-11.2$~eV), LW ($11.2-13.6$~eV), EUV1 ($13.6-24.6$~eV), EUV2 ($24.6-54.4$~eV), EUV3 ($>54.4$~eV). The unit of photon production rate is $[\text{s}^{-1}]$.}
    \addtolength{\tabcolsep}{0.5pt}
    \renewcommand{\arraystretch}{0.9090909090909090909}
    \label{uv:fit}
\end{table}

\begin{table}[h]
    \caption{Fit parameters for wind mass-loss rate and velocity.}
    \centering
    \addtolength{\tabcolsep}{-0.55pt}
    \renewcommand{\arraystretch}{1.1}
    \begin{tabular}{cccccccc}
        \hline
        Quantity & $Z\ [\rm Z_{\odot}]$ & $b_{0}$ & $b_{1}$ & $b_{2}$ & $b_{3}$ & $x_{\rm 0}$ \\
        \hline
        $\dot{m}_{\rm w}$ & $10^{-6}$ & $-19.66$ & $15.20$ & $-6.56$ & $1.06$ & $2$ \\
        $[\rm M_{\odot}\ yr^{-1}]$ & $10^{-5}$ & $-19.84$ & $15.47$ & $-6.47$ & $1.02$ & $2$ \\
        & $10^{-4}$ & $-20.13$ & $15.96$ & $-6.53$ & $1.01$ & $2$ \\
        & $10^{-2}$ & $-20.77$ & $16.94$ & $-6.69$ & $1.00$ & $2$ \\
        & $\ge 1$ & $-21.52$ & $17.98$ & $-6.88$ & $0.99$ & $2$ \\
        \hline
        $v_{\rm w}$ & $10^{-6}$ & $2.17$ & $0.49$ & $-0.14$ & $0.009$ & $2$ \\
        $[\rm km\ s^{-1}]$ & $10^{-5}$ & $2.39$ & $0.40$ & $-0.09$ & $0.0008$ & $2$\\
        & $10^{-4}$ & $2.58$ & $0.36$ & $-0.07$ & $-0.003$ & $2$\\
        & $10^{-2}$ & $2.92$ & $0.33$ & $-0.06$ & $-0.005$ & $2$\\
        & $\ge 1$ & $3.25$ & $0.33$ & $-0.06$ & $-0.005$ & $2$\\
        \hline
    \end{tabular}
    \tablefoot{The wind properties is calculated by \cite{2001A&A...369..574V,2021MNRAS.504.2051V} with the ZAMS stellar properties from \citetalias{2020MNRAS.495.4170T} including the extrapolated model for $Z=\rm Z_\odot$. We use the fit formula of Eq.~(\ref{qfit}). The units of mass-loss rate and wind velocity are $[\Msun\,\text{yr}^{-1}]$ and $[\text{km}\,\text{s}^{-1}]$, respectively.}
    \addtolength{\tabcolsep}{0.55pt}
    \renewcommand{\arraystretch}{0.9090909090909090909}
    \label{wind:fit_v}
\end{table}
\FloatBarrier

\section{Resolution of SN explosions}
\label{sec:SN_res}
\begin{figure}
	\includegraphics[width=\columnwidth]{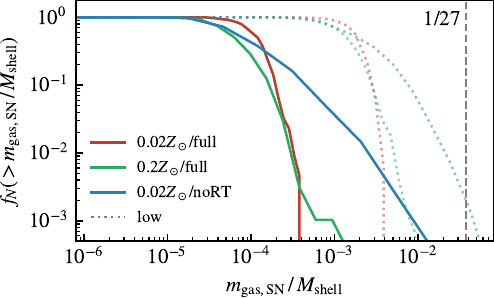}
    \caption{Cumulative distribution function of the ratio between the mass of the cell where the SNe injected explosion energy and the total swept-up mass at end of the energy conserving Sedov-Taylor phase (shell formation) given by equation~(\ref{equ:shell-formation-mass}). The red and green curves are the $1\,\Msun$ resolution results obtained with the fiducial RIGEL model, while the blue curve is obtained without radiative feedback (see Section~\ref{sec:isolateddwarf} for the naming conventions). The dotted curves are the results of the $10\,\Msun$ resolution simulations. The vertical dashed line denotes the ST-resolving criterion of $m_\text{gas,SN}\,/\,M_\text{shell}<1/27$ by \protect\cite{2015ApJ...802...99K}.}
    \label{fig:SN_resolution}
\end{figure}

To quantitatively assess the ability of our simulations to resolve the ST phase of SN explosion, we record the mass and gas density of the cell where SN energy is ejected. In Fig.~\ref{fig:SN_resolution}, we present a histogram of the ratio between the mass of the cell where the SN has injected explosion energy and the shell formation mass described by equation~(\ref{equ:shell-formation-mass}). 

As can be seen, the fiducial RIGEL model and the noRT model
at $1\,\Msun$ resolution
show a good ability to resolve the SNe. For the $10\,\Msun$ low-resolution simulations, the presence of radiative feedback ensures that all SN explosions occur in low-density gas (Section~\ref{sec:SN_Density}), where the $10\,\Msun$ resolution is sufficient to resolve all SNe. In the $0.02\Zsun$/noRT/low
simulation, a small fraction of SNe is marginally resolved, while they should have minor effects on the results.

\end{appendix}

\end{document}